\documentclass[prb,onecolumn,nofootinbib,citeautoscript,12pt,longbibliography,notitlepage]{revtex4-2}
\pdfoutput=1
\usepackage{setspace}
\usepackage{amsmath}
\usepackage{graphicx}
\usepackage{bigints}
\usepackage{dcolumn}
\usepackage{bm}
\usepackage{color}
\usepackage[tight]{subfigure}
\usepackage{array}
\usepackage[papersize={8.5in,11in}]{geometry}
\usepackage{xcolor}
\definecolor{darkblue}{rgb}{0.,0.,0.4}
\definecolor{darkred}{rgb}{0.5,0.,0.}
\definecolor{BlueViolet}{RGB}{138,43,226}
\definecolor{SkyBlue}{RGB}{30,144,255}
\definecolor{DarkGreen}{RGB}{0,100,0}
\usepackage[pdftex,colorlinks=true,linkcolor=darkblue,citecolor=red,urlcolor=darkred]{hyperref}
\usepackage{float}
\usepackage{physics}
\usepackage{cancel}
\usepackage[version=4]{mhchem}

\newcommand{\bk}{\mathbf{k}}
\newcommand{\bp}{\mathbf{p}}
\newcommand{\bq}{\mathbf{q}}

\newcommand{\beq}{\begin{equation}}
\newcommand{\eeq}{\end{equation}}
\newcommand{\bea}{\begin{eqnarray}}
\newcommand{\eea}{\end{eqnarray}}
\newcommand{\ee}{\end{equation}}

\def \nn{\nonumber \\}

\DeclareUnicodeCharacter{2212}{-}

\begin{document}

\title{Unconventional Superconductivity Mediated by  Nematic Fluctuations in a Multi-Orbital System - Application to doped FeSe}
\author{Kazi Ranjibul Islam}
\email{islam074@umn.edu}
\affiliation{School of Physics and Astronomy and William I. Fine Theoretical Physics Institute,
University of Minnesota, Minneapolis, MN 55455, USA}

\author{Andrey Chubukov*}
\email{achubuko@umn.edu}
\affiliation{School of Physics and Astronomy and William I. Fine Theoretical Physics Institute,
University of Minnesota, Minneapolis, MN 55455, USA}
\begin{abstract}
We analyze superconductivity in a multi-orbital fermionic system near the onset of a nematic order, using doped FeSe as an example. We associate nematicity with a spontaneous polarization between $d_{\text{xz}}$ and $d_{\text{yz}}$ orbitals (a Pomeranchuk-type order) and analyze the pairing mediated by soft nematic fluctuations.  Such a pairing gives rise to a highly anisotropic gap function whose structure strongly varies with temperature, and leads to strongly non-BCS behavior in thermodynamics, spectroscopy and transport.  We compute the specific heat and its directional variation  with a magnetic field,  magnetic susceptibility,  density of states, tunneling conductance, Raman intensity, superfluid stiffness and penetration depth
  without and with impurity scattering and for the latter computed also optical conductivity and $T_c$ variation.  We find good agreement with the existing data for FeSe$_{1-x}$S$_x$  and FeSe$_{1-x}$Te$_x$ and suggest new experiments.
\end{abstract}

\maketitle
\section{Introduction}
Unconventional superconductivity is one of the proliferate areas of
modern day research in the  condensed matter physics.
Compared to the conventional superconductors, where the pairing glue is provided by lattice vibrations a.k.a phonon \citep{maxwell1950isotope,serin1950mass,rowell1963image,reynolds1950superconductivity,giaever1960electron}, in unconventional superconductors the pairing mechanism is largely believed to be of electronic origin and in many cases can be thought of coming from  soft
fluctuations of a particle-hole order parameter in either spin or charge channel. The most common and well studied example here is pairing mediated by antiferromagnetic spin fluctuations. This mechanism works best when superconductivity develops near the onset of a spin density wave (SDW) order with  a large momentum $\mathbf{Q}$ ($\mathbf{Q}=(\pi,\pi)$ in 2D). Antiferromagnetic fluctuations originate from a screened Coulomb interaction and are nominally repulsive for pairing. Yet, because $\mathbf{Q}$ is large,  superconductivity does develop with a momentum transfer $\mathbf{Q}$ (the pair hopping from $(\bk, -\bk)$ to $\bk+\mathbf{Q}, -\bk-\mathbf{Q}$). This promotes a sign changing order parameter (gap function) between Fermi points at $\bk_F$ and $\bk_F + \mathbf{Q}$. For cuprates, this leads to $d_{x^2-y^2}$ pairing with four sign changes of the gap along the Fermi surface. For  Fe-pnictides, this mechanism promotes the strongest attraction in $s^{+-}$ channel, with a sign change between the gap function on hole and electron pockets. In both cases, there is still a repulsion at small momentum transfer, which is detrimental to any superconductivity. However, when spin fluctuations are strong, a repulsion at $Q$ is stronger than that at small momentum, and a sign-changing superconducting  order does develop.

 For Fe-pnictides,  superconductivity does develop in the vicinity of a $(\pi,\pi)$ antiferromagnetic  order (in the physical 2Fe Brillouin zone), which holds at weak hole/electron/isovalent doping \citep{fernandes2022iron}. The $s^{+-}$ gap symmetry in these materials has been verified by the observation of a spin resonance peak below $T_\text{c}$ \citep{christianson2008unconventional,lumsden2009two,chi2009inelastic,
 inosov2010normal,li2009spin,parshall2009spin}, and by STM measurements \citep{hanaguri2010unconventional}. In Fe-chalcogenide FeSe, long-range antiferromagnetic order is not present at ambient pressure, but the magnetic correlation length is large, and superconducting order parameter has the same $s^{+-}$ symmetry  as in Fe-pnictides.  It has been argued that superconductivity in FeSe is still mediated by spin fluctuations.  The three key evidences are (i) the correlation between NMR $1/T_1$ and superconducting $T_\text{c}$ \citep{imai2009does}, (ii) the consistency  between ARPES data on the gap anisotropy and calculations within spin fluctuation scenario \citep{liu2018orbital,hashimoto2018superconducting, classen2017interplay,xing2017competing}, and (iii)  the fact that a magnetic order rapidly develops under pressure\citep{glasbrenner2015effect,wang2016magnetic,sun2016dome}.

 Situation is different, however, in FeSe doped by either S or Te (FeSe$_{1-x}$S$_x$ and FeSe$_{1-x}$Te$_x$).
 A pure FeSe  possesses long-range nematic order below $T_p \sim 85K$. Upon doping, $T_p$ decreases and vanishes
 at critical $x_\text{c}$ (0.17 for S doping and 0.53 for Te doping).
  It has been argued \citep{hanaguri2018two,shibauchi2020exotic,sato2018abrupt,ishida2022pure,mukasa2023enhanced}
that superconductivity near and above
 $x_c$ is qualitatively different from the one in pure and weakly doped FeSe ($x < x_c$) and does not fit into
spin-fluctuation
scenario.
This argument is  based on a number of experimental/theoretical  evidences:
\begin{itemize}
\item
 The gap structure changes qualitatively across the nematic critical point. For pure and weakly doped FeSe, superconducting gap is nearly isotropic and have minima along the Fe-Fe bond direction \citep{xu2016highly,liu2018orbital,sprau2017discovery}.
Near and above $x_c$, the
 gap on the hole pocket becomes highly anisotropic with deep minima at $45^\circ$ with respect to the
 direction towards electron pockets and  maxima along the direction towards electron pockets \citep{nagashima2022discovery,
walker2023electronic,nag2024highly}
\item
 Doping dependence of $T_\text{c}$ in FeSe$_{1-x}$Te$_x$ is non-monotonic with maxima at $x =0$ and $x \geq
x_\text{c}$. The non-monotonicity becomes more prominent in the presence of a magnetic field: with increasing field \citep{mukasa2023enhanced}, the superconducting region  gradually shrinks to a dome centered slightly above $x_\text{c}$.
\item
The specific heat of FeSe$_{1-x}$S$_x$ shows highly unconventional behavior
 at $x>x_c$ \citep{sato2018abrupt,mizukami2021thermodynamics,mizukami2023unusual}.  The data clearly indicate that the jump of $C_v(T)$ at $T_\text{c} (x)$ decreases with increasing $x$ and vanishes at around $x_\text{c}$.  At smaller $T$,  The specific heat coefficient $\gamma (T) = C_v(T)/T$  passes through a maximum at around $0.8 T_\text{c}$ and then decreases nearly linearly apparently towards  a finite value at $T=0$. Although such a behavior has been depicted at finite $T$ down to roughly
$T \sim 0.1\, T_c$, it clearly indicates the presence of  low energy excitations deep inside the superconducting state. This is in sharp contract with a near-perfect BCS-like behavior of $\gamma (T)$ in pure FeSe. The  authors of Ref.\citep{hanaguri2018two} argued that non-BCS  behavior of the specific heat coefficient at $x>x_c$  is not  caused by disorder because residual resistivity does not exhibit a noticeable increase around $x_\text{c}$ (Ref. \citep{hosoi2016nematic}). Other experiments~\citep{coldea2019evolution} also indicated that the effects attributed to disorder get weaker with increasing $x$.
 \item
 Tunneling conductance  measurements~\citep{mizukami2021thermodynamics,nag2024highly}
 at $T \sim \, 0.1 T_c$ show  a large residual density of states (DOS), almost half of the normal state contribution, complimenting specific heat measurements.
\item
Measured superfluid density~\citep{matsuura2023two}  in FeSe$_{1-x}$S$_x$ for $x \geq x_\text{c}$ at low temperatures is significantly lower from the pure case, which is another evidence for low-lying excitations deep inside the superconducting state.
\item
 Recent $\mu$SR experiments~\citep{matsuura2023two,roppongi2023} presented evidence for time-reversal symmetry breaking in FeSe. The $\mu$SR signal is present below $T_\text{c}$  for all $x$, however in FeSe$_{1-x}$Te$_x$ it clearly increases above $x_\text{c}$.  This raises a possibility that the superconducting state at $x > x_\text{c}$ breaks time-reversal symmetry.
\end{itemize}

The presence of low-lying excitations at $T \ll T_c$ was interpreted by several group as a potential evidence of an exotic pairing that creates a Bogoliubov Fermi surface in a superconducting state~\citep{agterberg2017bogoliubov,setty2020topological,nagashima2022discovery}. Below we explore another scenario
 that superconductivity in doped FeSe near $x_\text{c}$ is mediated by nematic fluctuations.
We put forward this scenario recently in Ref. \citep{islam2024unconventional},
 where we analyzed the mechanism of nematic-fluctuation-mediated superconductivity (NFMS) and the gap structure.
 The central result of that study is that the gap structure is highly anisotropic. On a hole Fermi surface it is the largest in the direction towards electron pockets (hot spots on the Fermi surface) and is very small in between these directions. Right at the nematic critical point, the gap emerges at $T_\text{c}$ only at hot spots and extends at smaller $T$ into finite size arcs. The arcs length grows as $T$ decreases, but as long as $T$ is finite, there exist cold regions where the gap vanishes, i.e., the system preserves pieces of the original Fermi surface. At $T=0$, the gap opens everywhere except at the cold spots, but is still exponentially small near them.

In this communication we
  analyze the feedback of NFMS on the electronic properties in the superconducting state and obtain thermodynamic, spectroscopic and transport characteristics of such a superconductor, including
 the specific heat $C_v (T)$, the density of states $N_0(\mathbf{H})$ and its variation under a magnetic field,
Knight shift $\chi_s(T)$, superfluid density $\rho_s(T)$, and dynamical tunneling conductance $N(\Omega)$ and Raman intensity $I(\Omega)$ and compare the feedback from NFMS with that from conventional $s-$wave and $d-$wave  superconductivity. We also analyze the effect of  non-magnetic impurities and obtain impurity-induced variation of $T_c$ and optical conductivity $\sigma (\Omega)$.  We compare our results with the existing experiments and suggest new ones.

For the convenience of a reader we present the summary of the results which we obtain in this paper.

\subsection{Summary of the results}

\begin{itemize}
\item
Specific heat coefficient $C_\text{v}/T$, does not display a jump at $T_\text{c}$ and instead increases as $(T_\text{c}-T)^{1/2}$, passes through a maximum at $T \sim 0.8 T_\text{c}$, and  behaves at smaller  $T$ like there is a non-zero residual $C_\text{v}/T$ at $T \to 0$. In reality, $C_\text{v}/T$ vanishes at $T=0$, but nearly discontinuously, as $1/(\log(T_\text{c}/T))^{1/2}$ (see Fig. \ref{specific heat}).

\item
We study the Volovik effect~\citep{volovik1993superconductivity} at zero temperature by calculating the modulation of the residual density of states $N_0(H,\phi)$ with varying angle $\phi$ of an in-plane magnetic field $H$ with respect to the Fe-Fe bond (see Fig. \ref{sp heat with Magnetic field}(a,b)). For an anisotropic gap, measuring the Volovik effect reveals the interplay between the magnetic field direction $\phi$ and the gap structure, especially around its nodal positions. We find a qualitatively different angular modulation of residual DOS for NFMS from the d-wave gap. This is tied to the fact that near the cold spots, NFMS gap is highly flat, suppressed exponentially while for d-wave it varies linearly. For a d-wave superconductor, residual DOS is maximum when $\phi$ is directed along the hot spots  and minimum when it is along the cold spots. This is because in the first case, contribution for the DOS come from all cold spots equally  while in the second case, only two of the cold spots directed along the magnetic field contribute. For the nematic fluctuation mediated superconductor, the angular modulation of DOS with $\phi$ is more complex with maxima happening at both the hot and cold spots (hot spots with slightly higher DOS) and a minima present somewhere in between. In this case, extremely flat gap structure near the cold spots makes all of $4$ cold spots contributing equally to the DOS for any direction of magnetic field $\phi$.
\item
Similar behavior is also seen in static uniform magnetic susceptibility $\chi_s(T)$. Below the transition temperature, it decreases much slower compared to the s-wave and d-wave gaps and at smaller $T$ it behaves like there is a non-zero residual $\chi_s$, but falls abruptly to zero like $1/(\log(T_\text{c}/T))^{1/2}$ (see Fig. \ref{Spin Susceptibility fig}.
\item
 We compute the tunneling conductance and find that at low energy it behaves as if it saturates to a finite value. The magnitude of this seemingly residual DOS can be almost half or more than that of the normal state DOS depending on the value of the pairing strength. But in reality, DOS vanishes at zero energy, buy nearly discontinuously as  $1/(\log(T_\text{c}/T))^{1/2}$ (see Fig. \ref{zero bias fig}).

\item
The  superfluid density $\rho_s(T)$ also shows a highly exotic behavior. Contrary to the s- and d- wave, it increases much slower near the transition point as $(T_c-T)^{3/2}$, at low temperature it behaves as if it saturates at a value smaller than unity indicating that some fermions remain unpaired, but raises even at lower temperature abruptly almost like a vertical function and become equal to unity (see Fig. \ref{SF pure}(a,b)).

\item
 Raman intensity $I(\Omega)$ in $B_{1g}$ channel as a function of frequency $\Omega$ is shown in Fig. \ref{Raman figure}.  At large frequency, $\Omega \gg \Delta_0$, where $\Delta_0$ is the gap amplitude, the functional form of $I(\Omega)$  is qualitatively similar to s- and d-wave,  but the low frequency behavior is distinctly different. Namely, $I(\Omega)$  falls at a much lower rate compared to the s- and d- wave at small frequency and vanishes very fast as $ 1/(\log \dfrac{2\Delta_0}{\Omega})^{5/2}$. This is not surprising that the unusual behavior of these quantities at low temperature or low frequency is dictated by only one quantity: the density of states, especially its functional form at low energy $N(\Omega)$ which in turn is governed by the structure of the gap function near the cold spots.

\item Effect of non-magnetic impurity scattering on the nematic fluctuation mediated superconductor is discussed in detail. In contrast to an isotropic s-wave gap, the transition temperature of NFMS declines with increasing impurity scattering rate $\Gamma_0$, but never goes to zero, rather it saturates at high $\Gamma_0$ (Fig. \ref{Tc}). Impurity scattering also makes the gap function progressively less anisotropic compared to the clean NFMS and lifts the nodes at the cold spots completely. As a result, the behavior of NFMS resembles more like a conventional BCS s-wave, particularly at large values of scattering $\Gamma_0$. We capture this cross-over in the evolution of gap structure (Fig.~\ref{psi sol}), spectral function (Fig.~\ref{spectral function diagram}), superfluid density (Fig.~\ref{Sf with disorder}) and optical conductivity $\sigma(\omega)$(Fig. \ref{Conductivity Diagram}) with increasing disorder strength.
\end{itemize}

The structure of the paper is the following. In Sec.~\ref{Model Hamiltonian section}, we define our model Hamiltonian and briefly outline the derivation for the nematic susceptibility and nematic fluctuation mediated pairing interaction for a multi-pocket multi-orbital system. In Sec.~\ref{fingerprint section}, we focus on the pure case and compute the gap structure, specific heat, field induced modulation of the specific heat, DC magnetic susceptibility, tunneling conductance, density of states, superfluid density and Raman Intensity. In Sec.~\ref{disorder section}, we primarily focus on the impurity effect where we solve the gap equation in the presence of the impurity scattering, and use the solution to compute the correction to the transition temperature, low-energy excitation, spectral function, superfluid density and optical conductivity. In Sec.~\ref{comparison with experiment section}, we discuss our theoretical predictions and the current experimental results.
\section{Model Hamiltonian and Pairing Interaction}
 \label{Model Hamiltonian section}

 \subsection{Model}~~~The electronic structure of pure/doped FeSe in the tetragonal phase consists of two non-equal hole pockets, centered at $\Gamma$, and two electron pockets centered at $X = (\pi, 0)$ and $Y = (0, \pi)$ in the $1$FeBZ.  The hole pockets are composed of $d_\text{xz}$ and $d_\text{yz}$ fermions, the X pocket is composed of
$d_\text{yz}$ and $d_{xy}$ fermions, and the $Y$ pocket is composed of $d_\text{xz}$ and $d_{xy}$ fermions.
The inner hole pocket is quite small and likely does not play much role for nematic order and superconductivity. We assume that heavy $d_{xy}$ fermions also do not play much role and  consider an effective two-orbital model with a single $d_\text{xz}/d_\text{yz}$ circular hole pocket, and mono-orbital electron pockets ($d_\text{yz}$ X-pocket and $d_\text{xz}$ Y-pocket). We define fermionic operators for mono-orbital Y and X pockets  as $f_1$ and $f_2$, respectively
 ($f_{1,\bk,\sigma} = d_{xz, \bk+Y, \sigma}$, $f_{2,\bk,\sigma} = d_{yz, \bk+X, \sigma}$).
 The band operator for the hole pocket is \citep{fernandes2016low}  \begin{align}
 h_{\bk,\sigma} = \cos{\theta_\bk} d_{yz, \bk, \sigma} + \sin{\theta_\bk} d_{xz, \bk, \sigma}.
 \label{orbital to band}
 \end{align} The kinetic energy is quadratic in fermionic densities and have the following expression \begin{align}
     H_\text{Kin}= \sum_{\bk,\sigma} \xi_h(\bk)\, h_{\bk,\sigma}^\dagger h_{\bk,\sigma}+\xi_{Y}(\bk) f_{1,\bk,\sigma}^\dagger f_{1,\bk,\sigma}+\xi_{X}(\bk) f_{2,\bk,\sigma}^\dagger f_{2,\bk,\sigma},
 \end{align} where $\xi_i(\bk)$ is isotropic dispersion of the band index $i \,(=h,X,Y)$.

 There are $14$ distinct $C_4$-symmetric interactions involving low-energy $d_{xz}/d_{yz}$ orbital states near $\Gamma,X$ and $Y$ pockets. We present these interactions in Appendix \ref{Appendix A} in orbital and band representations.  These interactions
  can be separated into density-density, exchange and pair-hopping interactions.  In the band basis, density-density interactions are the ones with a small momentum transfer within a hole or an electron pocket, while the exchange and pair-hopping interactions are the ones with momentum transfer comparable to the distance between hole and electron pockets.  For our analysis of superconductivity near the end point of a $q=0$ nematic order, only density-density interactions are relevant. These 4-fermion interactions are described by
  \begin{align}
     H^\text{den}_\text{int}=\dfrac{1}{2}\sum_{i,j,\bk,\bp,\sigma,\sigma'} \rho_i(\bk,\bq) \rho_j(\bp,-\bq) V_{i,j}^\text{den}(\bk,\bk+\bq;\bp,\bp-\bq),
     \label{denisty-density interaction}
 \end{align} where $\rho_{i}(\bk,\bq)=c^\dagger_{i,\bk}c_{i,\bk+\bq}$  is the density operator for the $i^\text{th}$ band ($=h,X,Y$) and the interaction vertices  $V^\text{den}_{i,j}(\bk,\bk+\bq;\bp,\bp-\bq)$ are
 \begin{align}
 \label{Vhh}
V_{h,h}^{\text{den}}=& U_4  \Big\{\sin\theta_\bk \sin\theta_{\bk+\bq}\sin\theta_\bp\sin\theta_{\bp-\bq}+\cos\theta_\bk \cos\theta_{\bk+\bq}\cos\theta_\bp\cos\theta_{\bp-\bq}\Big\}\nn & +\tilde{U}_4\Big\{\sin\theta_\bk \sin\theta_{\bk+\bq}\cos\theta_\bp\cos\theta_{\bp-\bq}+\cos\theta_\bk \cos\theta_{\bk+\bq}\sin\theta_\bp\sin\theta_{\bp-\bq}\Big\}\nn & +\tilde{\tilde{U}}_4\Big\{\sin\theta_\bk \cos\theta_{\bk+\bq}\cos\theta_\bp\sin\theta_{\bp-\bq}+\cos\theta_\bk \sin\theta_{\bk+\bq}\sin\theta_\bp\cos\theta_{\bp-\bq}\Big\}\nn & + \bar{U}_4\Big\{\sin\theta_\bk \cos\theta_{\bk+\bq}\sin\theta_\bp\cos\theta_{\bp-\bq}+\cos\theta_\bk \sin\theta_{\bk+\bq}\cos\theta_\bp\sin\theta_{\bp-\bq}\Big\} ,\\
V_{h,Y}^{\text{den}}&= U_1 \, \sin\theta_\bk \sin\theta_{\bk+\bq}+\bar{U}_1\, \cos\theta_\bk \cos\theta_{\bk+\bq},\\
V_{h,X}^{\text{den}}&= U_1 \, \cos\theta_\bk \cos\theta_{\bk+\bq}+\bar{U}_1\, \sin\theta_\bk \sin\theta_{\bk+\bq},\\
V_{X,X}^\text{den}&=V_{Y,Y}^\text{den}=U_5,\quad V_{X,Y}^\text{den}=\tilde{U}_5,
\label{Vhx}
\end{align}
where  $U_i,\tilde{U}_i,\tilde{\tilde{U}}_i$ and $\bar{U}_i$, are the interactions in the orbital representation.
\subsection{ Nematic susceptibility}
\label{Nematic susceptibility section}
We associate the nematic order with the d-wave Pomeranchuk order. In the orbital basis, this order is an orbital polarization (densities of $d_\text{xz}$ and $d_\text{yz}$ fermions split). In the band basis, we introduce two $d-$wave order parameters on hole and electron pockets: $\phi_h = \sum_{\bk,\sigma}\langle h_{\bk,\sigma}^\dagger h_{\bk,\sigma} \rangle\cos 2 \theta_{\bk}$ and
 $\phi_e= \sum_{\bk}\langle f_{2,\bk,\sigma}^\dagger f_{2,\bk,\sigma} \rangle-\langle f_{1,\bk\sigma}^\dagger f_{1,\bk,\sigma}\rangle=\phi_x-\phi_y$, where $\phi_{x/y}= \sum_{\bk}\langle f_{1/2,\bk,\sigma}^\dagger f_{1/2,\bk,\sigma} \rangle$. We introduce an infinitesimally small $Z_2$ symmetry breaking external perturbation to the interaction Hamiltonian in the form of $\phi_0 \, \cos2\theta_\bk\, h^\dagger_\bk h_\bk$ and define nematic susceptibility as $\phi_h=\chi_{\text{nem}}\, \phi_0$. The onset of the nematic order is signaled by the divergence of $\chi_{\text{nem}}$. Within the mean field analysis, we find a set of two coupled self-consistent equations for $\phi_h$ and $\phi_e$ (see Appendix \ref{Appendix B} for the details),
 \begin{align}
\phi_h&= \phi_0 -\phi_h\,  U^d_h \, \Pi_h^d -\phi_e \, U^d_{he}\, \Pi_e, \label{phih equation1}\\
\phi_e&=-\phi_e\,  U^d_{e} \Pi_e-2 \phi_h \,  U^d_{he} \Pi_h^d.
\label{phie equation1}
 \end{align}
Here
\beq
U_h^d=(U_4-2\tilde{U}_4+\tilde{\tilde{U}}_4)/2, ~~U_e^d =U_5-2 \tilde{U}_5,~~  U_{he}^d= (U_1-\bar{U}_1)
\eeq~~
 are the d-wave components of the intra-pocket and inter-pocket density-density interactions for fermions near hole and electron pockets, and
 \beq
  \Pi_h^d =-\bigintss G_p^h \, G_p^h \cos^22\theta_\bp,~~ \Pi_e=-(1/2) \bigintss_p \left(G_p^X \, G_p^X + G_p^Y \, G_p^Y\right)
  \eeq
are the polarization bubbles for fermions at hole and  electron pockets. As defined, the polarization bubbles are positive $\Pi_h^d,\Pi_e>0$. Eq.~\eqref{phie equation1} is found from two other equations: $\phi_y=-\phi_y \, U_5 \, \Pi_e+\phi_h \, U_{he} \, \Pi_h^d-2\, \phi_x\,\tilde{U}_5 \, \Pi_e$ and $\phi_x=-\phi_x \, U_5 \, \Pi_e-\phi_h \, U_{he} \, \Pi_h^d-2\, \phi_y\, \tilde{U}_5 \, \Pi_e$. Combining Eqs.~(\ref{phih equation1}) and (\ref{phie equation1}), and evaluating further the nematic susceptibility at a small but finite momentum $q$, we  obtain the expression for the nematic susceptibility,
\begin{align}
\chi_{\text{nem}}(q)=\dfrac{1}{\left(1+U_h^d\,\Pi_h^d(q)\right)\left(1+U_e^d\, \Pi_e(q)\right)-2U_{he}^2 \Pi_h^d(q) \Pi_e(q)},
\label{nematic susceptibility}
\end{align}
 where $\Pi_{h,e}^d(q)$ is the polarization bubble with momentum $q$. As stated before the onset of the orbital order is set when
  $\chi_{\text{nem}}(0)$ diverges.
  This can happen when  $ U^d_{he}$ is strong enough, specifically when  $2\, (U_\text{he}^\text{d})^2\geq U_\text{h}^\text{d}\,  U^d_{e}$ \citep{xing2018orbital} assuming all interactions are repulsive. In this case, nematic order (known as $d^\pm$) with different signs of $\phi_h$ and $\phi_e$ develops. This is quite similar to sign-changing $s^{+-}$ order, which develops when pair hopping  exceeds intra-pocket repulsion in the particle-particle channel.

\subsection{ Pairing interaction}
\label{pairing interaction chapter}
 The analysis of the pairing interaction mediated by soft nematic fluctuations requires care as these fluctuations are at small momentum transfer and for this reason they do not affect the pair hopping term between hole and electron pockets, which is the key element for spin-mediated superconductivity.
  Also, the bare pairing interaction at small momentum transfer is repulsive, and if we just dress it  by  nematic fluctuations, we find that the repulsion only gets stronger. Despite all this,  the pairing interaction $V_{\text{eff}} (k,-k;p,-p)$, mediated by nematic fluctuations, does become attractive near $x_\text{c}$, although for a rather special reason.
Namely, as we just said, the driving force for a $d-$wave Pomeranchuk order is d-wave density-density interaction between hole and electron pockets, $ U^d_{he}$.
 By itself, $ U^d_{he}$  does not contribute to pairing, however taken at the second order, it produces an effective attractive interaction between fermion on the same pocket. The full pairing interaction,  proportional to the susceptibility for a $d-$wave Pomeranchuk order, is obtained by collecting ladder and bubble diagrams, which contain d-wave polarization bubbles at a  small momentum transfer. This can be done analytically (see Refs. \citep{ dong2023transformer,sm} for detail). Because $\bq$ is small, the dressed pairing interactions are between fermions on only  hole pocket or only electron pockets: $V_{\text{eff}}^\text{h}(\bk, \bq)  = V_{\text{eff}}^\text{h} (\bk + \bq/2, -(\bk + \bq/2); \bk -\bq/2, -(\bk - \bq/2)$, $V_{\text{eff}}^\text{e}(\bk, \bq)  = V_{\text{eff}}^\text{e} (\bk + \bq/2, -(\bk + \bq/2); \bk -\bq/2, -(\bk - \bq/2)$.  We find
\begin{align}
V_{\text{eff}}^\text{h}(\bk, \bq)&= A_h^{(0)} - A_h^{(1)} ( U^d_{he})^2 \, \cos^2 2 \theta_\bk\,  \chi_{\text{nem}} (\bq) +... \nonumber \\
V_{\text{eff}}^e(\bk,\bq)&=  A_e^{(0)} - A_e^{(1)} ( U^d_{he})^2\,  \chi_{\text{nem}}(\bq) + ...,
\label{electron pairing interaction1}
\end{align}
where $A_h^{(0)}=\frac{U^d_h}{1 + U^d_h \Pi^d_h (\bq)}$, $A_e^{(0)}=\frac{U^d_e}{1 + U^d_e \Pi_e (\bq)}$, $A_h^{(1)}=\dfrac{\Pi_e}{1+U_h^d\, \Pi_h^d (\bq)}$ and $A_e{(1)}=\dfrac{1}{2}\dfrac{\Pi_h^d}{1+ U^d_{e}\, \Pi_e (\bq)}$.
The dots stand for other terms which do not contain $\Pi_h^d (\bq)$ and $\Pi_e^d (\bq)$ and are therefore not sensitive to the nematic instability. We see that each interaction contains two terms. The first is the dressed intra-pocket pairing interaction.
It does get renormalized, but remains repulsive and non-singular at the nematic instability.  The second term is the interaction, induced by $ U^d_{he}$.  It is  attractive and  scales with the nematic susceptibility.
We emphasize that the attraction is induced by inter-pocket density-density interaction, despite that relevant nematic fluctuations are with small momenta and the pairing interactions involve fermions from the same pocket. Because the nematic susceptibility diverges at $x_\text{c}$,  the attractive component, induced by nematic fluctuations,  necessary  exceeds the bare intra-pocket repulsion in some range around $x_\text{c}$, i.e., the full intra-pocket pairing interaction
  becomes attractive. This attractive interaction  is, however,  rather peculiar on the hole pocket  because it inherits from $ U^d_{he}$  the $d-$wave form-factor
     $A_{\bk,\bp} = \cos^2{(\theta_\bk+\theta_\bp)}$, where $\theta_\bk$ and $\theta_\bp$ specify the location of the fermions.  A similar pairing interaction  has been earlier suggested for one-band models on
        phenomenological grounds \citep{lederer2015enhancement,schattner2016ising,lederer2017superconductivity,klein2018superconductivity}.  We emphasize that for our microscopic derivation, the presence of hole and electron bands is the necessary condition
         for a nematic-mediated superconductivity.

\subsection{Gap Structure}
\label{Solution of the gap equation}

To simplify the discussion, we neglect $A^{(0)}_h$ and  $A^{(0)}_e$ in Eqs. \eqref{electron pairing interaction1}, i.e.,
keep only the components of the pairing interaction proportional to the nematic susceptibility  and
 also assume that $A^{(1)}_h > A^{(1)}_e$, in which case the strongest  pairing interaction is for fermions on a hole pocket.
  We also treat $\chi_{nem} (q)$ as static. The full analysis, which we leave for further study, should involve the
  dynamical nematic $\chi_{nem} (\bq, \Omega_m)$, like it was done for a phenomenological one-band model in Refs. \citep{klein2018superconductivity,klein2019multiple}.

   For fermions on the hole pocket the pairing interaction is
$V_h(\bk,-\bk;\bp,-\bp) = V_h(\bk,\bp)=-
\cos^2(\theta_\bk+\theta_\bp)  \tilde{\chi}_\text{nem}(\bk-\bp)$,
 where
$\theta_{\bk(\bp)}$ is the angle between
$\bk (\bp)$  and the direction towards electron pockets ($\Gamma-X$ or $\Gamma-Y$ in 1Fe Brillouin zone).
 and $\tilde{\chi}_\text{nem}= A_h \left(U_\text{he}^\text{d}\right)^2\chi_\text{nem}$.
 A superconducting order parameter $\Delta_h(\bk, T)$ at temperature $T$ is obtained by solving  the non-linear gap equation
\begin{align}
    \Delta_h(\bk, T)&= -N_0\, \bigintss_{\Lambda}^{\Lambda} d\xi_\bp \bigintss_0^{2\pi} \dfrac{d\theta_\bp}{2\pi} \, V_h(\bk,\bp) \, \Delta_h(\bp, T) \dfrac{\tanh \beta E_\bk/2}{2 \, E_\bk}
    \label{Gap Equation}
\end{align}
where $E_\bk=\sqrt{\xi_\bk^2+|\Delta_h(\bk, T)|^2}$ is the quasi-particle excitation energy, $N_0$ is the density of states at the Fermi energy and $\Lambda$ is the ultra-violet cutoff for the pairing interaction.
 The presence of the $\cos^2(\theta_\bk+\theta_\bp)$ in the pairing interaction $V_h(\bk,\bp)$ combined with the fact that  $\tilde{\chi}_{\text{nem}}(\bk-\bp)$ is peaked at $\bk \approx \bp$ makes the gap structure rather exotic. We keep the momenta $\bk,\bp$ on the Fermi surface ($|\bk(\bp)|=k_F$) and model the nematic susceptibility as a Lorentzian $\tilde{\chi}_{\text{nem}}(\bk-\bp) = 2 V_0\, \delta /\left(\delta^2 + 4 \sin^2(\theta_\bk-\theta_\bp)/2\right)$, where $\delta$ is the distance to a nematic critical point in units of Fermi momentum. For this form,  $\int \tilde{\chi}_{\text{nem}}(\bk-\bp) d \theta_\bp$ tends to a finite value at $\delta \to 0$.

   The pairing potential $V_h(\theta_\bk,\theta_\bp)=-2 V_0\, \delta\, \cos^2(\theta_\bk+\theta_\bp)/\left(\delta^2 + 4 \sin^2(\theta_\bk-\theta_\bp)/2\right)$ is attractive in all
pairing channels associated with the irreducible representations of the $D_{4h}$ group for the square lattice (one-dimensional $A_{1g}$ $B_{1g}$, $B_{2g}$ and two-dimensional $E$, each with an infinite number of eigenfunctions).
    Furthermore, at small $\delta$, the pairing interaction is nearly identical for a large set of eigenfunctions within a given representation, and the interplay between these eigenfunctions makes the  gap function highly anisotropic and strongly peaked near particular points on the Fermi surface
    \footnote{There is a certain similarity between the present case and pairing near hot spots in a model of pairing mediated by soft antiferromagnetic fluctuations.}.

\begin{figure}
 \centering
   \subfigure[]{\includegraphics[width= 0.42 \textwidth]{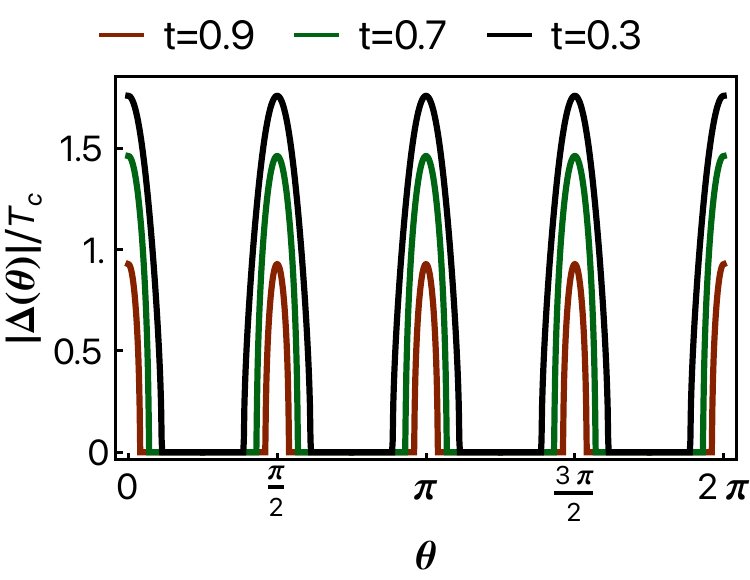}}
   \subfigure[]{\includegraphics[width= 0.44 \textwidth]{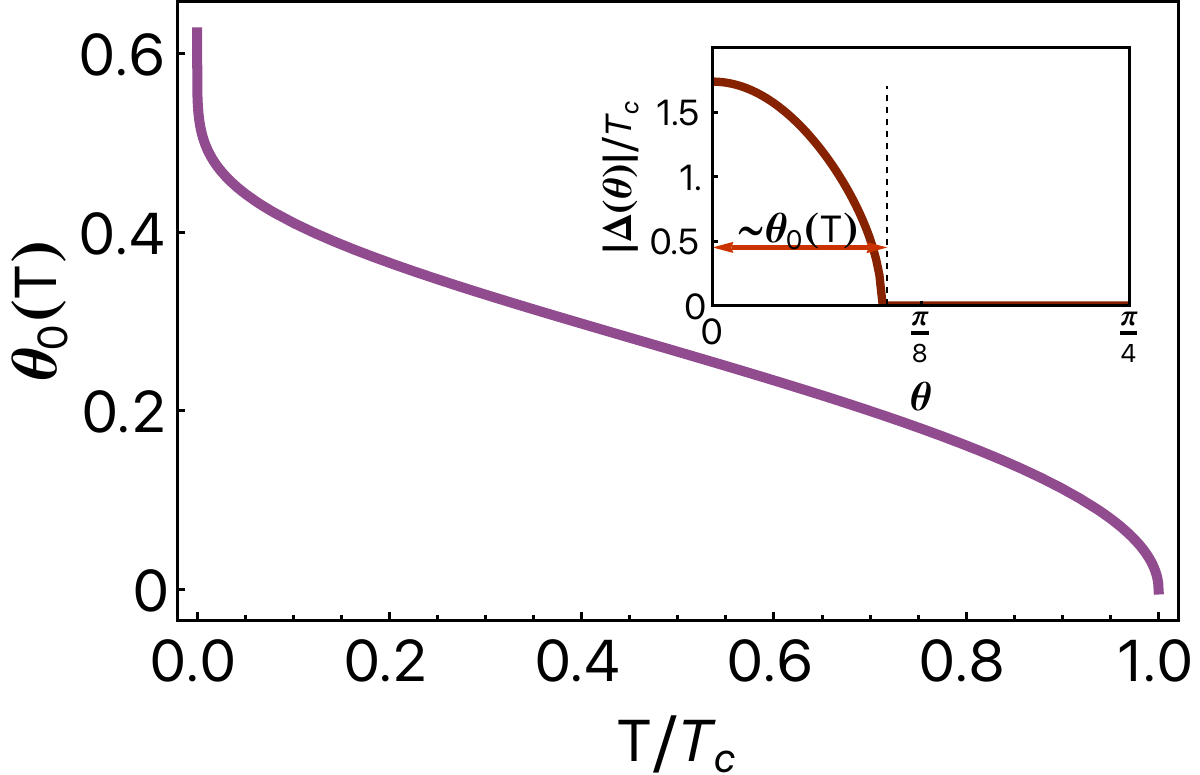}}
   \subfigure[]{\includegraphics[width= 0.36 \textwidth]{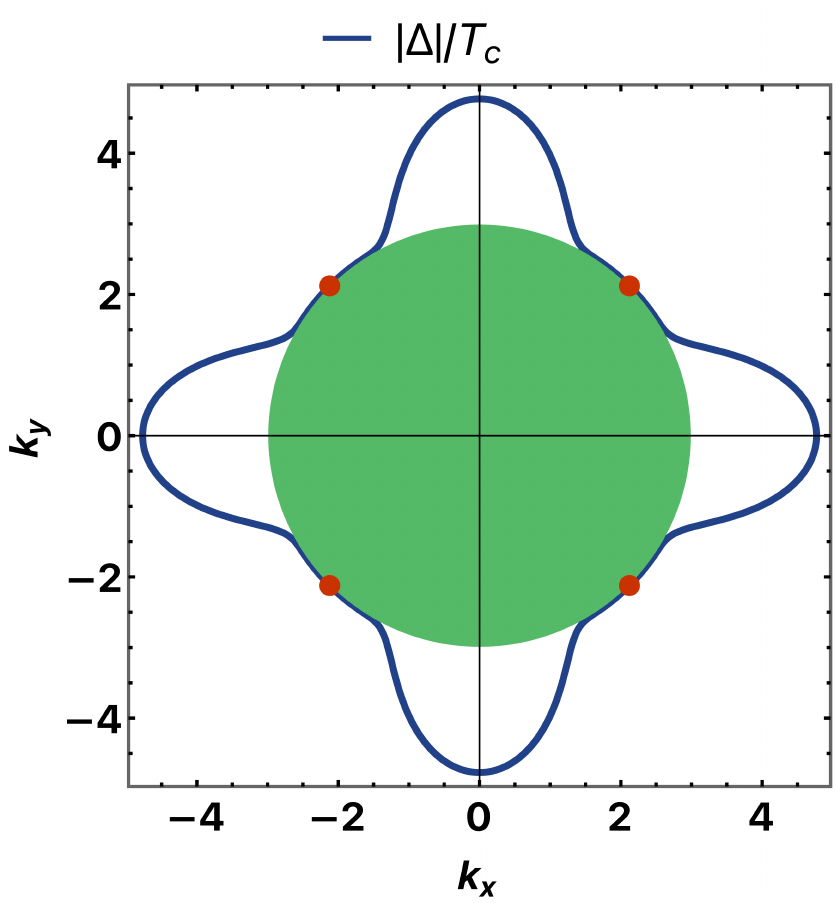}}
   \subfigure[]{\includegraphics[width= 0.42 \textwidth]{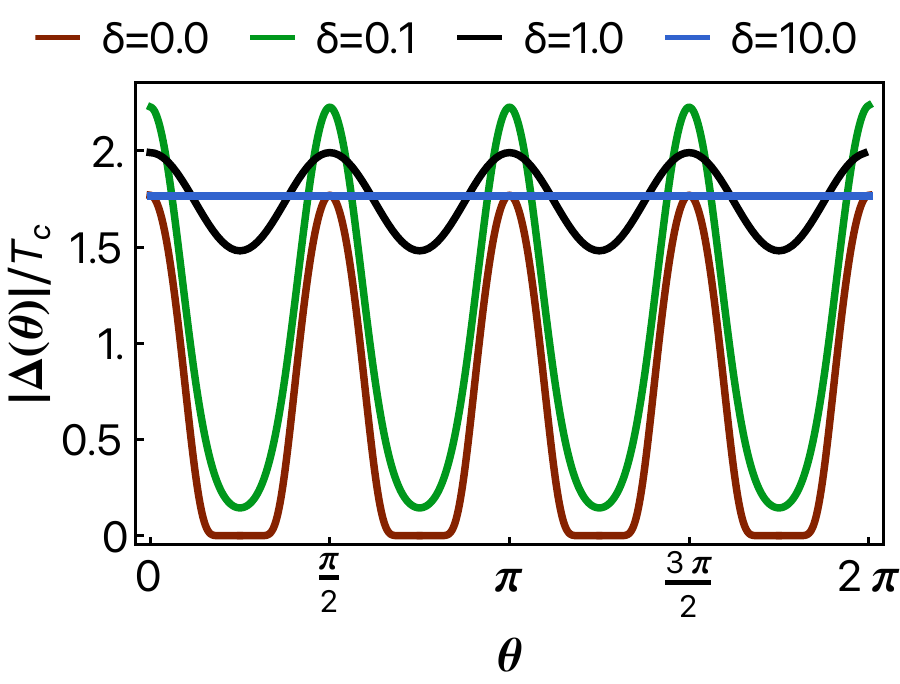}}
\caption{Gap function on a hole pocket for NFMS. (a) Angular variation  of $\Delta(\theta_\bk)$ on the hole pocket  for a set of $T/T_c$. (b) Temperature dependence of the angular width of the gap $\theta_0(T)$, specified in the inset
 (c) Polar plot of the  gap magnitude at $T=0$. In the cold region, the gap is non-zero except at points marked by red dots, but is exponentially small. (d) Angular  variation of  $\Delta(\theta_\bk, T=0)$ away from a nematic QCP,  for a set of deviations $\delta$.}
\label{evolution of gap}
 \end{figure}

   At vanishing $\delta$, the non-linear equation for $\Delta_h(\theta_\bk, T)$(\ref{Gap Equation}) becomes purely local with an angle-dependent coupling:
\begin{align}
1=g \, \cos^2 2\theta_{\bk} \bigintssss_0^\Lambda d\xi  \dfrac{\tanh\Bigr(\dfrac{\sqrt{\xi^2+|\Delta_h(\theta_\bk, T)|^2}}{2 T}\Bigr)}{\sqrt{\xi^2+|\Delta_h(\theta_\bk, T)|^2}}.
\label{int4}
\end{align}
where $g=N_0\, V_0$ is a dimensionless parameter.  Because the coupling is the largest at $\theta_\bk = n \pi/2, n=0-3$, the gap appears at  $T_\text{c} =1.13\Lambda \exp(-1/g)$ only at these points.  As $T$ decreases, the range, where the gap is non-zero, extends to four finite arcs (see Fig. \ref{evolution of gap}(a)) with the temperature dependent width $\theta_0 (x) = \theta_0 (T/T_c)$ given by
  \begin{align}
    \theta_0(x) = \dfrac{1}{2} \arctan{\sqrt{g\log{\dfrac{1}{x}}}}.
    \label{theta function}
\end{align}
We plot $\theta_0 (x)$ in  Fig. \ref{evolution of gap}(b).
 In the  areas between the arcs,
 the original Fermi surface survives. The gap amplitude, $|\Delta_h(\theta_\bk, T)|$ at a given momentum on one of the arcs  evolves like the s-wave gap function below the angle dependent transition temperature
 $T_c (\bk) =  1.13\, \Lambda \exp(-1/g \cos^22\theta_\bk)$.   We emphasize that the thermal evolution of $\Delta_h(\theta_\bk , T)$ is rather special
   both the amplitude and the angular dependence evolve with $T$. At $T=0$, the arcs ends merge at $\theta_c = n \pi/2 + \pi/4, n=0-3$ and the gap becomes non-zero everywhere except at the merging points (cold spots), shown as red dots in  Fig. \ref{evolution of gap}(c).
  An an arbitrary angle,
  \beq
  |\Delta_h(\theta_\bk , T=0)| = \Delta_h (\theta_\bk) = \Delta_0 \exp{-\, \tan^2 2\theta_\bk/g}
   \eeq
    where $\Delta_0 = 1.76 T_c$.
   This gap is highly anisotropic and exponentially suppressed near the cold spots.

   We note in passing that for the full dynamical pairing interaction, we expect $s$-wave gap
 to remain strongly peaked at the hot spots, but preserve a small
  finite value at $\theta_\bk =\theta_c$  even at a nematic QCP (Refs. \citep{klein2018superconductivity,klein2019multiple}).

Away from the nematic critical point, the gap equation is no longer local in momentum space and at $T_c$ the gap opens everywhere except nodal points set by gap symmetry.
The equivalence between different pairing channels also breaks down and in our model $s-$wave pairing wins -- the corresponding  condensation energy is the largest by magnitude.  We solved the gap equation, Eq.~\eqref{Gap Equation}, at a finite $\delta$  and show $s-$wave gap function at $T=0$ for several $\delta$ in Fig.  \ref{evolution of gap}(d).

For fermions on the electron pockets, the pairing interaction mediated by nematic fluctuation
 (\ref{electron pairing interaction1}) is $V_e(\bk,-\bk;\bp,-\bp) = -A^{(1)}_e/A^{(1)}_h\,  \tilde{\chi}_\text{nem}(\bk-\bp)$.
This interaction has no  momentum dependence, hence the gap on an electron pocket is independent of the position on the Fermi surface.  As we said, we assume that $A^{(1)}_e/A^{(1)}_h$ is smaller than one, hence the gap on the electron pocket develops at a smaller $T_c$ within our model, where we only include the pairing interaction mediated by nematic fluctuations.  In a more realistic model with pair-hopping interaction also present, the gap $\Delta_e$ is induced by $\Delta_h (\theta_{\bk}, T)$. In any case, because $\Delta_e$ is featureless, it will not contribute to the evolution of observables at low $T$, which we consider below.  To simplify the analysis we will  just neglect $\Delta_e$.

\section{Fingerprints of NFMS}
 \label{fingerprint section}

To highlight  novel features associated with NFMS, we chiefly focus on the system behavior right at the nematic quantum-critical point (QCP), where at an finite $T$ the gap on the hole pocket,
 $\Delta (\theta_\bk, T)$   is non-zero in a finite range around $\theta_k = n \pi/2, n= 0,1,2,3$ (we drop the subscript $h$ from $\Delta$ onward).  We also assume a parabolic hole dispersion and weak coupling. We  analyze below thermodynamic, spectroscopic and transport properties  of the NFMS. In Appendix~\ref{Appendix E}, we compare these properties  with those of ordinary s-wave and d-wave superconductors.

\subsection{Thermodynamic Properties}

 \subsubsection{Specific Heat}
The electronic contribution to a specific heat $C_v(T)$  of a superconductor at temperature $T$ is given by \citep{coleman2015introduction}
\begin{align}
C_v(T)&=\dfrac{2N_0 }{T}\bigintss \dfrac{d\xi_k d \theta_\bk}{(2\pi)}\Bigr(-\dfrac{\partial\,  n_\bk}{\partial E_ \bk}\Bigr) \left[E_\bk^2-\dfrac{T}{2} \dfrac{\partial |\Delta(\bk, T)|^2}{\partial T}\right]
\label{sp heat 1}
\end{align}
where $E_\bk=\sqrt{\xi^2_k +|\Delta(\theta_\bk, T)|^2}$ is the quasi-particle energy and  $n_\bk =1/(\exp(\beta E_\bk)+1)$
  with $\beta=1/k_B$ is the Fermi function. Because $\dfrac{\partial \, n_\bk}{\partial\, E_\bk}=-\dfrac{1}{4\, T}\sech^2\bigr(\beta \, E_\bk/2\bigr) $ is peaked near the Fermi surface, we split the momentum integration into
  integration over $\xi_\bk$ and $\theta_\bk$ and keep the density of states equal to
 $N_0$ at the Fermi surface.  We introduce dimensionless variables
      $t=T/T_c$, $\bar{\Delta}(\theta_\bk, t)=\Delta(\theta_{\bk}, T)/T_c$,  and dimensionless specific heat coefficient $\gamma(t)= C_v (t)/(TN_0) = (C_v(t)/t) (1/T_c N_0)$.  The expression for $\gamma (t)$ is
\begin{align}
 \gamma(t)=\dfrac{1}{t^3} \bigintss_{0}^\infty dx \bigintss_0^{2\pi} \dfrac{d\theta_\bk}{2\pi} \sech^2\Bigr(\dfrac{\sqrt{x^2+|\bar{\Delta}(\theta_\bk, t)|^2}}{2\,t}\Bigr) \Bigr[x^2+|\bar{\Delta}(\theta_\bk, t)|^2-\dfrac{t}{2} \dfrac{\partial |\bar{\Delta}(\theta_\bk, t)|^2}{\partial t}\Bigr].
\label{scaled sp heat 1}
\end{align}

\begin{figure}[H]
  \centering
 \includegraphics[scale=0.8]{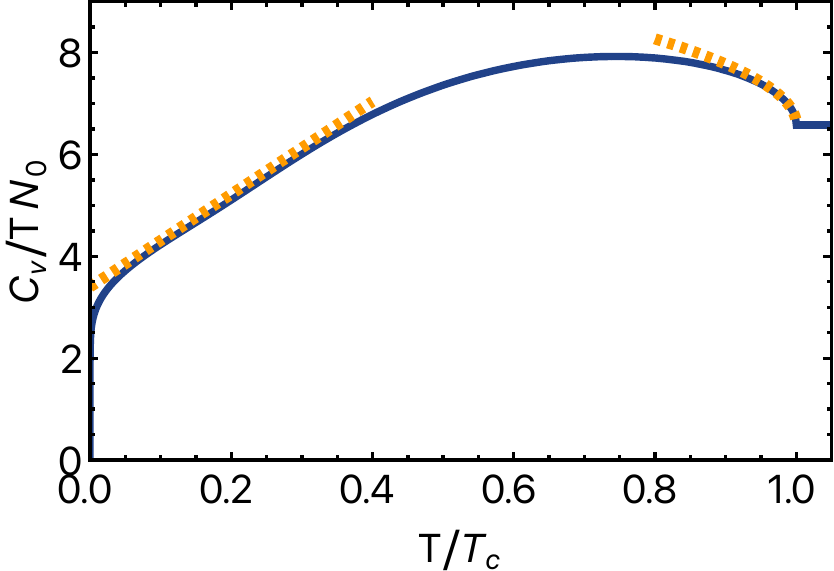}
  \caption{ Specific heat coefficient, $C_v/T\, N_0$ for a NFMS  as a function of  $T/T_c$. The orange dashed lines show the behavior near $T_c$ and the extrapolation from  small but finite  temperatures to $T=0$.}
\label{specific heat}
 \end{figure}

In the normal state  $\gamma (t) = 2\, \pi^2/3$ is temperature independent.
   Immediately below the transition point (at $t=1^-$) the gap $\bar{\Delta}(\theta_\bk, t)$ is infinitesimally small but its temperature derivative is not, and $\gamma (t)$ is generally supposed to jump by
\begin{align}
    \Delta\gamma=-\dfrac{\partial }{\partial t } \bigintss_0^{2\pi}\dfrac{d\theta_\bk}{2\pi} |\bar{\Delta}(\theta_\bk, t)|^2.
    \label{jump expression}
\end{align}
 Our case is special in this respect as
  the gap opens up at $T_c$ only at $4$ points on the Fermi-surface. In this situation,  $\Delta\gamma$ vanishes,
   i.e., there is no jump in the specific heat at $t=1^-$. (see Fig.  \ref{specific heat}).
At $t<1$, we split the specific heat coefficient into two contributions: $\gamma^\text{N}$  from the  part of the Fermi surface where there is no gap, and $\gamma^\text{SC}$ from the gapped region. The total
    $\gamma(t)=\gamma^\text{SC}(t)+ \gamma^\text{N}(t)$. The two contributions are
\begin{align}
\label{normal cv}
\gamma^\text{N}(t)&=\dfrac{8}{ t^3} \bigintss_{\theta_0(T)}^{\pi/4} \dfrac{d\theta_\bk}{2\pi}\bigintss_{0}^\infty dx \sech^2\Bigr( \dfrac{x}{2\,t}\Bigr) \,  x^2=\dfrac{8\, \pi}{3} \Big[\dfrac{\pi}{4}-\theta_0(T)\Bigr]
\end{align}
and
\begin{align}
\gamma^\text{SC}(t)&=
\dfrac{8}{t^3} \bigintss_{0}^\infty dx \bigintss_0^{\theta_0(t)} \dfrac{d\theta_\bk}{2\pi} \sech^2\Bigr(\dfrac{\sqrt{x^2+|\bar{\Delta}(\theta_\bk)|^2}}{2\,t}\Bigr) \Bigr[x^2+|\bar{\Delta}(\theta_\bk)|^2-\dfrac{t}{2} \dfrac{\partial |\bar{\Delta}(\theta_\bk)|^2}{\partial t}\Bigr]\nn & \approx \sqrt{\dfrac{g}{\pi}}\, \gamma_s(t)\, \text{Erfi}[\sqrt{|\log t|}],
\label{sc cv}
\end{align}
 where $\gamma_s (t)$ is the specific heat coefficient for an ordinary $s-$wave superconductor
 and $\text{Erfi}(z)$ is the imaginary error function (see Appendix \ref{Appendix C} for the derivation of Eq.~(\ref{sc cv}) and other details).
Near $t=1$, $\gamma^\text{N}(t) \approx 2\pi^2/3-4\pi\sqrt{g}\sqrt{1-t}/3$, while $\gamma^\text{SC}(t) \approx 2\, \sqrt{g}\, \gamma_s(1)\, \sqrt{1-t}/\pi$. Summing up the two contributions, we find that the specific heat coefficient  increases continuously at $T \leq T_c$, with a $\sqrt{T_c -T}$ temperature dependence.
At smaller $T$, $\gamma(t)$ passes through a maxima, present because
 $\gamma^{\text{SC}}(t)$ in Eq.~(\ref{sc cv}) is the product of a monotonically decreasing  $\gamma_s(t)$ and a monotonically increasing $\text{Erfi}(\sqrt{\log t})$.
 At low temperatures, $\gamma_s (t)$ is exponentially small, such that $\gamma (t)$ chiefly  comes from the normal part and behaves as
$\gamma(t) \approx  4  \,\pi/ (3\, \sqrt{|g \, \log t|})$.
  The specific heat coefficient does vanish at $t=0$, but very weakly, as $1/\sqrt{\log(t)}$.
 We plot  $\gamma(t)$ in Fig. \ref{specific heat}. We see that the $1/|\log t|^{1/2}$ downturn of
 $\gamma(t)$ occurs only at very low $t \sim 10^{-2}$. Extrapolating from higher $t$ one would obtain instead a constant offset, expected when some zero energy fermionic states remain at $T=0$.
\subsubsection{Change of a Specific Heat in a Magnetic Field}

\begin{figure}
 \centering
    \subfigure[]{\includegraphics[width= 0.45 \textwidth]{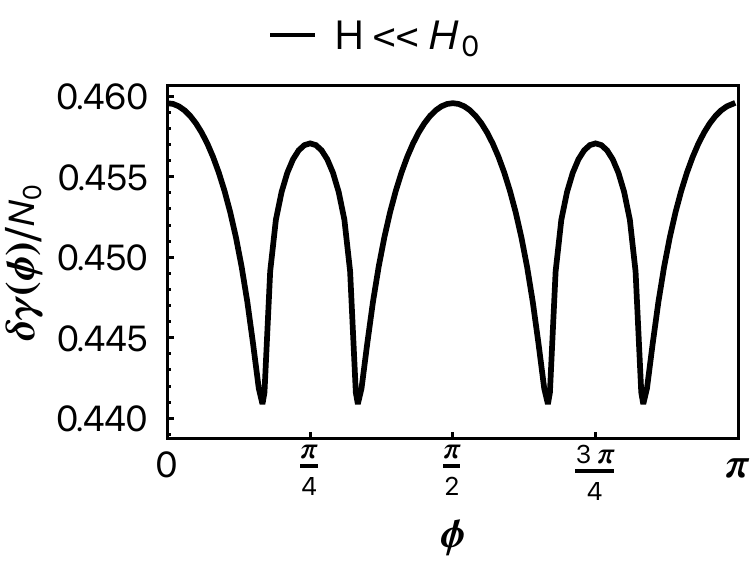}}\hspace{1 cm}
   \subfigure[]{\includegraphics[width= 0.45 \textwidth]{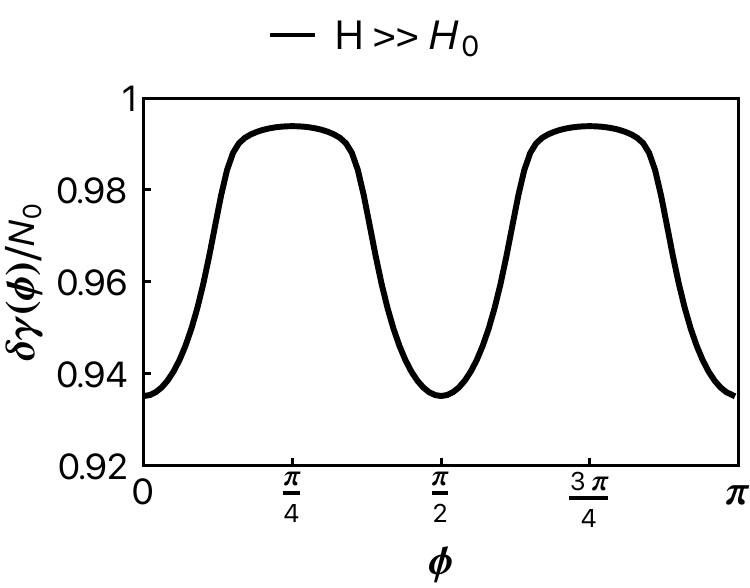}}
\caption{Modulation of the specific heat coefficient $\gamma=C_v/T N_0$  for NFMS  under an in-plane magnetic field $\mathbf{H}$ directed along $\phi$
 counted from
  $\Gamma-X$ axis.
  (a) -- Small field $H\ll H_\text{0}$, (b)-- large field $H\gg H_\text{0}$.  $H_\text{0}$ is defined in the text. }
  \label{sp heat with Magnetic field}
 \end{figure}

\begin{figure}
 \centering
   \subfigure[]{\includegraphics[width= 0.45 \textwidth]{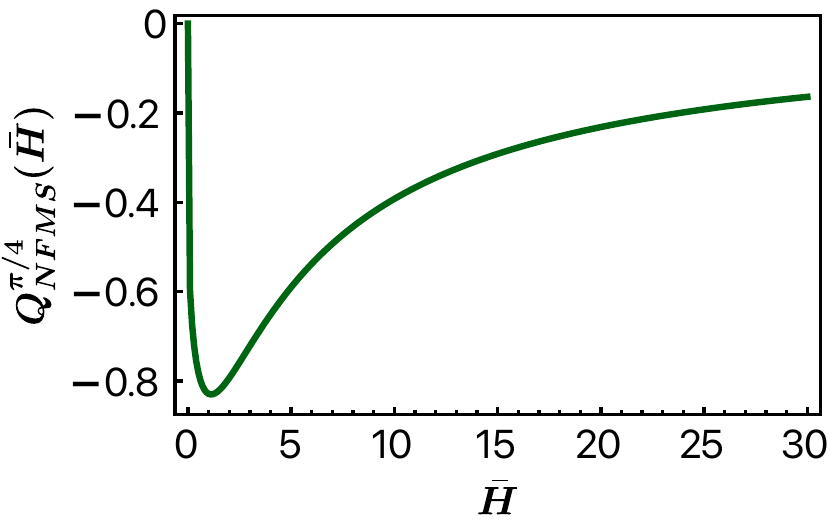}}
   \subfigure[]{\includegraphics[width= 0.45 \textwidth]{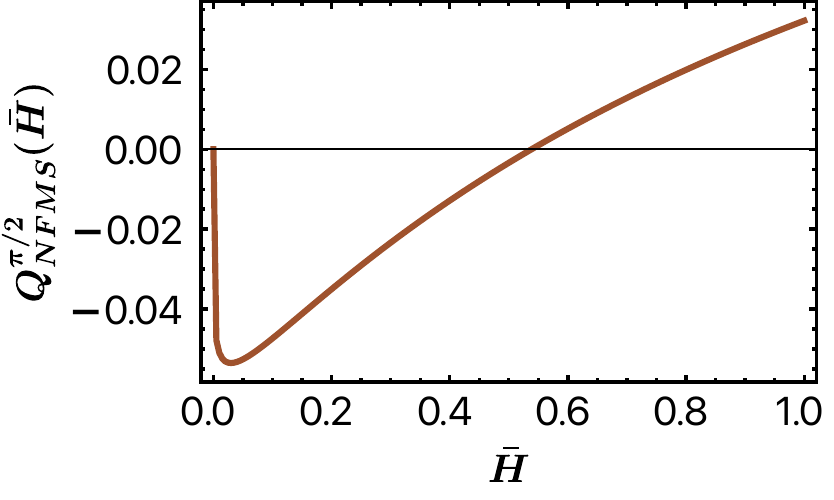}}
   \caption{The field strength ($H$) dependence of (a) $Q^{\pi/4}_\text{NFMS}(H)$, coefficient of $(\phi-\pi/4)^2$ term, and (b) $Q^{\pi/2}_\text{NFMS}(H)$, coefficient of $(\phi-\pi/2)^2$ term in Taylor expansion of $\delta\gamma(H,\phi)$ for a nematic fluctuation mediated superconductor (NFMS).}
   \label{QNFMS for hot and cold spot}
   \end{figure}

An important probe aimed to detect whether a superconductor posses nodes, and if yes, where the nodes are located and how the gap structure varies near these nodes, is to measure specific heat at low temperatures in the presence of an external magnetic field. The method explores the fact that a type-II superconductor (which most of the unconventional superconductors are) is  in a  vortex state when a magnetic field is above the critical $H_{c1}$. Scattering of quasiparticles by  vortices gives rise to a non-zero density of states (DOS) at zero energy
 and to the emergence of the field-induced linear in $T$ term in the specific heat,
 whose strength  depends on the angle the field makes with the position of the minima of
 $|\Delta (\theta_\bk, T)|$
 ~\citep{volovik1993superconductivity,vekhter1999anisotropic,vorontsov2006nodal,chubukov2010angle,xu1995nonlinear,yip1992nonlinear, vekhter2001thermodynamics,kubert1998quasiparticle}. The specific heat $C(T,\mathbf{H})$ is expressed via the density of states $N(T,\mathbf{H},\omega)$ at energy $\omega$ as
\begin{align}
    C(T,\mathbf{H}) & =\dfrac{1}{2 \, T^2} \bigintss_{-\infty}^{\infty} d\omega \, \omega^2\, N(T,\mathbf{H},\omega)\, \sech^2 \dfrac{\omega}{2 T}
\end{align}
We set the magnetic field $\mathbf{H}$ directed in the plane of a 2D material,  at an angle $\phi$ with respect to the $x-$ axis.  At low temperatures, $N(T,\mathbf{H},\omega)$ can be approximated as $N(0,\mathbf{H},0) = N_0 (\mathbf{H})$ and pulled out from the integral. We then have $C(T, \mathbf{H}) \approx (2\pi^2/3) T N_0 (\mathbf{H})$, i.e.,  at $T \to 0$ the
magnetic field induced contribution to the specific heat coefficient is independent on $T$ and is given by
  $\gamma(\mathbf{H}) = (2\pi^2/3) N_0(\mathbf{H})
\equiv (2\pi^2/3)  \delta \gamma (H,\phi)$. We follow Refs.(\citep{vekhter1999anisotropic,chubukov2010angle}) and  employ the formula for $N_0(\mathbf{H})$, obtained   using the quasi-classical Doppler shift approximation. Using it, we obtain
\begin{align}
\delta \gamma (H,\phi) =N_0\bigintss_0^{2\pi} \dfrac{d\theta_\bk}{2\pi} \dfrac{\alpha(\mathbf{H},\theta_\bk)}{\sqrt{\alpha(\mathbf{H},\theta_\bk)^2+|\Delta(\theta_\bk, T=0)|^2}}=N_0\, \bigintss_0^{2\pi} \dfrac{d\theta_\bk}{2\pi} \dfrac{1}{\sqrt{1+\dfrac{1}{\bar{H}} \dfrac{ f^2(\theta_\bk)}{\sin^2(\theta_\bk-\phi)}}}.
    \label{Sp mag 1}
\end{align} Here, $\alpha(\mathbf{H},\theta_\bk) = v_F \sqrt{H e} b  |\sin(\theta_\bk-\phi)|$, with $b=O(1)$,  is proportional to the Fermi velocity component normal to the field $v_F^\perp=v_F\, \sin(\theta_\bk-\phi)$,
$\bar{H}=H/H_0$ with $H_0= \Delta_0^2/b^2\, |e|\, v_F^2$ and $f(\theta_\bk)=\exp\left(-\tan^22\theta_\bk/g\right)$.
Note that $\delta \gamma (H,\phi)$ is $C_4$ symmetric, $\delta \gamma (H,\phi)=\delta \gamma (H,\phi+\pi/2)$,  because  $f^2(\theta_\bk)$ is invariant under $\theta_\bk \to \theta_\bk+ \pi/2$. We numerically integrated Eq.~(\ref{Sp mag 1}) and show our results for $\delta \gamma (H,\phi)$  in Fig.  \ref{sp heat with Magnetic field}(a,b).
 We see that the field-induced specific heat coefficient, viewed as a function of $\phi$, has a set of maxima and
 minima, and evolves between $H < H_0$ and $H > H_0$.  For small $H < H_0$, the maxima are at $\phi = n\pi/2$ and  at $\phi = \pi/4 (1+2n)$ $(n=0,1,2,3)$, i.e., along hot and cold directions for the gap function, and the minima are in between.  For large $H > H_0$, the maxima  at $\phi = (\pi/4) (1+2n)$ $(n=0,1,2,3)$ remain, but at $\phi = n\pi/2$ the function now has a minimum. This behavior can be understood analytically by expanding around  $\phi =\pi/4$ and $\phi = \pi/2$.
   In particular, expanding to quadratic order around  $\phi =\pi/4$ we obtain
 \begin{equation}
 \delta \gamma (H, \phi) = \delta \gamma (H, \pi/4) + \left(\phi-\frac{\pi}{4}\right)^2 Q^{\pi/4}_{NFMS} ({\bar H})
 \label{expansion near cold spot}
 \end{equation}
 where
   \begin{multline}
 Q^{\pi/4}_{NFMS}({\bar H}) = -\sqrt{\bar{H}}\int_0^{2\pi} \frac{d\theta}{2\pi} \frac{|\sin{\theta}|\exp\left(-2 \cot^22\theta\right)}{\left({\bar H} \sin^2{\theta} + \exp\left(-2\cot^22\theta\right)\right)^{5/2}} \left[\exp\left(-2\cot^22\theta\right)+\bar{H} \left(2+\cos2\theta\right) \right]\leq 0
 \label{qnfms2}
 \end{multline}
 for simplicity we set the coupling $g=1$.
 We plot this function in Fig.  \ref{QNFMS for hot and cold spot}(a).
  We see that it is finite and negative for all ${\bar H}$. Accordingly, $\delta \gamma (H, \phi)$  has a maximum at $\phi = \pi/4$ for all $H$.

  Performing a similar expansion near $\phi = \pi/2$ we obtain
  \begin{align}
      \delta\gamma(H,\phi)=\delta(\gamma,\pi/2)+\left(\phi-\dfrac{\pi}{2}\right)^2 Q^{\pi/2}_{NFMS}(\bar{H})
  \end{align}
  where
  \begin{align}
      Q^{\pi/2}_{NFMS}(\bar{H})=  -2 \sqrt{\bar{H}} \bigintss_0^{2\pi} \dfrac{d\theta}{2\pi}&\dfrac{\abs{\sin\theta}\sec^22\theta \exp\left(-2 \tan^22\theta\right)}{\left(\bar{H}\sin^2\theta+\exp\left(-2 \tan^22\theta\right)\right)^{5/2}} \left[2\bar{H}\sin^22\theta \left(-1+\tan^22\theta+4 \tan^42\theta\right) \right. \nn & \left. +\exp\left(-2 \tan^22\theta\right) \left(4-\left(5+\cos4\theta\right)\sec^42\theta\right)\right]
      \label{Qnfmspib2}
      \end{align}
  We plot this function in Fig.~\ref{QNFMS for hot and cold spot}(b) which shows that it is negative for $H < H_0$ and positive for $H > H_0$. Accordingly, $\delta\gamma(\pi/2,\bar{H})$ changes from a maximum to minimum at some
  value of $H_0$. We present a qualitative analysis of $\delta\gamma(\phi)$ for a generic field direction $\phi$ in Appendix~\ref{Appendix D}.

  \subsubsection{DC Magnetic Susceptibility}
\begin{figure}[H]
 \centering
 \includegraphics[scale=0.7]{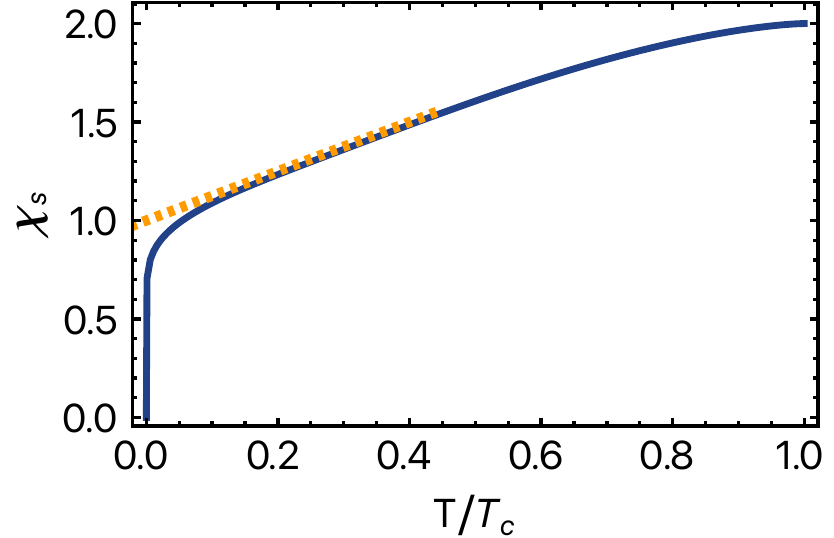}
   \caption{Uniform static spin susceptibility $\chi_s$ as a function of temperature $T$ for NFMS. The orange dashed line is the extrapolation from small but finite $T$.}
\label{Spin Susceptibility fig}
   \end{figure}

The static magnetic susceptibility, $\chi_s$ is expressed as \citep{coleman2015introduction}
\begin{align}
    \chi_s(T)=\dfrac{N_0}{T} \bigintss_0^{2\pi} \dfrac{d\theta_\bk}{2\pi}\bigintss_{|\Delta(\theta_\bk,T)|}^{\infty} dE\, E \dfrac{\sech^2\dfrac{E}{2\, T}}{\sqrt{E^2-|\Delta(\theta_\bk,T)|^2}}
    \label{susceptibility eq}
\end{align}
In the normal state $\chi_s (T) = 2N_0$. We numerically integrated Eq. (\ref{susceptibility eq}) using $\Delta(\theta_\bk, T)$ from Sec.\ref{Solution of the gap equation} and present the result in Fig.  \ref{Spin Susceptibility fig}. We also analyzed Eq.~(\ref{susceptibility eq}) analytically and obtained  $\chi_s(T)=1-(1-T_c)^{3/2}$ for $T \approx T_c$,and $\chi_s(T)=1-1/\sqrt{\log T/T_c}$ for $T\ll T_c$.  We discuss analytical integration in Sec. \ref{superfluid denisty section} below in the context of superfluid density, which is given by a similar formula.
\subsection{Transport and Spectroscopic Properties}
\subsubsection{Dynamical density of states and tunneling conductance}
\begin{figure}
 \centering
    \subfigure[]{\includegraphics[width= 0.49 \textwidth]{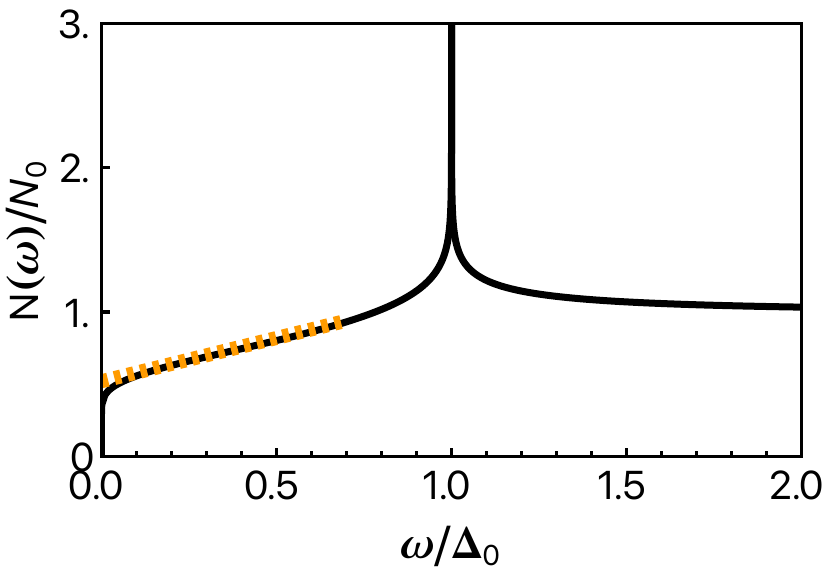}}
   \subfigure[]{\includegraphics[width= 0.48 \textwidth]{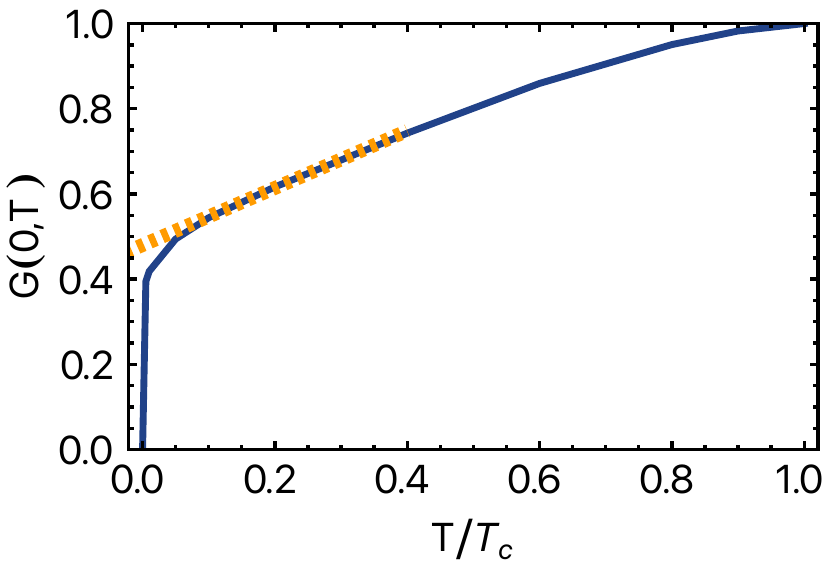}}
\caption{(a) Tunneling conductance  at zero temperature $G(\omega, 0)$ (the same as the dynamical single particle density of states, $N(\omega, 0)$) for  NFMS as a function of frequency  $\omega$. The orange dashed line is the extrapolation from small but finite frequencies. (b) Zero bias tunneling conductance $G(0,T)$ as a function of $T/T_c$ for NFMS. The orange dashed line is the extrapolation from small but finite $T$.}
\label{zero bias fig}
 \end{figure}

 The dynamical single particle density of state $N(\omega,T) =\bigintss \dfrac{d^2\bk}{(2\pi)^2} \delta(\omega-E_\bk(T))$ in a superconductor is given by
\begin{align}
    N(\omega,T )=N_0 \, \text{Re}\left[\bigintss_0^{2\pi} \dfrac{d\theta_\bk}{2\pi} \dfrac{\omega}{\sqrt{\omega^2-\Delta(\theta_\bk,T)^2}}\right].
    \label{DOS1 with T}
\end{align}
The tunneling conductance $G(V,T)$ at a bias voltage $V$ and temperature $T$ is related to $N(\omega, T)$ as \citep{coleman2015introduction,tinkham2004introduction}
 \begin{align}
     G(V,T)=-\bigintss_{-\infty} ^{\infty}d\omega\, N(\omega,T)\, \dfrac{\partial n(\omega-V)}{\partial \omega},
 \end{align}
 where $n(x)=1/(\exp(x/T)+1)$ is the Fermi Function.  We will analyze  two quantities: $(1)$ tunneling conductance at zero temperature which is the zero temperature density of states, and $(2)$ zero bias tunneling conductance as a function of temperature.
   The first coincides with $N(\omega, T=0)$:
 \begin{align}
    G(\omega,0)& =N_0 \, \text{Re}\left[\bigintss_0^{2\pi} \dfrac{d\theta_\bk}{2\pi} \dfrac{\omega}{\sqrt{\omega^2-\Delta(\theta_\bk, 0)^2}}\right] \label{DOS1}
 \end{align}
 where, we remind, $\Delta (\theta_\bk, 0) = \Delta_0\, \exp(-\tan^22\theta_\bk/g))$, and the
  second is
 \begin{align}
    G(0,T)&=\dfrac{1}{2T}\bigintss_0^\infty d\omega \, N(\omega,T)\, \sech^2\dfrac{\omega}{2 T} \label{Zero Bias}.
\end{align}
At $T=0$ and  $\omega\ll \Delta_0$, we have analytically
\begin{align}
    G(\omega,0)=8\,N_0 \bigintss_{\theta_0(\omega)}^{\pi/4} \dfrac{d\theta_\bk}{2\pi} \dfrac{\omega}{\sqrt{\omega^2-\Delta(\theta_\bk, 0)^2}} \approx 8\,N_0 \bigintss_{\theta_0(\omega)}^{\pi/4} \dfrac{d\theta}{2\pi}=8\,N_0 \left(\pi/4-\theta_0(\omega)\right)\propto1/\sqrt{g \log \dfrac{\Delta_0}{\omega}}.
     \label{DOS3}
\end{align}
where  $\theta_0(\omega)$ is determined by $\Delta (\theta_0 (\omega), 0) = \omega$ and  at the smallest $\omega$ is approximately
$\pi/4 - 1/\sqrt{g \log{\dfrac{\Delta_0}{\omega}}}$.   Close to $\omega = \Delta_0$, the contribution to $G(\omega,0)$ comes from frequencies near the gap maxima at $\theta_\bk=n\, \pi/2, n=\{0,3\}$. Expanding $\Delta(\theta_\bk, 0)$ as $\Delta_0(1-4 \theta_\bk^2/g)$, we find that $G(\omega, 0)$  diverges as $\log(1-\omega/\Delta_0)$. We show the result of full numerical evaluation of $G(\omega, 0)$ in Fig. \ref{zero bias fig}(a). We show the numerical result for the zero bias tunneling conductance $G(0,T)$ in  Fig. \ref{zero bias fig}(b). The behavior is similar to that for the susceptibility, but note that the extrapolation from small but finite $T$ yields a finite value, which is almost half of the normal state $G(0,T)$.
\subsubsection{Superfluid Density}
\label{superfluid denisty section}
\begin{figure} [H]
 \centering
    \subfigure[]{\includegraphics[width= 0.44 \textwidth]{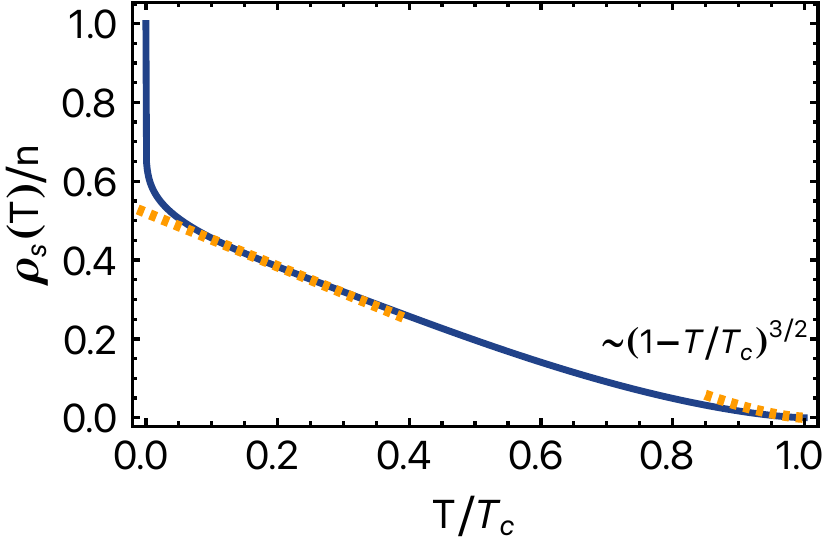}}
   \subfigure[]{\includegraphics[width= 0.5 \textwidth]{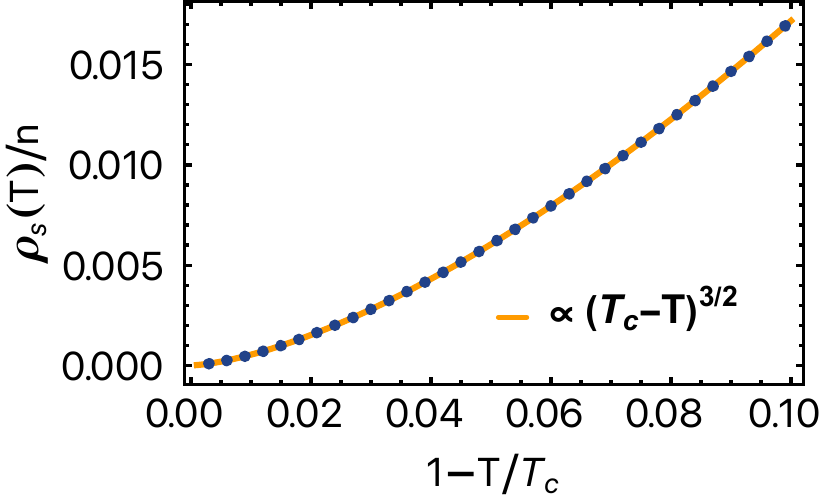}}
\caption{(a) Superfluid density, $\rho_s$ normalized by the total fermion density $n$, as a function of $T/T_c$ for NFMS. Orange dashed lines show (i) the extrapolation to $T =0$ from small but finite $T$ and (ii) the  power-law behavior $\rho_s \propto (1-T/T_c)^{3/2}$ near $T_c$. The power-law behavior is shown explicitly in (b).}
\label{SF pure}
 \end{figure}

\begin{figure}[H]
 \centering
 \includegraphics[scale=0.8]{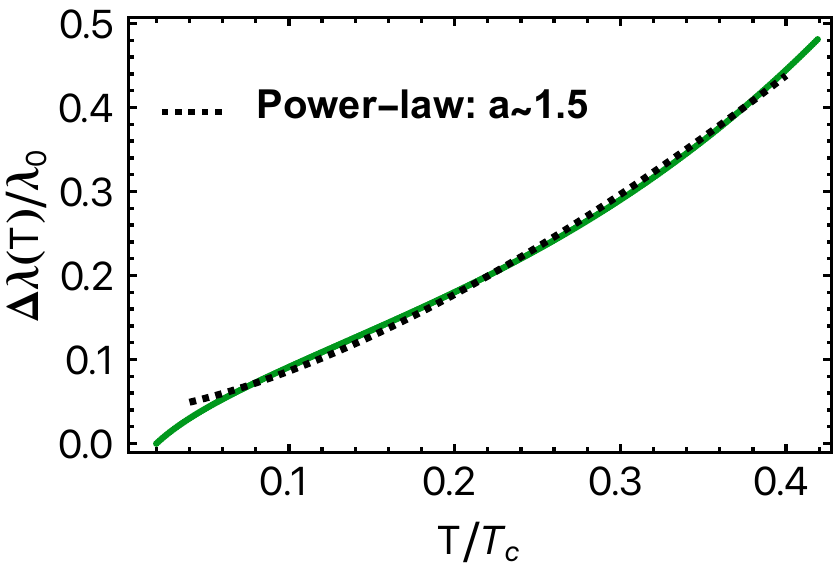}
\caption{Temperature variation of change of penetration depth  $\Delta\lambda(T)$ scaled by zero temperature penetration depth
 $\lambda_0$. The orange dashed line is a power-law fit at intermediate temperatures.}
\label{Pen pure fig}
 \end{figure}

Superfluid density $\rho_s(T)$ is related to  the current-current correlation function at $\omega = 0+$ and is
expressed as   \citep{coleman2015introduction,arikosov1963methods}
\begin{align}
    \dfrac{\rho_s(T)}{\rho_s(0)}=\pi\,T\, \bigintss_0^{2\pi}\dfrac{d\theta_\bk}{2\pi}\sum_{\omega_n}\dfrac{|\Delta(\theta_\bk,T)|^2}{(|\Delta(\theta_\bk,T)|^2+\omega_n^2)^{3/2}}=1-\bigintss_0^{2\pi} \dfrac{d\theta_\bk}{2\pi} \bigintss _{|\Delta(\theta_\bk,T)|}^\infty dE\, \dfrac{\sech^2 \dfrac{E}{2 T}}{\sqrt{E^2-\Delta(\theta_\bk,T)^2}}\dfrac{E}{2 T}.
    \label{SF eq}
\end{align}
Here, $\rho_s(0)$ is superfluid density at zero temperature, equal to the total fermion density $n$, and $\omega_n=(2\, n+1)\,\pi\, T$ is the Matsubara frequency. Above  $T_c$,  the first term in the r.h.s. of Eq.~(\ref{SF eq}) cancels the contribution  from the second term, and $\rho_s =0$. Below $T_c$, we split the integral in Eq.~(\ref{SF eq}) into two parts: superconducting part for $\theta_\bk \in[0,\theta_0(T))$ and the normal part for $\theta_\bk \in [\theta_0(T),\pi/4]$, where $\Delta(\theta_\bk)$ vanishes. Combining the two terms, we get,
\begin{align}
    \dfrac{\rho_s(T)}{n} &=1-8 \left[\dfrac{\frac{\pi}{4}-\theta_0(T)}{2\pi}+\bigintss_0^{\theta_0(T)} \dfrac{d\theta_\bk}{2\pi}\bigintss_{|\Delta(\theta_\bk,T)|}^\infty  dE \dfrac{\sech^2 \dfrac{E}{2 T}}{\sqrt{E^2-\Delta(\theta_\bk,T)^2}}\dfrac{E}{2 T}\right]=\bigintss_0^{\theta_0(t)} \dfrac{d\theta}{2\pi} f_\rho(t_{\theta_\bk})
    \label{SF 2}
\end{align}
where
\begin{align}
    f_\rho(t_{\theta_\bk})= 8\left(1-\bigintss_{\tilde{\Delta}(t_{\theta_\bk})}^\infty dE  \dfrac{\sech^2 \dfrac{E}{2 t_{\theta_\bk}}}{\sqrt{E^2-\tilde{\Delta}(t_{\theta_\bk})^2}}\dfrac{E}{2 t_{\theta_\bk}}\right).
\end{align}
 Here we introduced $t_{\theta_\bk}=T/T_c({\theta_\bk})$, $\tilde{\Delta}(t_{\theta_\bk})=\Delta(\theta_\bk)/T_c(\theta_\bk)$, and $T_c(\theta_\bk)=T_c \exp{-\tan^2 2\theta_\bk/g}$. Because
  $\tilde{\Delta}(t_{\theta_\bk})$ is not explicitly angle dependent (see Appendix \ref{Appendix C} where this is explained in detail),  the scaling function $f_\rho(t_{\theta_\bk})$ is equal to $8$ times the superfluid density of an s-wave superconductor, $\rho_s^\text{s-wave} (t_{\theta_\bk})$ at an angle dependent temperature $t_{\theta_\bk}$.  This gives
\begin{align}
    \dfrac{\rho_s (t)}{n}=\bigintss_0^{\theta_0(t)} \dfrac{d\theta_\bk}{2\pi} f_\rho(t_{\theta_\bk})=\dfrac{\sqrt{g}}{\pi} \bigintss _t^1 dx \dfrac{\rho_s^\text{s-wave}(x)}{x\, \left(1+g\, \log \dfrac{x}{t}\right)\, \sqrt{\log \dfrac{x}{t}} }\approx   \sqrt{\dfrac{g}{\pi}}\, \rho_s^\text{s-wave}(t)\, \text{Erfi}[\sqrt{g |\log t|}].
\end{align}
 where $t = T/T_c$.
 Near the transition point ($t\approx1$), $\rho_s^\text{s-wave} \propto (1-t)$ and $\text{Erfi}[\sqrt{|\log t|}] \approx \sqrt{1-t}$. Combining, we find that near $T_c$, the superfluid density  increases as $\rho_s(t)\propto (1-t)^{3/2}$. At low temperature $t\ll 1$, $\rho^\text{s-wave} \approx 1$, while $\text{Erfi}[\sqrt{|\log t|}] \approx 1/\sqrt{\log t}$. As a result, $\rho_s (t)$ approaches $n$ such that  $1- \rho_s/n \sim 1/\sqrt{\log(t)}$. To find the full temperature dependence of  $\rho_s(T)$ at $T < T_c$,  we numerically integrate Eq.~(\ref{SF eq}). We show the result in Fig.  \ref{SF pure}(a,b). We clearly see $(T_c-T)^{3/2}$ behavior near $T_c$ and rapid increase of $\rho_s$ towards $n$ at the lowest $T$.  At the same, if one extrapolates $\rho_s (T \to 0)$  from small but finite $T$, one would obtain $\rho_s (T\to 0)$ at a fraction of $n$. The London penetration depth $\lambda(T)$ is expressed via $\rho_s(T)$ as $\lambda(T)=1/\sqrt{\rho_s(T)}$. We plot $\Delta\lambda(T)=\lambda(T)-\lambda(T=0)$ in Fig.  \ref{Pen pure fig} for $t <0.4$.
 The black solid line shows a power-law fit $\Delta\lambda(T) \sim T^a$. We find that $a\approx 1.5$ captures the low temperature behavior of $\Delta\lambda(T)$ quite accurately.

  A word of caution.  In the calculations above we assumed the validity of local electrodynamics.  It has been pointed out~\cite{kosztin1997nonlocal} that in the case of a nodal gap, there is a range near the nodes where the angle-dependent coherence length $\xi (\theta_\bk) = v_F/\pi |\Delta (\theta_\bk)|$ becomes larger that London penetration depth $\lambda_0 = (mc^2/(4\pi n e^2)^{1/2}$.  Once this happens, the electrodynamics becomes non-local and the temperature dependence of the penetration depth changes. This certainly happens in our case as the gap near cold spots is exponentially small. However, we argue that the temperature below which non-local effects become relevant is very small. The reasoning is that, at zeroth-order approximation, our gap function can be viewed as discontinuous -- existing in finite ranges around hot spots, where it is  weakly angle-dependent, and vanishing  in the regions near cold spots.  In the case, the temperature dependence of the penetration depth is comes from the ranges where $\xi << \lambda_0$. Beyond zero-order approximation, $\Delta (\theta_\bk)$ is a continuous function of the angle at $T=0$, and $\xi (\theta_\bk) >\lambda_0$ in cold regions.  However, the value of $\Delta (\theta_\bk)$ is very small in the whole cold region, and the crossover from cold to hot regions is rather sharp.  As a result, we expect that
 non-local electrodynamics will be relevant only at very low $T$, in the range where in Fig. \ref{SF pure}(a)
 $\rho_s$ rapidly goes up and in Fig. \ref{Pen pure fig} $\Delta \lambda (T)$ dives down.

\subsubsection{Raman Intensity}
\begin{figure}
 \centering
    \includegraphics[scale=.8]{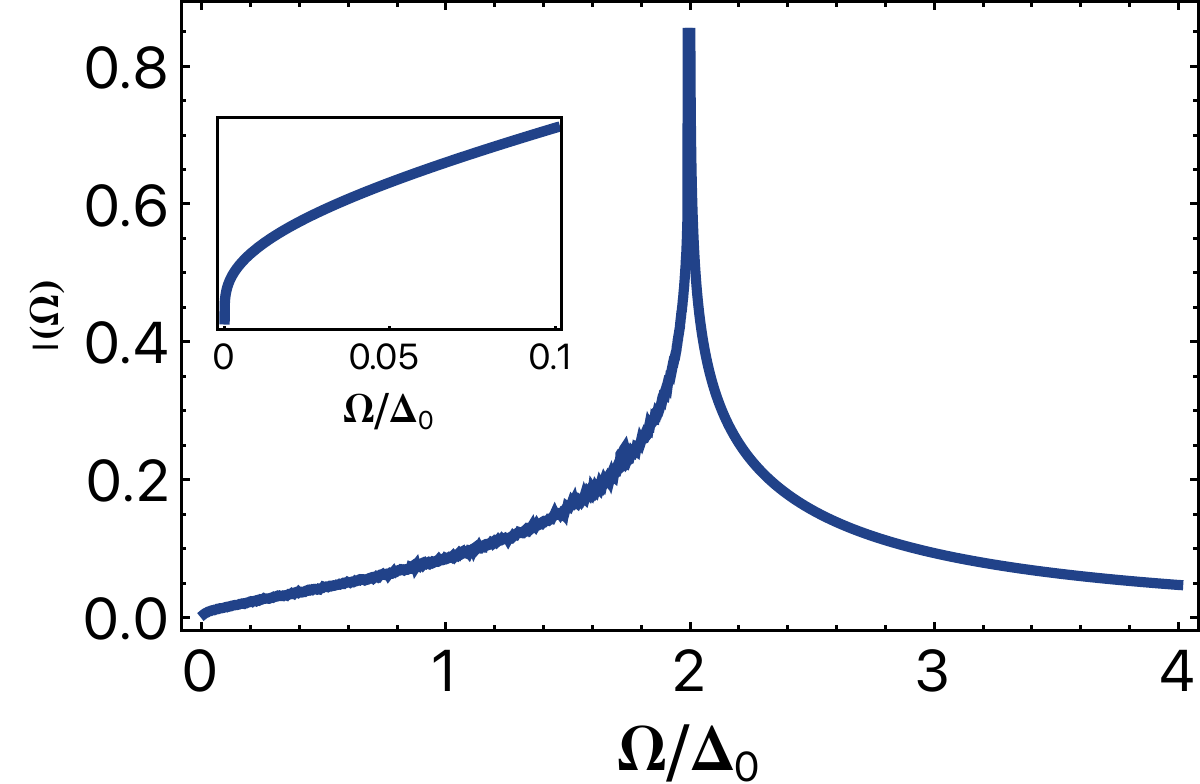}
\caption{$B_{1g}$ Raman Intensity $I$ for NFMS as a function of frequency $\Omega$ at $T=0$. The low frequency behavior of $I$ is shown in the inset. }
\label{Raman figure}
 \end{figure}

  Raman intensity $I(\Omega)$ is related to the dynamic density-density like correlation function with finite frequency $\Omega$ and vanishing momenta $\bq$, and weighted with symmetry related form factors a.k.a Raman vertices. We compute Raman intensity with B$_{1g}$ vertices ($\cos 2\theta$) which at zero temperature takes the following expression (we ignore vertex correction in B$_{1g}$ channel) \citep{devereaux2007inelastic,chubukov2009theory,klein1984theory,chubukov2000relative,chubukov1999electronic,chubukov2006resonance}
  \begin{align}
        I(\Omega)= -\dfrac{N_0}{\Omega} \text{Im}\left[ \bigintss_0^{2\pi} \dfrac{d\theta_\bk}{2\pi} \cos^22\theta_\bk  \dfrac{\Delta^2(\theta_\bk,0) \sin^{-1} \dfrac{\Omega}{2\, \Delta(\theta_\bk,0)}}{\sqrt{\Delta^2(\theta_\bk,0)-\left(\dfrac{\Omega}{2}\right)^2}}\right]=\dfrac{4 N_0}{\Bar{\Omega}} \bigintss_{\theta_0(\Bar{\Omega}/2)}^{\pi/4} d\theta_\bk\dfrac{\cos^22\theta_\bk f^2(\theta_\bk)}{\sqrt{\Bar{\Omega}^2-4 f^2(\theta_\bk)}}
        \label{Raman 2}
    \end{align}
where we remind that $\Delta (\theta_\bk, 0) = \Delta_0\, f(\theta_\bk)$ is the zero temperature superconducting gap with angular dependence $f(\theta_\bk) =\exp(-\tan^22\theta_\bk/g)$, $\Bar{\Omega}=\Omega/\Delta_0$ and the lower integration cut-off $ \theta_0(x)$ is defined in Eq.~\eqref{theta function}.  We numerically integrate Eq.~(\ref{Raman 2}) and show our result in Fig. \ref{Raman figure}.
 At low frequency $\Bar{\Omega}\ll 1$, we expand the integrand near the cold spot $\theta=\pi/4$ and define $\delta\beta =\pi/4-\theta_0(\bar{\Omega}/2)\approx 1/2\sqrt{\log 2\Delta_0/\Omega}$. This gives us analytically (let's take $g=1$ for the simplicity)
    \begin{align}
        I(\Omega)=& \dfrac{4\, N_0}{\Bar{\Omega}} \bigintss_{\theta_0(\Bar{\Omega}/2)}^{\pi/4} d\theta_\bk \, \dfrac{\cos^2 2\theta_\bk\, f^2(\theta_\bk)}{\sqrt{\Bar{\Omega}^2-4 f^2(\theta_\bk)}} \approx \dfrac{16 N_0}{\Bar{\Omega}^2} \bigintss_0^{\delta\beta} d\tilde{\theta} \tilde{\theta}^2\exp^{-1/2\tilde{\theta}^2}\propto \left[\log(\dfrac{2 \, \Delta_0}{\Omega})\right]^{-5/2}.
        \label{Raman 3}
    \end{align}
We note that the power $5/2$  of $1/\log(2\Delta_0/\Omega)$ at low frequency regime is a signature of considering $B_{1g}$ vertices. For $A_{1g}$ channel (ignore the vertex correction to consider the particle number conservation), we would get $I(\Omega) \propto 1/\sqrt{\log(2\Delta_0/\Omega)}$. Close to $\Omega=2 \Delta_0^+$, the contribution to Eq.~\eqref{Raman 2} comes from the gap maxima (hot spots: $\theta_\bk=n\, \pi/2, n=\{0,3\}$).
We expand the gap function near these points where $\Delta(\theta_\bk)=\Delta_0(1-4\, \theta_\bk^2)$  and $\cos^22\theta_\bk \approx 1$ and find
\begin{align}
   I(\Bar{\Omega})\propto \bigintss_0^{W}\dfrac{1}{\sqrt{\Bar{\Omega}-2(1- 4\, \theta_\bk^2)}}\propto \log \left(\dfrac{\Omega}{2\Delta_0}-1\right)
\end{align}
    where $W=O(1)$ is some arbitrary upper cut-off. At large frequency $\Bar{\Omega}\gg 2$, $I(\Bar{\Omega})\propto 1/\Bar{\Omega}^2$.

 \section{Effects of Disorder}
\label{disorder section}
In this Section we study the effects on NFMS from scattering by non-magnetic impurities.
We analyze how the impurity scattering affects the transition temperature, gap structure and superfluid density.
  We also compute optical conductivity.
   We consider a uniform distribution of $n_\text{imp}$ impurity centers with a static s-wave scattering potential $U_\text{dis}(\bk-\bp)=U_0$, and treat  electron-impurity scattering within the self consistent Born approximation~\citep{arikosov1963methods}.
   To simplify the formulas, we again write $\Delta (\bk, T)$ as $\Delta (\bk) = \Delta (\theta_k)$.
     The translationally invariant normal ($G$) and anomalous ($F$) Matsubara Green's functions, averaged over
      the impurity positions, have the form
\begin{align}
    G(\bk,\omega_m)&=\dfrac{i\, \omega_m+\Sigma_1(\omega_m)+\xi_\bk}{\left(i\, \omega_m+\Sigma_1(\omega_m)\right)^2-\xi_\bk^2-\left(\Delta(\bk)+\Sigma_2(\omega_m)\right)^2} \nn
    F(\bk,\omega_m) &=\dfrac{\Delta(\bk)+\Sigma_2(\omega_m)}{\left(i\, \omega_m+\Sigma_1(\omega_m)\right)^2-\xi_\bk^2-\left(\Delta(\bk)+\Sigma_2(\omega_m)\right)^2},
    \label{impure Green's function}
\end{align}
where  $\Sigma_1 (\omega_m)$ and  $\Sigma_2(\omega_m)$ are normal and anomalous components of the impurity-induced self energy.
 Both are calculated self-consistently:  $\Sigma_1(\omega_m)=-n_\text{imp} \bigintss \dfrac{d^2\bp}{(2\pi)^2} |U_\text{dis}(\bk-\bp)|^2\, G(\bp,\omega_m)$ and $\Sigma_2(\omega_m)=-n_\text{imp} \bigintss  \dfrac{d^2\bp}{(2\pi)^2} |U_\text{dis}(\bk-\bp)|^2\, F(\bp,\omega_m) $.
  %
  Using the expressions for the Green's functions (\ref{impure Green's function}), and approximating $\bigintss  \dfrac{d^2\bp}{(2\pi)^2}$ by $N_0 \bigintss \dfrac{d\theta_\bp}{2\pi} \bigintss d\xi_\bp$,
    we obtain the standard expressions
\begin{align}
   \Sigma_1(\omega_m) & =\Gamma_0 \bigintss_0^{2\pi} \dfrac{d\theta_\bp}{2\pi} \dfrac{i\, \Tilde{\omega}_m(\omega_m)}{\sqrt{\Tilde{\omega}_m^2(\omega_m)+\Tilde{\Delta}^2(\theta_\bp,\omega_m)}} \label{normal self energy}\\
   \Sigma_2(\omega_m)&=\Gamma_0 \bigintss_0^{2\pi} \dfrac{d\theta_\bp}{2\pi} \dfrac{\Tilde{\Delta}(\theta_\bp,\omega_m)}{\sqrt{\Tilde{\omega}_m^2(\omega_m)+\Tilde{\Delta}^2(\theta_\bp,\omega_m)}}
    \label{anamolous self energy}
\end{align}
where $\Gamma_0=n_\text{imp}\, N_0\, U_0^2\, \pi$ is the impurity scattering rate, and  $\tilde{\omega}_m$ and $\tilde{\Delta}$ are the dressed Matsubara frequency and the gap function $i\, \tilde{\omega}_m(\omega_m)=i\, \omega_m+\Sigma_1(\omega_m)$ and $\tilde{\Delta}(\theta_\bk,\omega_m)=\Delta(\theta_\bk)+\Sigma_2(\omega_m)$, which have to be computed self consistently. The gap equation in the presence of  impurities becomes
\begin{align}
    \Delta(\theta_\bk)&=
    -T \pi N_0 \sum_{\omega_m=-\Lambda}^{\Lambda}\bigintss_0^{2\pi} \dfrac{d\theta_\bp}{2\pi}\, V(\theta_\bk,\theta_\bp) \dfrac{\Tilde{\Delta}(\theta_\bp,\omega_m)}{\sqrt{\Tilde{\omega}_m^2(\omega_m)+\Tilde{\Delta}^2(\theta_\bp,\omega_m)}}
    \label{impurity gap equation}
\end{align} where, we remind, that $V(\theta_k, \theta_p)$ is the pairing interaction, and  $\Lambda$ is the upper cut-off for the theory.
  It is convenient to define an auxiliary function $\psi(\theta_\bk,\omega_m)=\omega_m\,\dfrac{\Tilde{\Delta}(\theta_\bk,\omega_m)}{\Tilde{\omega}_m(\omega_m)}$ and re-express Eqs.(\ref{normal self energy}-\ref{impurity gap equation})  as a set of two self-consistent equations for $\psi$ and $\Delta$:
 \begin{align}
     \psi(\theta_\bk,\omega_m)&=\dfrac{\Delta(\theta_\bk)+\Gamma_0\, \bigintss_0^{2\pi} \dfrac{d\theta_\bp}{2\pi}\dfrac{\psi(\theta_\bp,\omega_m)}{\sqrt{\psi^2(\theta_\bp,\omega_m)+\omega_m^2}}}{1+\Gamma_0\, \bigintss_0^{2\pi} \dfrac{d\theta_\bp}{2\pi}\dfrac{1}{\sqrt{\psi^2(\theta_\bp,\omega)+\omega^2}}},\label{Eq 1}\\
     \Delta(\theta_\bk)&=-T\, \pi\, N_0 \sum_{\omega_m=-\Lambda}^{\Lambda}\bigintss_0^{2\pi} \dfrac{d\theta_\bp}{2\pi}\, V(\theta_\bk,\theta_\bp) \dfrac{\psi(\theta_\bp,\omega_m)}{\sqrt{ \omega_m^2+\psi^2(\theta_\bp,\omega_m)}},\label{Eq 2}
 \end{align}
For an isotropic s-wave gap, transition temperature and  gap function are unaffected by the impurity scattering \citep{anderson1959theory}. For this case,  $V(\theta_\bk,\theta_\bp)=-V_0$, and  $\psi_s(\theta_\bk)=\Delta_s(\theta_\bk)=\Delta_s$. This  make the gap equation (\ref{Eq 2}) invariant of scattering rate $\Gamma_0$. However, for a highly anisotropic NFMS gap, this is not the case as one can't ignore the angular integration.
Below we  solve Eqs. (\ref{Eq 1}-\ref{Eq 2}) for NFMS for various values of the scattering rate $\Gamma_0$ and use the solutions, $\psi(\theta_\bk,\omega_m)$ and $\Delta(\theta_\bk)$ to compute the self energy, spectral function, superfluid density and optical conductivity.
\subsection{Transition Temperature}
As before, we  focus on the pairing right at the nematic QCP, when the pairing interaction
$ V(\theta_\bk,\theta_\bp) = -(g/N_0) \cos^2 2\theta_\bk \, \delta(\theta_\bk-\theta_\bp)$.
  We  discuss the effect of impurity on the transition temperature and whether
  the impurity scattering breaks the degeneracy between $s,p$, and $d-$channels and uniquely determine the gap function.  We label the superconducting transition temperature for the clean and disorder cases as $T_c^0$ and $T_c$, respectively, and denote their ratio as $\eta=T_c/T_c^0$. To compute $T_c$, we linearize Eq.(\ref{Eq 1}) in $\psi$ and  plug it into Eq.(\ref{Eq 2}) for $\Delta $ as $\psi(\theta_\bk,\omega_m)=\left(\Delta(\theta_\bk)+\Gamma_0\, \langle \Delta \rangle /|\omega_m|\right)/\left(1+ \Gamma_0/|\omega_m|\right)$ where $\langle\Delta\rangle=\bigintss_0^{2\pi} \dfrac{d\theta_\bp}{2\pi} \Delta(\theta_\bp)$ is the gap function averaged over the Fermi surface.  This gives
  \begin{align}
     \Delta(\theta_\bk)&=2 T_c \pi g \cos^22\theta_\bk \, \sum_{\omega_m>0}^{\Lambda}\, \dfrac{\Delta(\theta_\bk)+\Gamma_0\, \dfrac{\langle \Delta \rangle}{\omega_m}}{ \omega_m+\Gamma_0 }=\cos^22\theta_\bk \left[\Delta(\theta_\bk)\, I_1(\eta,\bar{\Gamma}_0)+\langle \Delta \rangle \, I_2(\eta,\bar{\Gamma}_0)\right],
     \label{gap linear}
 \end{align}
  with
\begin{align}
      I_1(\eta,\bar{\Gamma}_0) &= 2 T_c g\pi \sum_{\omega>0}^{\Lambda} \dfrac{1}{\omega+\Gamma_0} =g\left[\Psi\left(\dfrac{1}{2}+\dfrac{\bar{\Gamma}_0}{\eta}+\dfrac{\bar{\Lambda}}{\eta}\right)-\Psi\left(\dfrac{1}{2}+\dfrac{\bar{\Gamma}_0}{\eta}\right)\right]\approx g \left[\Psi\left(\dfrac{1}{2}+\dfrac{\bar{\Lambda}}{\eta}\right)-\Psi\left(\dfrac{1}{2}+\dfrac{\bar{\Gamma}_0}{\eta}\right)\right]\label{I1 equation}\\
      I_2(\eta,\bar{\Gamma}_0) &=2 T_c g\pi \sum_{\omega>0}^{\Lambda} \dfrac{\Gamma_0}{\omega\,\left(\omega+\Gamma_0\right)}= g\left[\Psi\left(\dfrac{1}{2}+\dfrac{\bar{\Lambda}}{\eta}\right)-\Psi\left(\dfrac{1}{2}\right)-\Psi\left(\dfrac{1}{2}+\dfrac{\bar{\Gamma}_0}{\eta}+\dfrac{\bar{\Lambda}}{\eta}\right)+\Psi\left(\dfrac{1}{2}+\dfrac{\bar{\Gamma}_0}{\eta}\right)\right] \nn
      & \approx g \left[\Psi\left(\dfrac{1}{2}+\dfrac{\bar{\Gamma}_0}{\eta}\right)-\Psi\left(\dfrac{1}{2}\right)\right]
      \label{I2 equation}
 \end{align}
 where $\Psi(z)$ is the De-Gamma function\citep{coleman2015introduction} and $\bar{\Gamma}_0=\Gamma_0/2\pi T_c^0$  and $\bar{\Lambda}=\Lambda/2\pi T_c^0$ are  the rescaled pair breaking parameter and the cutoff.
   In the clean case, $\bar{\Gamma}_0=0, \eta=1$ and $I_2=0$.
 We  see that because of $\langle \Delta \rangle$ in the r.h.s. of Eq.(\ref{gap linear}), the equation for the gap with an  s-wave structure  is impacted differently from non-s-wave gap functions. For d-wave and p-wave gaps,  $\langle \Delta \rangle=0$, the gap equation (\ref{gap linear}) is fully local in momentum space, and a superconducting instability  develops at hot spots when
 \begin{align}
     I_1(\eta,\bar{\Gamma}_0)=1.
     \label{tc for non s}
 \end{align}
  We note in passing that this condition is the same as for a BCS s-wave superconductor with magnetic impurities.

\begin{figure}
     \centering
     \includegraphics[scale=1]{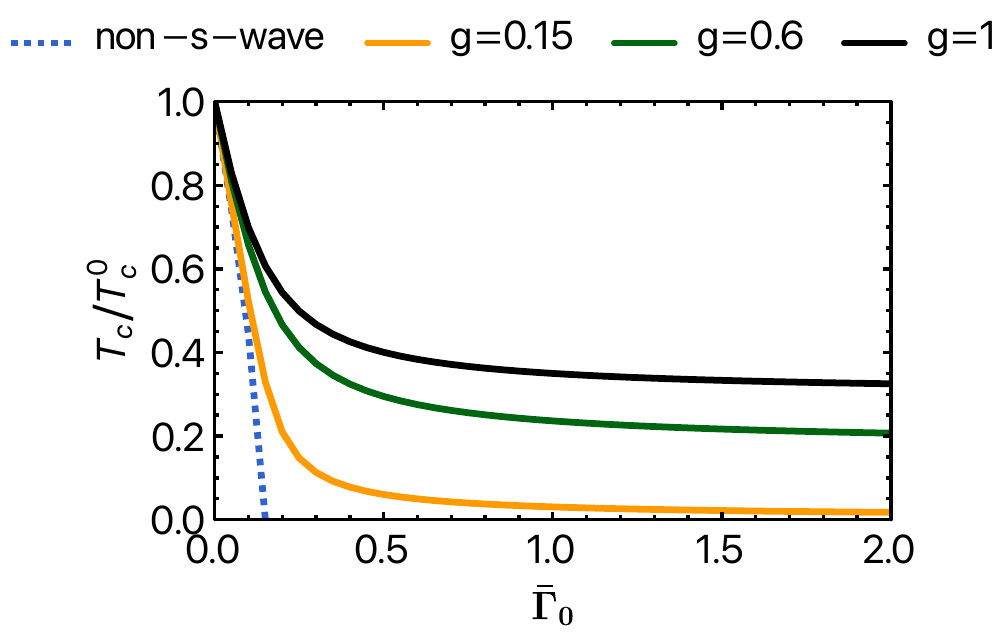}
     \caption{The transition temperature $T_c$  for NFMS in the presence of $s-$wave impurity scattering   as a function of the impurity scattering rate $\bar{\Gamma}_0=\Gamma_0/2\pi T_c^0$ ( $T_c^0$ is $T_c$ in the clean case). The  yellow, green and black lines are for s-wave gap function for different $g$, blue  dashed line is for a p-wave or d-wave gap (for them, $T_c$ does not depend on $g$).  }
     \label{Tc}
 \end{figure}

  For s-wave gap function, we compute $\langle \Delta \rangle $ using Eq.(\ref{Eq 2})  and find
 \begin{align}
     \langle \Delta \rangle&=-2T_c \pi N_0 \sum_{\omega>0}^{\Lambda} \bigintss_0^{2\pi} \dfrac{d\theta_\bk}{2\pi} \bigintss_0^{2\pi} \dfrac{d\theta_\bp}{2\pi} V(\theta_\bk,\theta_\bp)\dfrac{\Delta(\theta_\bp)+\dfrac{\Gamma_0\, \langle \Delta \rangle}{\omega_m}}{\omega_m+\Gamma_0}\\
    &= \bigintss_0^{2\pi} \dfrac{d\theta_\bp}{2\pi} \cos^22\theta_\bp \left[\Delta(\theta_\bp)\, I_1(\eta,\bar{\Gamma}_0)+\langle \Delta \rangle \, I_2(\eta,\bar{\Gamma}_0))\right]\\
    &=\dfrac{\langle \Delta \rangle}{2}\, I_2(\eta,\bar{\Gamma}_0)+ I_1(\eta,\bar{\Gamma}_0)\bigintss_0^{2\pi} \dfrac{d\theta_\bp}{2\pi} \cos^22\theta_\bp\, \Delta(\theta_\bp).
 \end{align}
 Substituting  $\Delta(\theta_\bp)$ in terms of $I_1$ and $I_2$ using Eq.(\ref{gap linear}),
  we obtain the equation for $T_c$ in the s-wave channel in the form
 \begin{align}
     1=\dfrac{I_1\, I_2}{1-I_2/2} \bigintss_0^{2\pi} \dfrac{d\theta_\bp}{2\pi} \dfrac{\cos^42\theta_\bp}{1- I_1\, \cos^2 2\theta_\bp}= \dfrac{I_1\, I_2}{1-I_2/2}\left[\dfrac{-2+2\sqrt{1-I_1}+I_1+I_1^2}{2(1-I_1)I_1^2}\right],
     \label{self consistency 1}
 \end{align}
 where we have suppressed the $\eta$ and $\bar{\Gamma}_0$ dependence of $I_1$ and $I_2$ for the simplicity of the notation.  The last equation can be re-expressed as
 \begin{align}
     (I_1+I_2)\sqrt{1-I_1}=I_2.
     \label{Tc for s}
 \end{align}
 We  solve Eqs.(\ref{tc for non s}) and (\ref{Tc for s}) numerically and show our results for $T_c/T_c^0$  as a function of impurity scattering rate $\bar{\Gamma}_0$  in Fig.  \ref{Tc} for various values of dimensionless interaction $g$.
 We  see that (i) an s-wave impurity scattering breaks the degeneracy between  s-wave and non-s-wave channels and favors s-wave as the leading instability, although in the weak disorder limit, the fall of $\log T_c/T_c^0$ is the same in all pairing channels;  (ii) The transition temperature, $T_c$ in all channels decreases with the  increase of the impurity scattering rate $\Gamma_0$. For non s-wave gap functions, $T_c$ vanishes at a critical scattering strength, while for an s-wave gap it saturates at large disorder strength; (iii) the suppression of $T_c$ is in the s-wave channel is stronger at smaller $g$. This is the consequence of the exponential dependence of the gap width $\Delta_0(\theta)=\exp{-\tan^2 2\theta/g}$ on the coupling strength $g$ in the clean case. The width is  smaller at smaller $g$, and this leads to stronger suppression of $T_c$ by disorder; (iv) below the transition temperature, the gap opens up everywhere on the Fermi surface except the cold spots.\newline
 The behavior of $T_c/T_c^0$ in  different  channels in the weak and strong disorder limits can be understood analytically. For $\bar{\Gamma}_0\ll 1  \, (T_c \approx T_c^0, \, I_1 \gg I_2)$, we expand Eqs.(\ref{I1 equation},\ref{I2 equation})
  to the quadratic order in  $ \bar{\Gamma}_0$ and obtain
 \begin{align}
     I_2 &= g \left[\dfrac{\pi^2}{2} \bar{\Gamma}_0+ \dfrac{\text{PolyGamma}(2,\dfrac{1}{2})}{2} \bar{\Gamma}_0^2\right] \label{I2 approx}\\
     I_1 &= 1-g \log \dfrac{T_c}{T_c^0}-I_2 \label{I1 approx}.
 \end{align}For non s-wave channels, Eq. (\ref{tc for non s}) for $T_c$ gives
  \begin{align}
      \log \dfrac{T_c}{T_c^0}=-\dfrac{\pi^2}{2} \bar{\Gamma}_0- \dfrac{\text{PolyGamma}(2,\dfrac{1}{2})}{2} \bar{\Gamma}_0^2.
      \label{d wave tc}
  \end{align}
 For the s-wave channel, Eq.(\ref{Tc for s}) for $T_c$  gives
 \begin{align}
      \log \dfrac{T_c}{T_c^0}=-\dfrac{\pi^2}{2} \bar{\Gamma}_0- \bar{\Gamma}_0^2 \left(\dfrac{\text{PolyGamma}(2,\dfrac{1}{2})}{2} -g \dfrac{\pi^4}{4}\right).
       \label{s wave tc}
 \end{align}
Comparing Eq.(\ref{d wave tc}) with Eq.(\ref{s wave tc}), we see
that the transition temperature for the s-wave gap is higher because of an extra  $(\pi^2/4) g \, \bar{\Gamma}_0^2 $ in the r.h.s. of (\ref{s wave tc}).
At $T_c$, the s-wave gap varies as
 \begin{align}
    \Delta(\theta_\bk)\propto \dfrac{I_2\, \cos^2 2\theta_\bk}{1-I_1\, \cos^22\theta_\bk}=\dfrac{I_2}{\tan^22\theta_\bk+I_2^2}.
\end{align} The gap has a  maximum at the hot spots, has a width set by $I_2$(\ref{I2 approx}), and vanishes at the cold spots. \newline In the opposite limit of  strong disorder ${\bar \Gamma}_0 \gg 1 (\Gamma_0 \gg T_c)$, we expand $I_1$ and $I_2$ in $1/{\bar \Gamma}_0$ using   Eqs.(\ref{I1 equation}-\ref{I2 equation}) and obtain
\begin{align}
    I_1 = g\, \log(1+\dfrac{\Lambda}{\Gamma_0}),~~ I_2= 1- g\, \log \dfrac{T_c}{T_c^0}-g\, \log\left(1+\dfrac{\Lambda}{\Gamma_0}\right).
\end{align}
 We see that $I_2 \gg I_1$. We then expand Eq(\ref{Tc for s}) for the s-wave gap  in $I_1/I_2$ and
  obtain
\begin{align}
    \dfrac{T_c}{T_c^0}=\dfrac{\Gamma_0}{\Lambda+\Gamma_0}\, e^{-1/g}
\end{align}
We see that $T_c$ for s-wave gap saturates at a finite value, as we found numerically.

 \subsection{Gap Structure: $\psi(\theta,\omega_m)$ and $\Delta(\theta_\bk)$}
 \label{gap structure section}
\begin{figure}[H]
 \centering
   \subfigure[]{\includegraphics[width= 0.46 \textwidth]{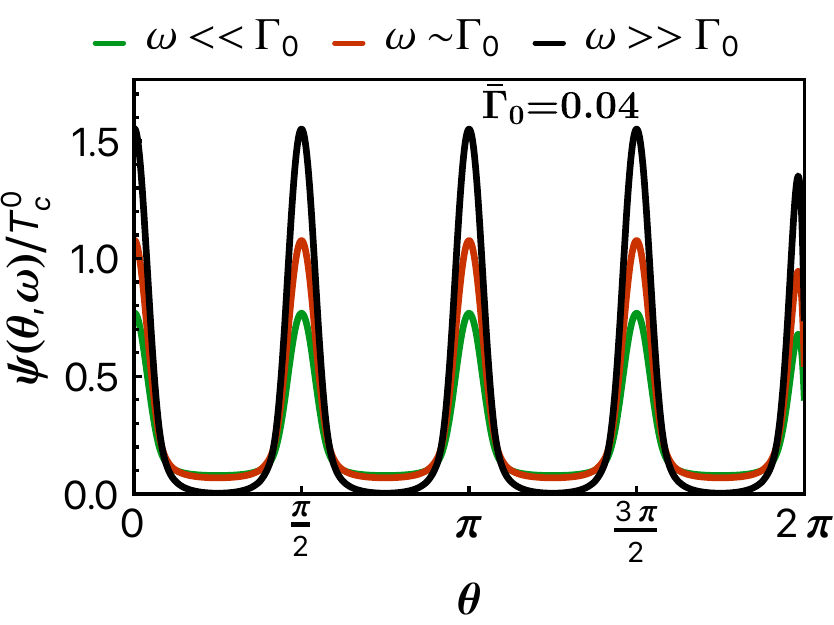}}\hspace{1 cm}
   \subfigure[]{\includegraphics[width= 0.46 \textwidth]{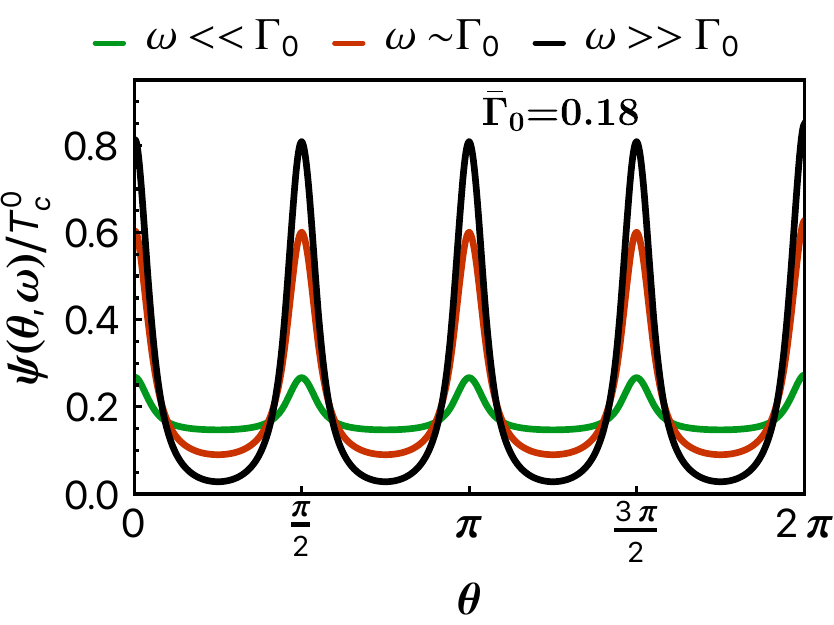}}
   \subfigure[]{\includegraphics[width= 0.46 \textwidth]{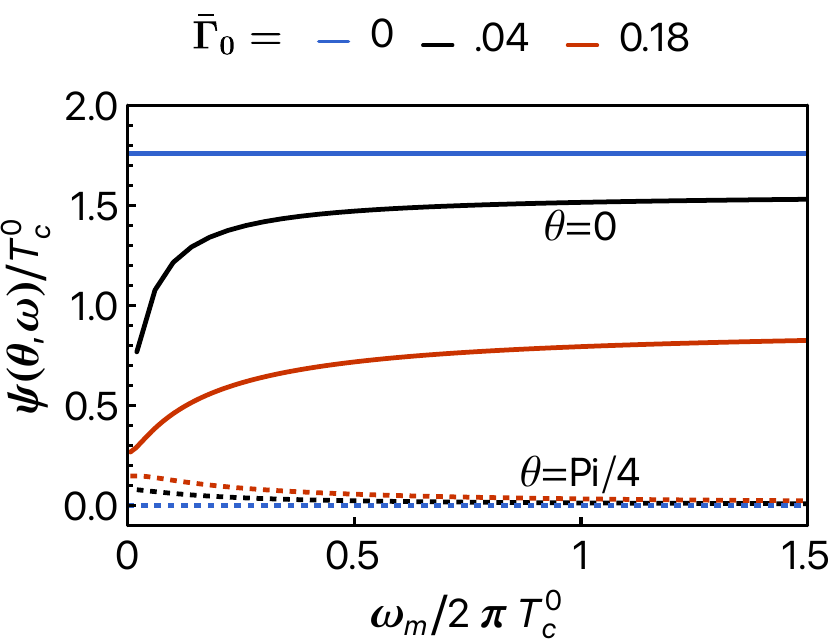}}\hspace{1 cm}
   \subfigure[]{\includegraphics[width= 0.46 \textwidth]{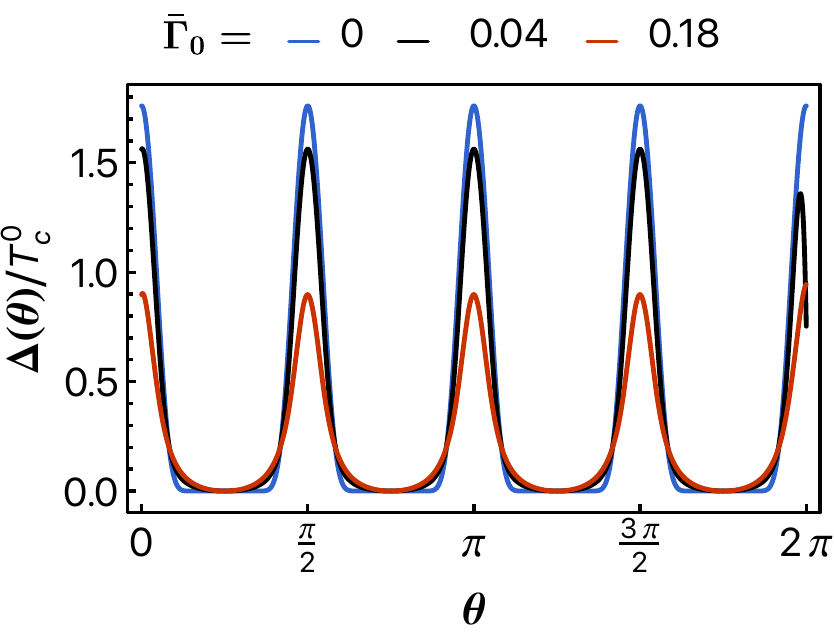}}
\caption{\label{psi sol}
  (a and b) Variation of the auxiliary function $\psi(\theta_\bk,\omega_m)$ with an angle $\theta_\bk$ in  different frequency ranges depending on  $\omega_m/\Gamma_0$, where $\Gamma_0$ is the impurity scattering rate
     for $\bar{\Gamma}_0 = \Gamma_0/2\pi T_c^0 =0.04$ in (a)  and 0.18 in (b).   (c) Variation of $\psi(\theta_\bk,\omega)/T_c^0$ as a  function of $\omega_m$ at a hot spot (solid curve) and a cold spot (dashed curve) for the clean case $T_c = T_c^0$ and two different $T_c/T_c^0$ in the presence of impurity scattering. (d) Angular variation of the superconductor gap function $\Delta(\theta_\bk)$ for the clean case (blue line) and the cases of weak disorder (black line) and strong disorder (red line). We set $T=0$ and  $g=0.15$.}
 \end{figure}

 We numerically solved the non-linear equations (\ref{Eq 1}-\ref{Eq 2})  for the superconducting order parameter $\Delta(\theta_\bk,T)$ and the auxiliary function $\psi(\theta_\bk,\omega_m,T)$  for various impurity scattering rates $\Gamma_0$. In the clean case, $\psi$ is frequency independent and equal to $\Delta_0(\theta_\bk)$. We scale both
  $\Delta$ and $\psi$ by $T_c^0$ and plot the angular and frequency dependence of $ \psi/T_c^0$ and $\Delta/T_c^0$ in Fig.~\ref{psi sol} for the reduced temperature $t=T/T_c=0.05$  and for different disorder strengths
specified by $T_c/T_c^0$. In the presence of  impurity scattering, $\psi (\theta_\bk,\omega_m,T)$ develops strong dependence on both the angle and the frequency. It gets reduced near the hot spots compared to $\Delta_0 (\theta_\bk)$, and increases near  the cold spots. The effect is more prominent at the low frequencies $\omega \ll \Gamma_0$ where, with increasing disorder strength, $\psi(\theta_\bk)$ becomes progressively more isotropic (green curve in Fig.  \ref{psi sol} (a,b)). At  larger frequencies, $\psi$ remains highly anisotropic( red and black curves
   in Fig. \ref{psi sol}(a,b)). In Fig.  \ref{psi sol}(c), we plot the frequency dependence of $\psi$ at hot  and cold spots ((solid and dashed curves) for two different disorder strengths and compare it with the clean  case
   (the blue curve). We again see that the effects of disorder are the strongest at small frequencies.
    in Fig. \ref{psi sol}(d) we  plot the angular variation of $\Delta(\theta_\bk)/T_c^0$ for weak and strong disorder and compare it with the clean case.
 \subsection{Fermionic  Self Energy}
 \label{impurity self energy section}
The normal and anomalous self energies $\Sigma_1(\omega_m)$ and $\Sigma_2(\omega_m)$ are given by Eqs. (\ref{normal self energy}-\ref{anamolous self energy}). In terms of the auxiliary function $\psi$, they are
  \begin{align}
      \Sigma_1(\omega_m) &= i\, \Gamma_0\, \omega \bigintss_0^{2\pi} \dfrac{d\theta\bp}{2\pi}\dfrac{1}{\sqrt{\omega_m^2+\psi^2(\theta_\bp,\omega_m)}}\label{Mats self energy 1}
      \end{align}
      \begin{align}
       \Sigma_2( \omega_m) &=  \Gamma_0\,  \bigintss_0^{2\pi} \dfrac{d\theta\bp}{2\pi}\dfrac{\psi(\theta_\bp,\omega_m)}{\sqrt{\omega_m^2+\psi^2(\theta_\bp,\omega_m)}}.
       \label{Mats self energy 2}
  \end{align}
 We recall that for an s-wave BCS superconductor,
$\Sigma_1^\text{BCS}(\omega_m)=i\, \Gamma_0 \, \omega_m/\sqrt{\omega_m^2+\Delta_s^2}$ and $\, \Sigma_2^\text{BCS}(\omega_m)= \Gamma_0 \, \Delta_s/\sqrt{\omega_m^2+\Delta_s^2}$. At small frequencies $\omega_m \ll \Delta_s$, $\Sigma_1^\text{BCS} \propto \omega_m$ with a slope $\Gamma_0/\Delta_s$ and $\Sigma_2^\text{BCS}\approx\Gamma_0$, while at large frequencies $\omega_m \gg \Delta_s$ $\Sigma_1^\text{BCS} \approx i\Gamma_0$ and $\Sigma_2^\text{BCS}$ falls as $1/\omega_m$.  For real frequencies, the retarded self-energies are
$\Sigma_{1R}^\text{BCS}=\Gamma_0\, \omega/\sqrt{\Delta_s^2-(\omega+i\, \delta)^2}$ and $ \Sigma_{2R}^\text{BCS}=\Gamma_0\, \Delta_s/\sqrt{\Delta_s^2-(\omega+i\, \delta)^2}$.

\begin{figure}
 \centering
   \subfigure[]{\includegraphics[width= 0.45 \textwidth]{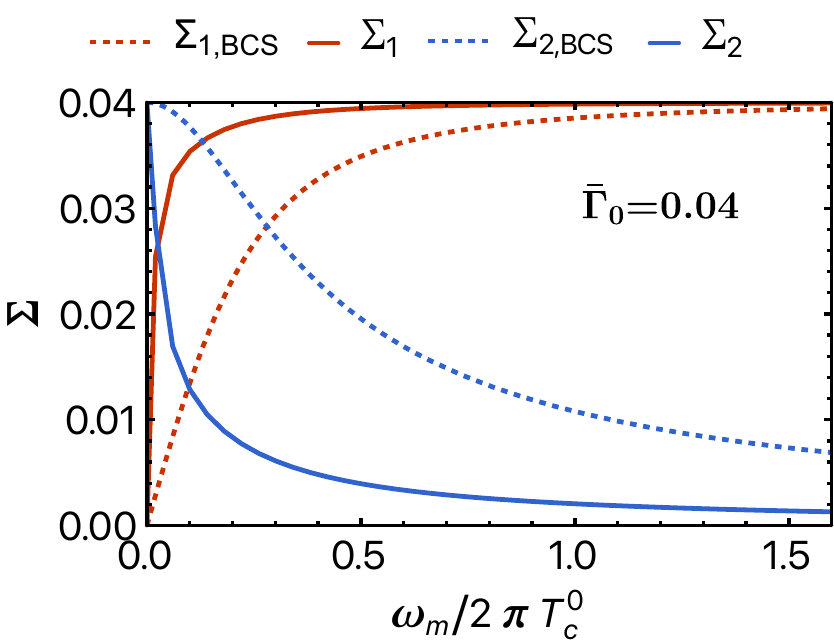}}\hspace{1 cm}
   \subfigure[]{\includegraphics[width= 0.45 \textwidth]{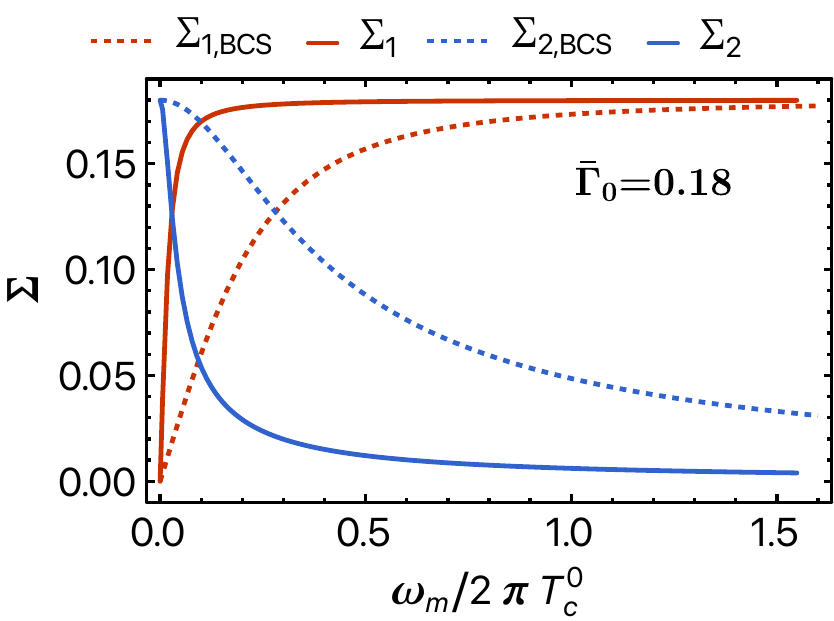}}
   \subfigure[]{\includegraphics[width= 0.45 \textwidth]{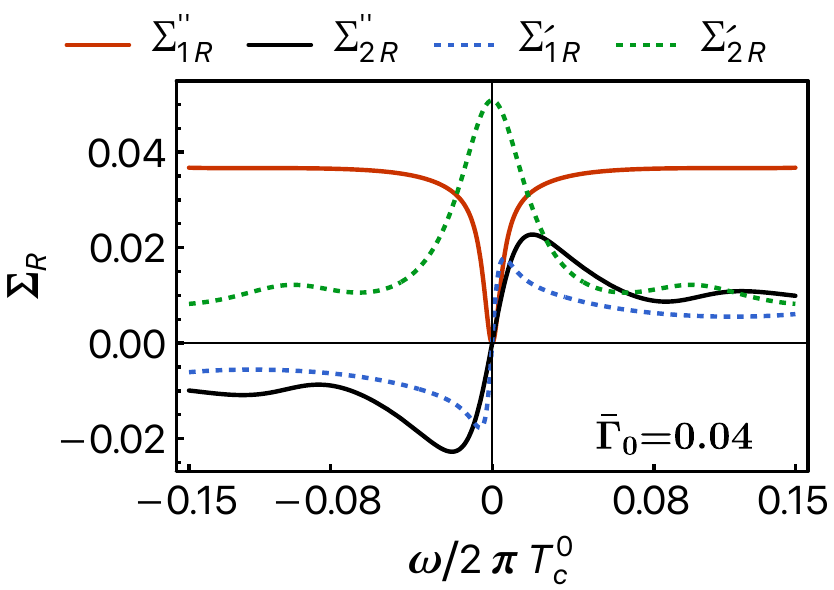}}\hspace{1 cm}
   \subfigure[]{\includegraphics[width= 0.45 \textwidth]{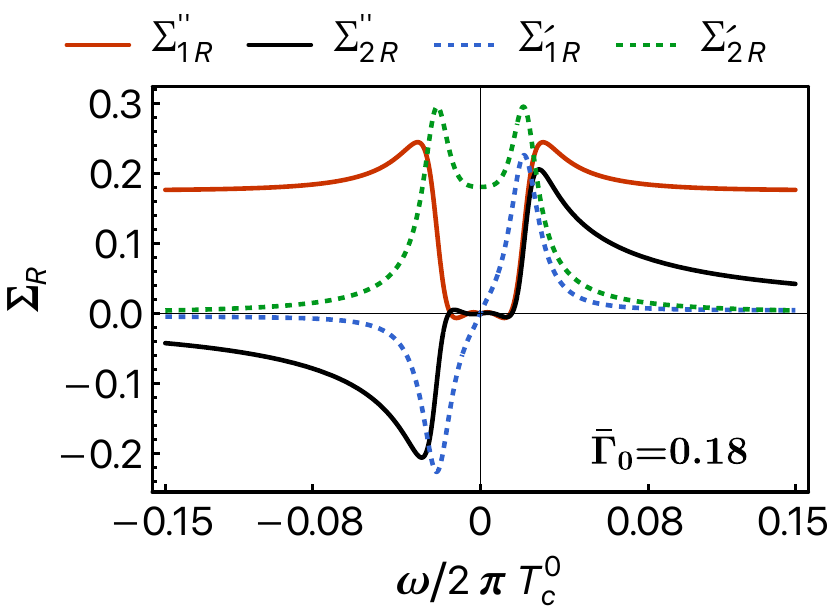}}
\caption{\label{self energy diagram}
   Panels (a) and (b) --frequency dependence of the normal $(\Sigma_1)$ and anomalous ($\Sigma_2$) Matsubara self energies (red and blue curves, respectively)  for  NFMS (solid curves) and a dirty BCS superconductor (dashed curves). Panel  (a) -- weak disorder, panel (b) -- strong disorder. Panels (c) and (d) --  frequency dependence of the real and imaginary  components of $\Sigma_{1R}$ and $\Sigma_{2R}$  for NFMS (solid and dashed curves, respectively). The frequency is in units $T_c^0$ in the clean limit.}
 \end{figure}

For our case, we plug the solution of the gap equation (\ref{Eq 1}-\ref{Eq 2}): $\psi(\theta_\bk,\omega)$ into Eqs.(\ref{Mats self energy 1}-\ref{Mats self energy 2}) and compute the disorder self energies numerically at zero temperature. We show the results in Fig. \ref{self energy diagram} (a,b) (solid curve) for two sets of scattering rates. For comparison, we also plot $\Sigma^\text{BCS}(\omega_m)$  (dashed curve).
  For the normal self energy $\Sigma_1(\omega_m)$ (which is purely imaginary) we find that at low frequency $\Sigma_{1}$ in our case rises much faster than $\Sigma_{1,\text{BCS}}$.  This can be understood qualitatively by noticing that in our case  the dominant contribution to $\Sigma_1( \omega_m)$  at low frequency comes from the range near cold spots, where $\psi (\pi/4, \omega_m) \ll \Delta_s$. We also see that the  anomalous self energy $\Sigma_2$ falls much faster than $\Sigma_{2,\text{BCS}}$. This again is the consequence of the smallness of $\psi (\pi/4, \omega_m)$ compared to
   $\Delta_s$.  To find the retarded self energy from the data for $\Sigma (\omega_m)$,  we use the Pade approximation technique \citep{schott2016analytic}.  We  plot real and imaginary parts of $\Sigma_{1R}(\omega)$ and $\Sigma_{2R}(\omega)$  as  functions of frequency in Fig.  \ref{self energy diagram}(c,d) for two values of ${\bar \Gamma}_0$. A simple analysis of these plots shows that the effects of disorder in our case are much stronger than for a dirty BCS superconductor.  We see in particular that  with increasing disorder,
   low energy excitations  becomes gaped and nodes are lifted. This is a consequence of
    the fact that  $\psi(\theta_\bk,\omega_m)$
     becomes non-zero even at cold spots, $\theta_\bk = \pi/4 + n \pi/2$, and the magnitude of  $\psi(\pi/4,\omega_m)$ increases with ${\bar \Gamma}_0$ (see Fig. \ref{psi sol}(b)).

 \subsection{Spectral Function at zero temperature}
 \label{impurity effect on gap function}
\begin{figure}
 \centering
   \subfigure[]{\includegraphics[width= 0.45 \textwidth]{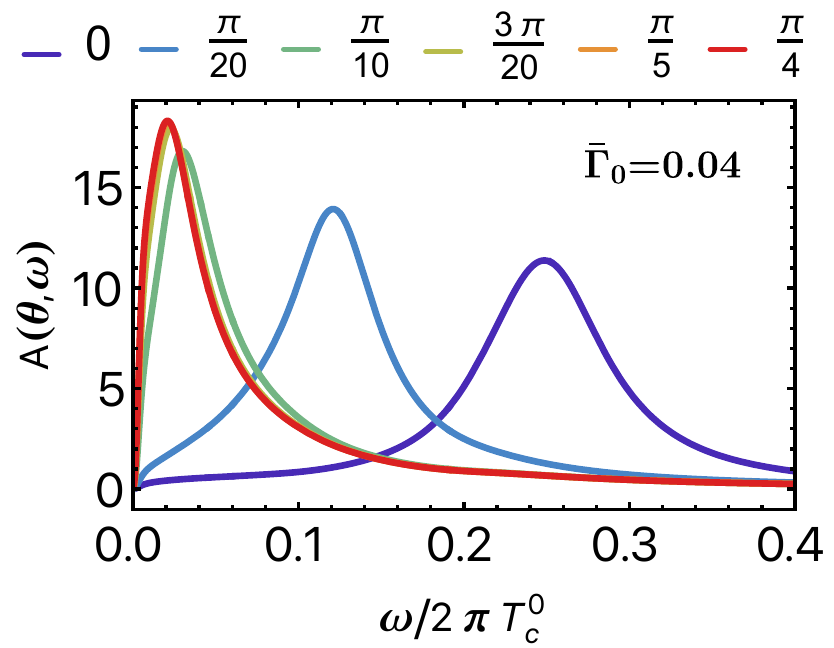}}\hspace{1 cm}
   \subfigure[]{\includegraphics[width= 0.45 \textwidth]{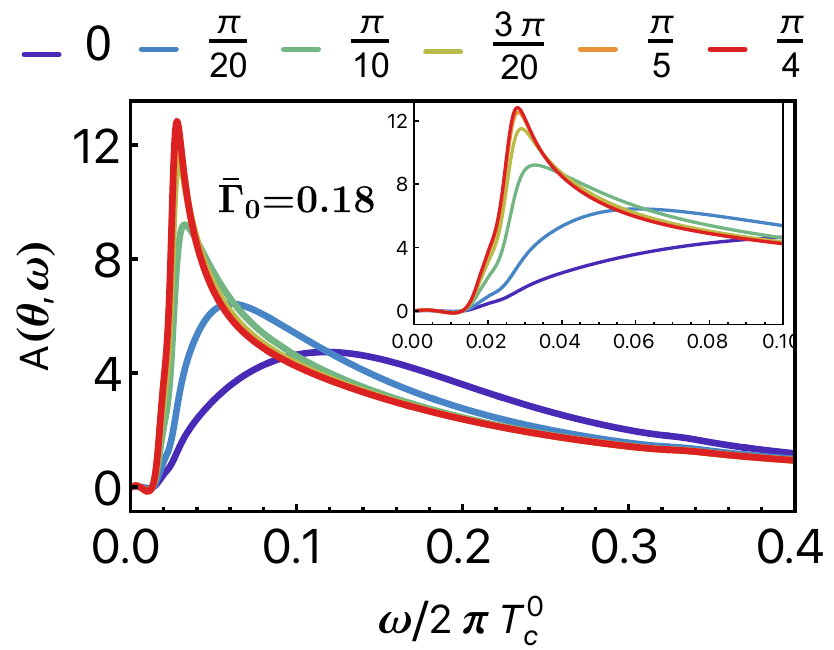}}
\caption{\label{spectral function diagram}
 Frequency dependence of the  spectral function $A(\theta_\bk,\omega)$ for NFMS at $T=0$ for a set of angles on the Fermi surface between the hot spot ($\theta_\bk=0$) and the cold spot ($\theta_\bk=\pi/4$)  for the weak and strong impurity scattering rate (panels (a) and (b), respectively). In the inset in (b), we zoomed the spectral function at low frequencies.}
 \end{figure}

 With the self-energy $\Sigma^R (\omega)$ at hand, we are in position to compute the spectral function, $A(\bk,\omega)
 -\dfrac{1}{\pi} \text{Im} G^R(\theta_\bk,\omega)$ as a function of momentum $\bk$ and real frequency $\omega$. We keep the momentum on the Fermi surface, $|\bk|=k_F$ and vary $\theta_\bk$ along the Fermi surface. The retarded Green's function is
   \begin{align}
      G^R(\theta_\bk,\omega) &=\dfrac{\omega+\Sigma_{1 R}}{(\omega+i\, \delta+\Sigma_{1R}(\omega))^2-(\Delta(\theta_\bk)+\Sigma_{2R}(\omega))^2} \nn
      & =\dfrac{1}{2}\left[\dfrac{1}{(\omega+i\, \delta+\Sigma_{1R}(\omega))-(\Delta(\theta_\bk)+\Sigma_{2R}(\omega))}+\dfrac{1}{(\omega+i\, \delta+\Sigma_{1R}(\omega))+(\Delta(\theta_\bk)+\Sigma_{2R}(\omega))}\right].
      \label{GR equation}
  \end{align}
 In the clean case,  $\Sigma$ vanishes and the spectral function $A(\bk,\omega)=\delta(\omega-\Delta(\theta_\bk))$. In the presence of impurity scattering the spectral function broadens and the frequency dependence of the self-energy becomes important. For a dirty $s-$wave superconductor,
  \begin{align}
    A(\theta_\bk, \omega) =  A(\omega)=\Theta(|\omega|-\Delta_s)\dfrac{\Gamma_0\, \omega}{\sqrt{\omega^2-\Delta_s^2}\, \left(\omega^2-\Delta_s^2+\Gamma_0^2\right)}
 \end{align}
  where $\Theta (x) = 1$ for $x >0$.
  In our case, the behavior of $A(\theta_\bk, \omega)$ is more complicated.  We computed the spectral function numerically using the results for the self-energy. We plot $A(\theta_\bk, \omega)$ in Fig. (\ref{spectral function diagram})
   as a function of frequency for a set of angles and
   for two values of ${\bar \Gamma}_0$, corresponding to weak and strong disorder.
    We list several characteristics of the spectral functions:
  (i) The width of $A(\theta_\bk, \omega)$ expectedly increases with ${\bar \Gamma}_0$. However,  unlike the s-wave case, the width strongly depends on the position of ${\bf k}$ on the  Fermi surface. The width is the largest at  hot spots (blue curve) and  the smallest at cold spots (red curve) implying  that single particle excitations live longer near cold spots; (ii)  For larger disorder strength, the excitations become more isotropic along the Fermi surface --
     the distance between the peaks of $A(\theta_\bk, \omega)$ at hot and cold spots decreases with increasing  ${\bar \Gamma}_0$; (iii) The emergence of a gap in the excitation spectrum, which we discussed above, is clearly visible in the spectral function, particularly  at larger ${\bar \Gamma}_0$, see the inset to  Fig.\ref{spectral function diagram}(b). The threshold frequency is determined by the value of the auxiliary function $\psi$ at the cold spot;

  The position of the peak of the spectral function, $\omega = \omega^*$,   at a given point on the Fermi surface is determined by ($\omega^*>0$)
     \begin{align}
         \omega^*=\Delta(\theta_\bk)+\Sigma_{2R}^\prime(\omega^*)-\Sigma_{1R}^\prime(\omega^*).
         \label{peak position}
     \end{align}
     Using the results for the self-energies,  we find that for $\theta_\bk$ close  hot spots, the contribution from the self energies is smaller than $\Delta(\theta_\bk)$ and to first approximation can be ignored, i.e.,
       $\omega^* \approx \Delta(\theta_\bk)$. The situation changes as one gets close to the cold spot. In this range of angles  $\Delta(\theta_\bk)$ is small and can be ignored compared to the self-energies.
        As a result, the  peak position is determined $\Sigma_{2R}^\prime - \Sigma_{1R}^\prime $ and become independent of $\theta_\bk$, i.e., the  excitations  above the minimum gap are flat.

\subsection{Optical conductivity}

We calculate the optical conductivity, $\sigma_\text{SC}^l(\Omega,T)$, of our superconductor  at a real frequency $\Omega$ and temperature $T$ within the standard linear response theory \cite{coleman2015introduction,arikosov1963methods}:
\begin{align}
    \sigma_\text{SC}(\Omega,T)=\dfrac{\pi e^2}{m} \, \rho_s(T)\, \delta(\Omega)- \dfrac{1}{\Omega}\text{Im}\,\Pi^R(\Omega,T),
    \label{cond equation}
\end{align}
where $e$ and $m$ are the charge and mass of the electron, $\rho_s(T)$ is the superfluid density and $\Pi^R(\Omega, T)$ is the retarded current-current correlation function given by
\begin{align}
 \Pi(\Omega_m,T)=  \dfrac{2\, e^2}{m^2}\,T \sum_{\omega_m} \bigintss \dfrac{d^2\bk}{(2\pi)^2}  \bk_x^2 \left[G(\bk,\omega_m+\Omega_m)\, G(\bk,\omega_m)+ F(\bk,\omega_m+\Omega_m)\, F(\bk,\omega_m)\right],
 \label{Current Current Correlation}
\end{align}
 where $G(\bk,\omega_m)$ and $F(\bk,\omega_m)$ are given by
 Eqs.(\ref{impure Green's function}).
  We assume that relevant $\bk$ are near the Fermi surface and replace $\int d^2k/(2\pi)^2$ by $m/(2\pi)^2 d \xi_k d \theta_\bk$, $k^2_x$ by $k^2_F/2 = n \pi$, and impose the upper cutoff of the integration over $\xi_k$ at $|\xi_k| = \Lambda$. Within this scheme, the superfluid density, $\rho_s(T)$ is related to  $\Pi$  as \citep{coleman2015introduction,arikosov1963methods}
\begin{align}
    \rho_s(T) &= n + \dfrac{m}{e^2} \Pi(0,T),
    \label{SF equation}
\end{align}
where $n = K^2_F/(2\pi)$ is the electron density.  One can easily make sure that in the normal state
$\dfrac{m}{e^2} \Pi(0,T) =-n$ and $\rho_s =0$.
The retarded current-current correlation function, $\Pi^R(\Omega,T)$ at a finite $\Omega$ is obtained from  $\Pi(\Omega,T)$ by analytic continuation $i\, \Omega_m \rightarrow \Omega+i\, \delta$.

\subsubsection{Superfluid density and penetration depth}
\begin{figure}
     \centering
     \includegraphics[scale=0.8]{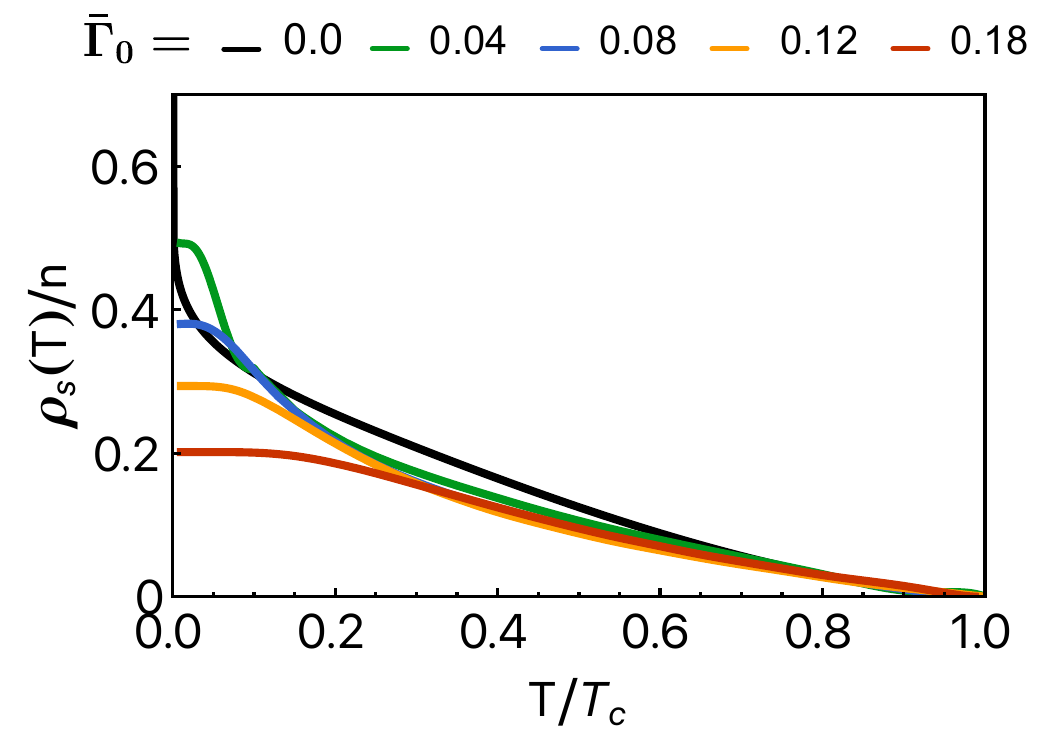}
     \caption{Superfluid density as a function of $T/T_c$ for a set of  impurity scattering rates $\bar{\Gamma}_0$, ranging from  $\bar{\Gamma}_0=0$ to  strong disorder  $\bar{\Gamma}_0 \approx .2$}
     \label{Sf with disorder}
 \end{figure}

 To compute the superfluid density in the presence of disorder, we substitute the expressions for the Green's functions, Eqs.(\ref{impure Green's function}), in Eq.(\ref{SF equation}). This yields
  \begin{align}
    \rho_s(T)= n\left[1+T \sum_{\omega}\bigintss_{-\Lambda}^{\Lambda} d\xi_k\bigintss_0^{2\pi} \dfrac{d\theta_\bk}{2\pi}\dfrac{\Tilde{\omega}^2-\xi^2_k+(\Tilde{\Delta} (\theta_k, \omega_m))^2}{\left(\Tilde{\omega}^2+\xi^2_k+(\Tilde{\Delta} (\theta_k, \omega_m))^2\right)^2}\right].
    \label{SD in dis 1}
\end{align}
 where, we remind, ${\tilde \omega} = \omega_m - i \Sigma_1 (\omega_m)$ and
 ${\tilde \Delta} (\theta_k, \omega_m) = \Delta (\theta_k) + \Sigma_2 (\omega_m)$. Integrating over $\xi_k$, we obtain
 \begin{align}
    \rho_s(T) &= n  \pi \, T \sum_{\omega} \bigintss_0^{2\pi}  \dfrac{d \theta_\bk}{2\pi}\dfrac{(\Tilde{\Delta}(\theta_\bk, \omega_m))^2}{\left((\Tilde{\Delta}(\theta_k, \omega_m))^2+\Tilde{\omega}^2\right)^{3/2}}=n  \pi \, T \sum_{\omega} \dfrac{\bigintss_0^{2\pi}  \dfrac{d \theta_\bk}{2\pi} \dfrac{\psi^2(\theta_\bk,\omega)}{\left(\psi^2(\theta_\bk,\omega)+\omega^2\right)^{3/2}}}{1+\Gamma_0 \bigintss_0^{2\pi}\dfrac{d\theta'_\bk}{2\pi}\dfrac{1}{\sqrt{\psi^2(\theta'_\bk,\omega)+\omega^2}}}.
    \label{SF dis 2}
\end{align}
In the clean limit one can compute $\rho_s (T)$ analytically, and  the result is  Eq.(\ref{SF eq}). In the presence of the disorder, full analytical analysis is quite involved and we  compute the superfluid density  numerically. We show the results in Fig.  \ref{Sf with disorder}.
for various impurity scattering rate $\bar{\Gamma}_0$.

\begin{figure}
 \centering
    \includegraphics[scale=0.8]{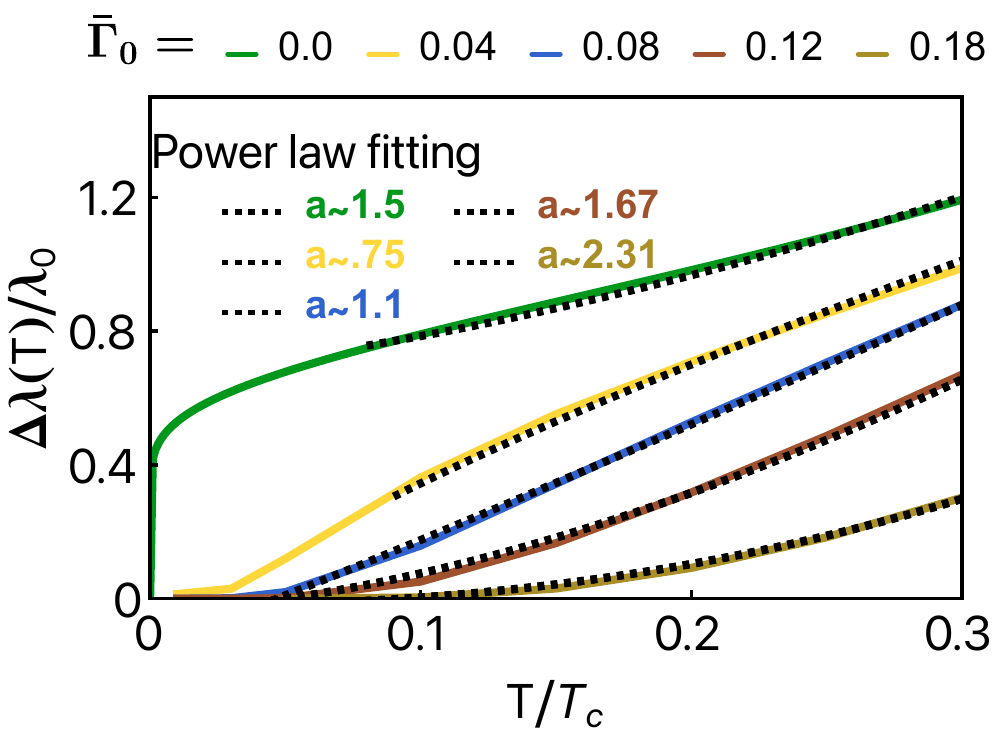}
   \caption{Deviation of the penetration depth from its zero temperature value, $\Delta \lambda(T,\bar{\Gamma}_0)=\lambda(T,,\bar{\Gamma}_0)-\lambda(0,,\bar{\Gamma}_0)$  as a function of  $T/T_c$  for a set of ${\bar \Gamma}_0$   ranging from  $\bar{\Gamma}_0=0$ to strong disorder $\bar{\Gamma}_0\approx .2$. The black dashed line is a power-law fit $T^a$.}
\label{Pen depth}
 \end{figure}

It is instructive to compare these results with the clean limit (black curve in Fig. \ref{Sf with disorder}).
 In the clean case the superfluid density varies as $(1-T/T_c)^{3/2}$ near the transition temperature
   and increases drastically at  low temperatures.  The situation is very different at strong impurity scattering(red curve in Fig. \ref{Sf with disorder}). In this limit, superfluid density in our case
    displays the behavior similar to that in a conventional dirty s-wave superconductor. Namely, it
   increases roughly linearly near $T_c$  and saturates at low temperatures.
   The temperature range where $\rho_s (T)$ saturates increases with increasing impurity scattering (from black curve to red curves in Fig. \ref{Sf with disorder}).
    The low $T$  behavior can be understood by noticing that
the leading contribution to the superfluid density in Eq.(\ref{SF dis 2}) comes from the range where  $\psi(\theta_\bk,\omega) \approx \omega$. At low temperatures, this is the region near cold spots $\theta_\bk = \pi/4 + n \pi/2$. As we already found, in the presence of impurity scattering, $\psi(\pi/4,\omega)$ becomes finite and its value  increases with increasing impurity strength
(see Fig. \ref{psi sol}). This accounts for saturation of $\rho_s$ at $T \to 0$ at a progressively smaller value as ${\bar \Gamma}_0$ increases.
In Fig. \ref{Pen depth} we show the temperature variation of the penetration depth $\Delta\lambda(T,\bar{\Gamma}_0)=1/\sqrt{\rho_s(T,\bar{\Gamma}_0)}-1/\sqrt{\rho_s(0,\bar{\Gamma}_0)}$.  We see that
 at low temperatures
$\Delta\lambda$ gets progressively suppressed  with increasing $\bar{\Gamma}_0$. This is the consequence of the
 saturation of the superfluid density.
 The black dashed lines in Fig.~\ref{Pen depth} show a power-law fit  $\Delta\lambda(T,\bar{\Gamma}_0)\sim T^{a}$  at intermediate temperatures. The exponent $a$ varies with $\bar{\Gamma}_0$. In the clean limit,
   $a\sim 1.5$, while at large $\bar{\Gamma}_0$, $a>2$. The last behavior is again similar to that  of a dirty  s-wave superconductor.

\subsubsection{Optical conductivity}

\begin{figure}[H]
 \centering
   \subfigure[]{\includegraphics[width= 0.45 \textwidth]{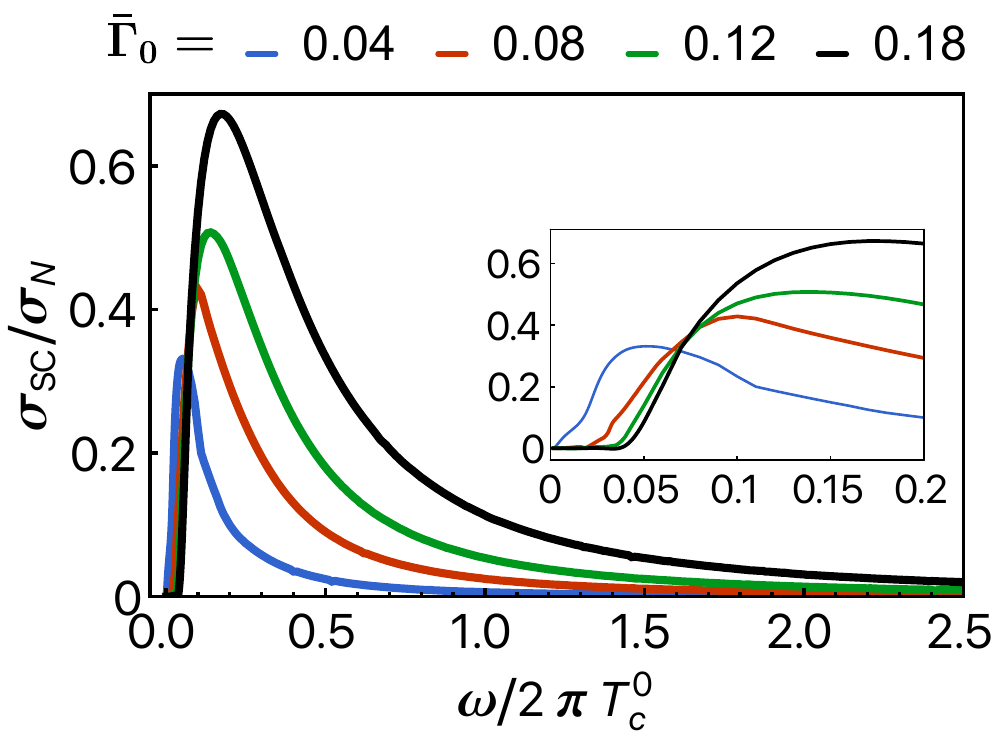}}
   \subfigure[]{\includegraphics[width= 0.49 \textwidth]{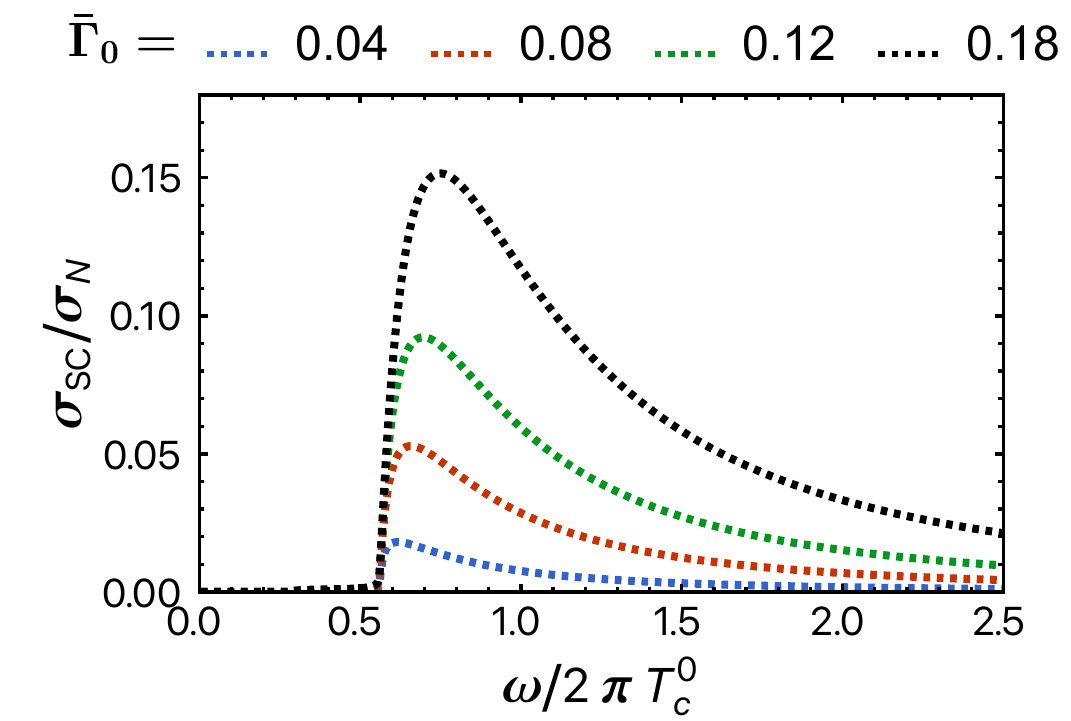}}
    \caption{Panel (a) -- frequency dependence of the optical conductivity $\sigma_\text{SC}(\omega)$ for MFMS
     for various  ${\bar \Gamma}_0$. In the inset we zoom the optical conductivity at low frequencies.
     Panel (b) - the same for a dirty BCS superconductor (for comparison).
The conductivity is normalized to its value in the normal state, $\sigma_\text{N}$.}
\label{Conductivity Diagram}
\end{figure}

We next present the results for optical conductivity, $\sigma_\text{SC}(\Omega)=-\text{Im}\Pi^R(\Omega)/\Omega$.
 We obtained $\text{Im}\Pi^R(\Omega)$ by using the spectral representation
 \begin{align}
     \text{Im}\, \Pi^R(\Omega)= \dfrac{2\, e^2}{m^2}\,\bigintss \dfrac{d^2\bk}{(2\pi)^2} \, \bigintss_{-\infty}^\infty \dfrac{d\omega}{\pi}\, &  \bk_x^2 \left[\text{Im}\,G^R(\bk,\omega+\Omega)\, \text{Im}\,G^R(\bk,\omega)+  \text{Im}\,F^R(\bk,\omega+\Omega)\, \text{Im}\,F^R(\bk,\omega)\right] \nn
      &\times  \left(n_F(\omega+\Omega)-n_F(\omega)\right)
    \label{im retarded pi}
 \end{align}
 where $n_F(\omega)=1/\exp(\omega/T)+1$ is the Fermi function and using the
 retarded Green's functions
 \begin{align}
     G^R(\bk,\omega)&=\dfrac{\omega+\Sigma_{1R}(\omega)+\xi_\bk}{(\omega+\Sigma_{1R}(\omega))^2-\xi_\bk^2-(\Delta(\bk)+\Sigma_{2R}(\omega))^2} \label{Retarded G},\\
      F^R(\bk,\omega)&=\dfrac{\Delta(\bk)+\Sigma_{2R}(\omega)}{(\omega+\Sigma_{1R}(\omega))^2-\xi_\bk^2-(\Delta(\bk)+\Sigma_{2R}(\omega))^2}, \label{Retarded F}
\end{align} with retarded  $\Sigma_{1R}(\omega)$ and $\Sigma_{2R}(\omega)$, which we obtained in Sec.{\ref{impurity effect on gap function}}.  As before, we replace the momentum integration, $\int d^2\bk/(2\pi)^2$ by $N_0\, \int d\xi_k \int \, d\theta_\bk/(2\pi)$ and $\bk_x^2$ by $\bk_F^2/2  = n\pi$. We measure $\sigma_\text{SC}$ in units of $\sigma_N=n e^2/2 \,m\, \Gamma_0$. At $T \to 0$, the ratio $\sigma(\Omega)=\sigma_\text{SC}/\sigma_N$  is expressed as
\begin{align}
    \sigma(\Omega)=\dfrac{2\, \Gamma_0}{\Omega} \bigintss_{-\infty}^{\infty} d\xi\bigintss_0^{2\pi} \dfrac{d\theta}{2\pi} \bigintss_{-\Omega}^0 \dfrac{d\omega}{\pi} \left[\text{Im}\,G^R(\bp,\omega+\Omega)\, \text{Im}\,G^R(\bp,\omega)+ \text{Im}\,F^R(\bp,\omega+\Omega)\, \text{Im}\,F^R(\bp,\omega)\right].
    \label{scaled conductivity}
\end{align}
We  compute $\sigma(\Omega)$ numerically and show the results in Fig. \ref{Conductivity Diagram}(a) for several  ${\bar \Gamma}_0$. For comparison, in Fig. \ref{Conductivity Diagram}(b) we  plot the conductivity of a dirty s-wave superconductor. The latter is zero below $2\Delta_0$ and has a peak at a larger $\omega$.   In our case,
  the conductivity is non-zero down to much smaller frequencies, particularly in the weak disorder limit, due to  the presence of  low-lying excitations near cold spots.  With increasing ${\bar \Gamma}_0$, these excitations get gapped, and the frequency dependence of the conductivity becomes qualitatively similar to that for a dirty s-wave superconductor, although the threshold frequency is set by ${\bar \Gamma}_0$.   At  large frequencies $\Omega \gg \Delta_0$, the behavior of conductivity in our case does not differ much from that in a dirty s-wave superconductor.

\section{Comparison With Experiments}
\label{comparison with experiment section}
We argued in this work is that  pairing in  doped FeSe
   near a nematic QCP is mediated by nematic fluctuations rather than by spin fluctuations.  This is generally consistent with the observations in Ref. \citep{shibauchi2020exotic,ishida2022pure,hanaguri2018two} of two distinct pairing states in pure FeSe and in doped FeSe$_{1-x}$S$_x$ and FeSe$_{1-x}$Te$_x$ at $x \geq x_\text{c}$.
     More specifically, one can distinguish between magnetic and nematic pairing scenarios by measuring the angular dependence of the gap along the hole $d_\text{xz}/d_\text{yz}$ pocket.
   We argued that a nematic-mediated pairing gives rise to an anisotropic gap, with maxima along $k_\text{x}$ and $k_\text{y}$ directions.  Within spin-fluctuation scenario,  the gap  $\Delta_\text{h} (k) = a + b \cos{4 \theta}$  appears to be the largest along the diagonal  directions  $k_\text{x} \pm k_\text{y}$ ($b <0$, see e.g., Ref.~\citep{graser2010spin}).
      The angular dependence of the gap in pure and doped FeSe has been extracted from ARPES and STM data in  Refs. \citep{xu2016highly,liu2018orbital,sprau2017discovery,nagashima2022discovery,
      walker2023electronic,nag2024highly}.

\begin{figure}[H]
 \centering
    \includegraphics[scale=0.8]{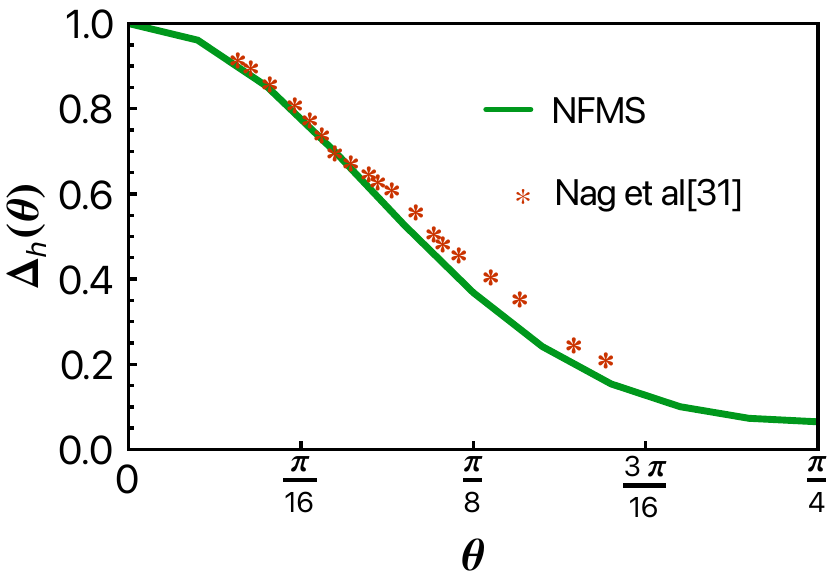}
    \caption{The line is the theoretical result for the angular variation of the superconducting gap $\Delta_h(\theta_\bk)$ on the hole pocket, obtained by numerically solving the gap equation slightly away from the nematic quantum critical point ($\theta_\bk$ is with respect to $\Gamma-X$ axis).  The points marked "*" are the STM data from~\citep{nag2024highly}.}
\label{Theory with Experiment: gap structure}
\end{figure}

       For pure and weakly doped FeSe, an extraction of $\cos{4\theta}$ dependence is complicated because superconductivity co-exists with long-range nematic order, in which case the gap additionally has $\cos{2 \theta}$ term due to nematicity-induced mixing of $s-$wave and $d-$wave components~\citep{hashimoto2018superconducting,kushnirenko2018three}.  Still, the fits of the ARPES data in Refs.~\citep{xu2016highly,liu2018orbital} yielded a negative $b$, consistent with spin-fluctuation scenario. A negative $b$  is also consistent with the flattening of the gap on the  hole pocket near $\theta =\pi$, observed in the STM study~\citep{sprau2017discovery}. A negative prefactor for $\cos{4 \theta}$ term was also reported for Fe-pnictides, e.g.,  Ba$_0.24$K$_{0.76}$Fe$_2$As$_2$, Ref. \citep{ota2014evidence}.   In contrast, gap maximum along $k_\text{y}$ has been reported in a recent laser ARPES study of
   FeSe$_{0.78}$S$_{0.22}$ (Ref. \citep{nagashima2022discovery}). Further, recent  STM data for FeSe$_{0.81}$S$_{0.19}$ (Ref. \citep{walker2023electronic, nag2024highly})  detected a  clear gap maxima along  $k_\text{x}$ and $k_\text{y}$. In Fig.~\ref{Theory with Experiment: gap structure} we compare our theoretical result for the angular variation of the  theoretical gap function $\Delta_h(\theta_\bk)$ on the hole pocket as a function of angle $\theta_\bk$ with the gap extracted from the STM experiment of Ref.~\citep{nag2024highly}. We find a good qualitative match between theory and experiment.
   For Te-doped FeSe, STM data for
tetragonal FeSe$_{0.45}$Te$_{0.55}$ (Ref. \citep{sarkar2017orbital}) also found the maximal
gap along  $k_\text{x}$ and $k_\text{y}$ directions, consistent with the pairing
 by nematic fluctuations. This STM data are in variation with earlier ARPES data for the same material, which reported a near-isotropic gap on the hole pocket~\citep{miao2012isotropic}. However, the gap magnitude in FeSe$_{0.45}$Te$_{0.55}$ is only 2 meV, a bit too small for a conventional ARPES. Perhaps, laser ARPES measurements will be useful in this regard.  Taken together, these data strongly support the idea about different pairing mechanisms in pure FeSe and in doped ones at $x \geq x_\text{c}$, and are consistent with the change of the pairing glue from spin fluctuations at $x <x_\text{c}$
       to nematic fluctuations at $ x \geq x_\text{c}$.
         The gap anisotropy
         has also been analyzed in Ref. \citep{zeng2010anisotropic} for FeSe$_{0.45}$Te$_{0.55}$, using angle-resolved specific heat data, but the authors
 of that work  focused on the gap on the electron  pockets, which, according to our theory, should not show  spectacular variation with the angle.

 We argued in this communication  that right at a nematic QCP,  the vanishing or  near-vanishing of the gap   in the cold regions on the Fermi surface leads to highly unconventional behavior of thermodynamic, spectroscopic and transport observables and presented theoretical results.  We now show that these results are consistent with the data.

  \begin{itemize}
  \item
   The specific heat of FeSe$_{1-x}$S$_x$ has been measured in Refs.~\citep{sato2018abrupt,mizukami2021thermodynamics}. The data clearly indicate that the jump of $\gamma_\text{c} (T)$ at $T_\text{c}$ decreases with increasing $x$ and vanishes at around $x_\text{c}$.  At smaller $T$,  $\gamma_\text{c} (T)$  passes through a maximum at around $0.8 T_\text{c}$ and then decreases nearly linearly towards  apparently  a finite value at $T=0$.
     The  authors of Ref.\citep{hanaguri2018two} argued that this behavior  is not  caused by fluctuations
     because
residual resistivity does not exhibit a noticeable increase around $x_\text{c}$ (Ref. \citep{hosoi2016nematic}). Other experiments~\citep{coldea2019evolution} also indicated that fluctuation effects  get weaker with increasing $x$. The behavior of $\gamma_\text{c} (T)$ around $x_\text{c}$ was first interpreted first as potential BCS-BEC crossover~\citep{shibauchi2020exotic} and later as a potential evidence of an exotic pairing that creates a Bogoliubov Fermi surface in the superconducting state~\citep{agterberg2017bogoliubov,setty2020topological,nagashima2022discovery}. For a direct comparison with the experiments we plot our theoretical result for the specific heat coefficient $\gamma(t)$ (green curve) in Fig~\ref{Theory with Experiment: specific heat} along with the data from  Ref.~\citep{sato2018abrupt}.
 \begin{figure}[H]
 \centering
   \includegraphics[scale=0.8]{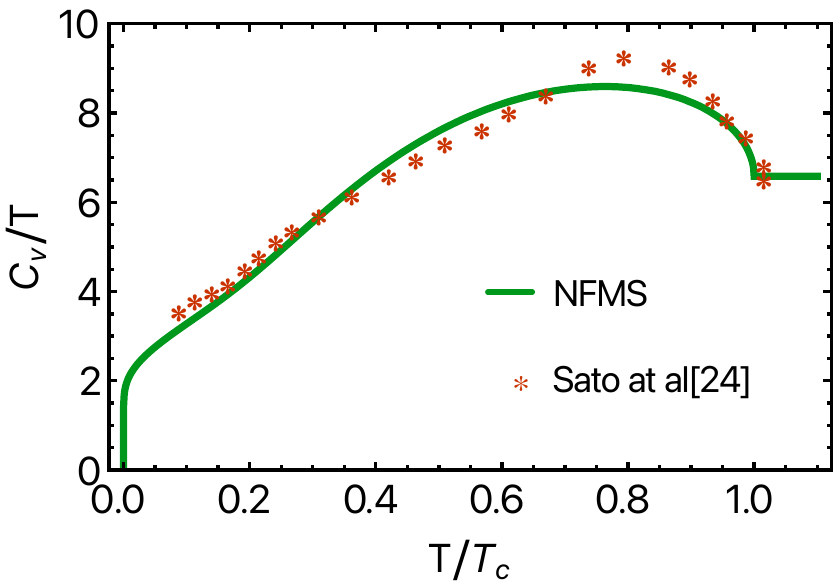}
    \caption{Line-- theoretical temperature dependence of the specific heat coefficient $\gamma(T)=C_v(T)/T N_0$ below the transition temperature $T_c$. The points "*" are the  data from  \citep{sato2018abrupt}.}
\label{Theory with Experiment: specific heat}
\end{figure}
\item
 The tunneling conductance of FeSe$_{1-x}$S$_x$ has been measured in Refs.\citep{nag2024highly,mizukami2021thermodynamics}. The data show that the zero bias tunneling conductance at low temperatures is almost half of the normal state value.  In  Fig.  \ref{Theory with Experiment: zero bias conductance} we  compare our result for the zero bias conductance $G(0,T)$ (green curve) with the data \citep{nag2024highly}.
\begin{figure}
 \centering
   \includegraphics[scale=0.8]{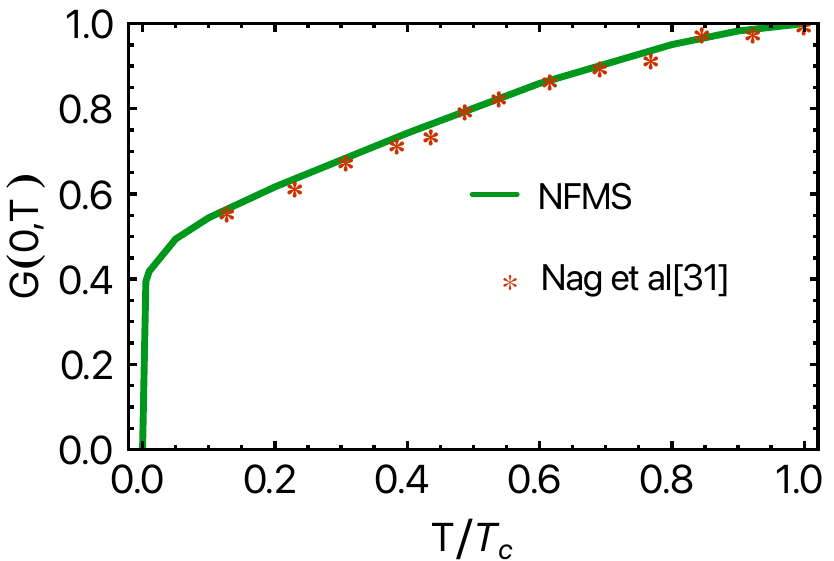}
    \caption{Line -- theoretical temperature dependence of the zero bias tunneling conductance
      $G(0,T)$ below  $T_c$. The points "*" are the data from \citep{nag2024highly}.}
\label{Theory with Experiment: zero bias conductance}
\end{figure}
\item
 The superfluid density, $\rho_s(T)$, and the change in penetration depth $\Delta\lambda(T)$ for the pure FeSe$_{1-x}$S$_x$ has been measured in Refs.  \citep{matsuura2023two,nagashima2024lifting}. The data in Ref.\citep{matsuura2023two} show a depletion of superfluid density at low temperature at $x\sim x_c$, which suggests that some fermions remain unpaired. This is consistent with NFMS  as we showed that at not extremely low temperatures the superfluid density is smaller than the fermion density $n$ (see Fig. \ref{SF pure}(a)). Ref.\citep{nagashima2024lifting} shows the temperature dependence of the normalized superfluid density and change in the penetration depth $\Delta\lambda(T)$. A power law fit of the low temperature behavior  $\Delta\lambda(T) \propto T^a$ gives $a=1.5$ for $x=0.18$ and $n=1.62$ for $x=0.25$. We find the data consistent with NFMS, particularly for $x=0.18$.  Our exponent $a$ (Fig.  \ref{SF pure}(c)) is quite close to $1.5$, even though it slightly depends on the pairing potential $g$ and the temperature range of fitting.
 \begin{figure}
 \centering
   \subfigure[]{\includegraphics[width= 0.45 \textwidth]{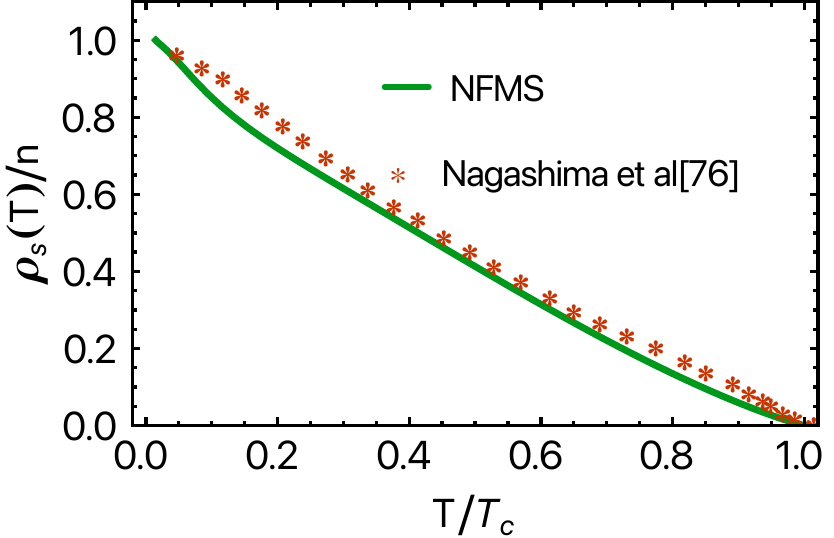}}
   \subfigure[]{\includegraphics[width= 0.45 \textwidth]{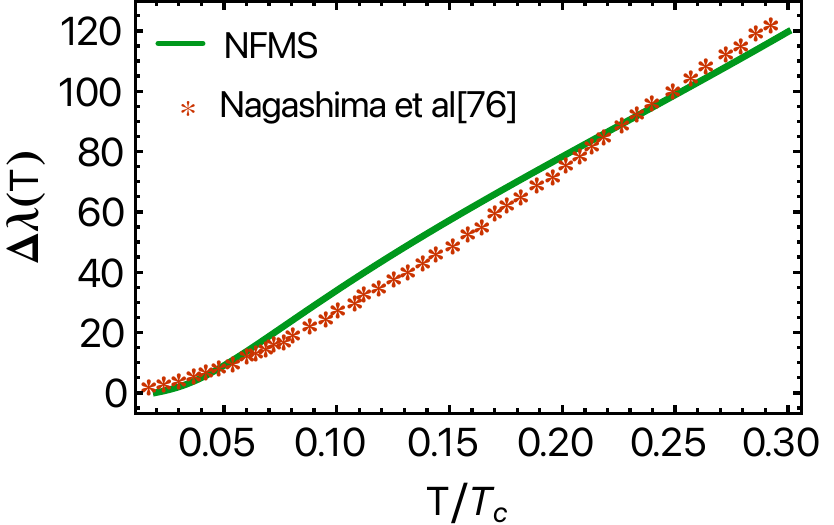}}
    \caption{Lines -- theoretical temperature dependence of the normalized superfluid density $\rho_s(T)/n$ (a)  and penetration depth $\Delta\lambda(T)$ (b)  slightly away from the nematic critical point. The points "*" are  the  data from  \citep{nagashima2024lifting}.}
\label{Theory with Experiment: superfluid}
\end{figure}
\item The uniform magnetic susceptibility $\chi_s(T)$ for  FeSe$_{1-x}$S$_x$ has been  measured in Ref. \citep{yu2023spin, wiecki2018persistent} for a range of $x \geq  x_c$.  The data indicate that  over a range of $T < T_c$, $\chi_s(T)$ is almost  temperature independent.This is also consistent with NFMS scenario, for which we found  that $\chi_s(T)$ changes very slowly below $T_c$  (see Fig. \ref{Spin Susceptibility fig}).
\item  The effect of non-magnetic impurities on superconductivity in FeSe$_{1-x}$S$_x$ for $x \approx x_c$ has been analyzed experimentally in Ref.\citep{nagashima2024lifting} via electron irradiation. The data show that  
    the transition temperature $T_c$ decreases monotonically with increasing impurity scattering, but the suppression of $T_c$  is slower than in the  Abrikosov-Gorkov theory.
      We argue that within NFMS, $T_c$ goes down in the presence of impurity scattering (see Fig. \ref{Tc}), but the suppression  rate is slower than the AG theory, and $T_c$ eventually saturates.  For quantitative comparison with the data, we computed the ratio $T_c/T_c^0$ within our theory as a function of the same pair-breaking parameter as in Ref.~\citep{nagashima2024lifting}: $g_\text{pair}=1/\tau_{tr} T_c^0$ where $\tau_\text{tr}$ is the transport lifetime. We relate $\tau_{tr}$ to our scattering rate $\Gamma_0=1/2\tau_{imp}$ as $\Gamma_0=1/4 \tau_{tr}$, assuming $\tau_{imp}=2 \tau_{tr}$.
      We plot our theoretical result in Fig. \ref{Theory with Experiment: disorder} along with the data from Ref.\citep{nagashima2024lifting}. They match quite well.

      \item
      The data of Ref.\citep{nagashima2024lifting} also show that a power-law fit of  the penetration depth $\Delta\lambda(T)$ at  low temperatures shows a non-monotonic dependence on the impurity scattering, $\Delta\lambda(T) \propto
     T^a$. In the clean limit, $a\approx 1.5$. With increasing scattering rate $\Gamma_0$,  $a$ increases, exceeds $a=2$ at intermediate $\Gamma_0$  and at larger $\Gamma_0$  approaches  $a=2$ from above. In our case, $a\approx 1.5$ in the clean case, consistent with the data.  For finite $\Gamma_0$,  $a$ initially drops a bit, but then start increasing and at some $\Gamma_0$ exceeds $a=2$.  This is again consistent with the data. We attribute an increase of $a$ with $\Gamma_0$  to the fact that impurity scattering lifts the nodes at the cold spots and  makes the gap function less anisotropic such that  at strong scattering the system behavior resembles that of a dirty BCS superconductor, for which  a format power-law fit yields $a\gg 2$. We, however, didn't find a re-entrant behavior towards $a=2$ at even larger $\Gamma_0$.
           
\begin{figure}
 \centering
    \includegraphics[scale=0.8]{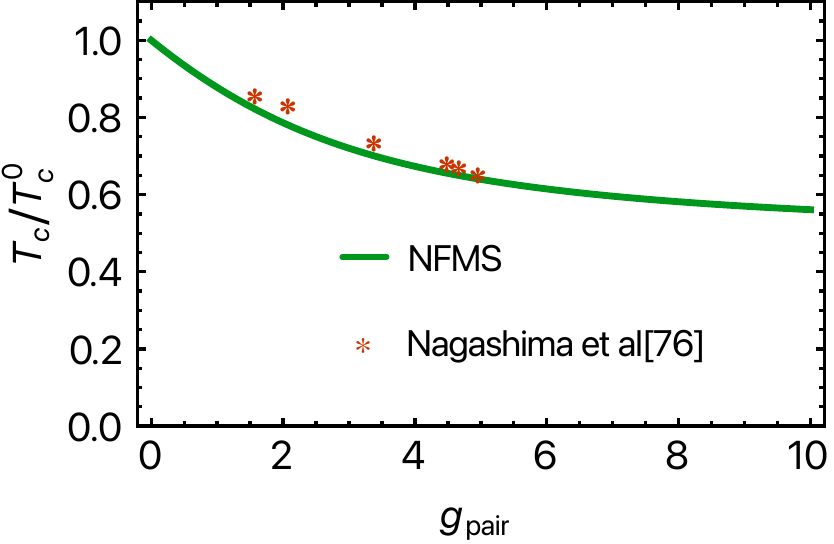}
    \caption{Line -- theoretical result for the variation of the superconducting transition temperature $T_c$  as a function of pair-breaking parameter $g_\text{pair}=s \Gamma_0/T_c^0$, where $s \approx 4 $ is set to match with the pair-breaking parameter defined in Ref.~\citep{nagashima2024lifting}. The red points "*" are
       the data from ~\citep{nagashima2024lifting}.}
\label{Theory with Experiment: disorder}
\end{figure}

\item
  We did't find experimental results for the Raman intensity measurement, field induced specific heat measurement and optical conductivity data. These experiments are called for.
  \end{itemize}
Two more points.  First,  $\mu$SR experiments~\citep{matsuura2023two,roppongi2023} presented evidence for time-reversal symmetry breaking in FeSe.  The $\mu$SR signal is present below $T_\text{c}$  for all $x$, however in FeSe$_{1-x}$Te$_x$ it clearly increases above $x_\text{c}$.  This raises a possibility that the superconducting state at $x > x_\text{c}$ breaks time-reversal symmetry, at least in FeSe$_{1-x}$Te$_x$.   Within our nematic scenario, this would indicate a $p-$wave pairing with $k_\text{x} \pm i k_\text{y}$ gap structure. We argued that $p-$wave pairing, mediated by nematic fluctuations, is  a strong competitor to $s^{+-}$ pairing.
 Second, there is one recent data set, which we cannot explain within NFMS.  Laser ARPES study of FeSe$_{0.78}$S$_{0.22}$ (Ref. \citep{nagashima2022discovery}) detected superconducting gap in the polarization of light, which covers momenta near  the $X$ direction, but no gap in polarization selecting momenta near $Y$.  Taken at a face value, this data  implies that superconducting order strongly breaks $C_4$ symmetry.  In our scenario, a pure $k_\text{x}$ (or $k_\text{y}$) order is possible, in principle,  but has a smaller condensation energy than $k_\text{x} \pm i k_\text{y}$.  More analysis  is needed to resolve this issue.

 \section{Conclusions}

In this work,  we  analyzed in detail the behavior of observables in a superconductor with pairing mediated by soft charge-nematic fluctuations.  We argued that such a pairing is difficult do realize in a one-band/one-orbital system, but it develops naturally near the onset of a nematic order in a multi-orbital/mulri-band  system.   We used doped FeSe as an example.  The electronic structure of FeSe consists of hole and electron pockets, and excitations near a hole Fermi surface are made out of $d_{xz}$ and $d_{yz}$ orbitals.
A pure FeSe develops a nematic order at $85K$ and superconductivity at $8K$. Upon doping by either $S$ or $Te$, the onset temperature of a nematic order decreases an vanishes at some critical doping, but superconducting $T_c$ remains finite. Numerous experimental data cited above have demonstrated that superconducting state near and above critical doping is very different from the one in pure and weakly doped FeSe and probably has different origin.

The point of departure for our study is the assumption that while the pairing in pure and weakly doped FeSe is mediated by spin fluctuations, much like in Fe-pnictides,  the pairing mechanism near the end point of the nematic order is the exchange of soft charge nematic fluctuations.  Such a mechanism is difficult to realize in a one-band/one-orbital system, but we showed that it can be realized in a multi-band/multi-orbital FeSe near a nematic quantum critical point under the assumption that the nematic order is a spontaneous polarization between $d_{xz}$ and $d_{yz}$ orbitals and the polarization changes sign  between hole and electron pockets. Even though all the interactions are repulsive, we showed (following the earlier work \citep{islam2024unconventional}) that the renormalized pairing interaction mediated by soft d-wave nematic fluctuations is attractive.  This interaction is, however, a rather unconventional one as for fermions on a hole pockets, where, we argued, the pairing interaction is the strongest, its strength depends on the location of a fermion on a Fermi surface.  This leads to highly anisotropic gap function and to a highly unconventional thermal evolution of the gap structure below $T_c$.
Specifically, we found that right at the nematic quantum-critical point,  the gap opens up at $T_c$ only at $4$ points on the hole Fermi surface, along the direction towards electron pockets. As $T$ decreases, the range where the gap is non-zero widens, but there are still regions on the Fermi surface where  the gap vanishes.  At $T=0$,
the gap opens everywhere except special points, but remains highly anisotropic, creating hot and cold regions.
 Symmetry-wise, such a gap can be either $s-$wave, or $p-$wave or $d-$wave - the condensation energies for all three cases are the same.   Away from a critical point and/or in the presence of impurity scattering the degeneracy is broken and $s-$ wave gap function wins.  In our analysis we assumed $s-$wave gap, but we caution that by adding extra interactions, which we neglected, one can tip the balance towards $p-$wave.

 In the bulk of this work we analyzed the effects of the non-trivial gap structure on the behavior of thermodynamic, spectroscopic and transport properties: the specific heat and its directional variation  with a magnetic field,  magnetic susceptibility,  density of states, tunneling conductance, Raman intensity, superfluid stiffness and penetration depth. We found clear signatures of strong gap anisotropy and its unconventional thermal evolution in all observables that we calculated.  In general terms, there is a strong non-BCS behavior near $T_c$ because the gap only develops in narrow momentum ranges, and the behavior at $T < T_c$ indicates the presence of un-gapped portions of the Fermi surface.   In our theory,  such a behavior holds down to very small $T ~10^{-2} T_c$ as the gap in some momentum ranges remains exponentially small even at $T=0$.  We compared the theory with the existing data and found a reasonably good agreement. We made several predictions for future experiments, e.g., for measurements of a directional variation of the specific heat under an applied magnetic field (a Volovik effect).
  We also studied the effects how superconductivity mediated by nematic fluctuations is affected by  an s-wave scattering by non-magnetic impurities. We analyzed the evolution of the behavior of the  same observables that we   studied in the clean limit, and also computed optical conductivity and the variation of $T_c$.  We found that with increasing impurity scattering, the gap becomes less anisotropic, but this process  holds differently at different temperatures and  frequencies. We again found a good agreement between our theory and the existing data and made suggestions for future experiments.

 As we wrote in the Introduction, another theoretical proposal for superconductivity in doped FeSe is an exotic pairing that creates a Bogoliubov Fermi surface  within a superconducting state~\citep{agterberg2017bogoliubov,setty2020topological,nagashima2022discovery}.  It would be interesting to compare direct predictions for the observables within our and Bogoliubov Fermi surface scenarios and compare both with the experiments.

  \section{Acknowledgements}

We acknowledge with thanks useful discussions with
D. Agterberg, E. Berg, B. Büchner, P. Canfield, P. Coleman, A. Coldea, Z.
Dong, R. Fernandes, I. Fisher, Y. Gallais, E. Gati, R. Hackl, T. Hanaguri,  P.
Hirschfeld, S. Kasahara, B. Keimer, A. Klein, H. Kontani, A. Kreisel, L. Levitov, A.
Pasupathi, I. Paul, A. Sacuto, J. Schmalian, E. da Silva Neto, T. Shibauchi,
and R. Valenti. This work was supported by U.S. Department
of Energy, Office of Science, Basic Energy Sciences,
under Award No. DE-SC0014402.

\appendix
\section{Derivation of the Interaction in the band basis}
\label{Appendix A}
 We collect all $C_4$-symmetric  interactions involving the low-energy $d_{xz}/d_{yz}$ orbital states near $\Gamma,X$ and $Y$ pockets. This gives $14$ distinct terms in the orbital representation
 \begin{align}
H_{\text{int}}&=\dfrac{U_4}{2} \sum  \left[d_{xz,\sigma}^\dagger d_{xz,\sigma}d_{xz,\sigma'}^\dagger d_{xz,\sigma'}+d_{yz,\sigma}^\dagger d_{yz,\sigma}d_{yz,\sigma'}^\dagger d_{yz,\sigma'}\right]+\tilde{U}_4 \sum \, d_{xz,\sigma}^\dagger d_{xz,\sigma}d_{yz,\sigma'}^\dagger d_{yz,\sigma'}  \nn &
 +\tilde{\tilde{U}}_4 \sum  \, d_{xz,\sigma}^\dagger d_{yz,\sigma}d_{yz,\sigma'}^\dagger d_{xz,\sigma'} +\dfrac{\bar{U}_4}{2} \sum \left[d_{xz,\sigma}^\dagger d_{yz,\sigma}d_{xz,\sigma'}^\dagger d_{yz,\sigma'}+d_{yz,\sigma}^\dagger d_{xz,\sigma}d_{yz,\sigma'}^\dagger d_{xz,\sigma'}\right] \nn & +U_1 \sum\left[f_{1,\sigma}^\dagger f_{1,\sigma} d_{xz,\sigma'}^\dagger d_{xz,\sigma'}+f_{2,\sigma}^\dagger f_{2,\sigma} d_{yz,\sigma'}^\dagger d_{yz,\sigma'}\right] +\bar{U}_1 \sum\left[f_{1,\sigma}^\dagger f_{1,\sigma} d_{yz,\sigma'}^\dagger d_{yz,\sigma'}+f_{2,\sigma}^\dagger f_{2,\sigma} d_{xz,\sigma'}^\dagger d_{xz,\sigma'}\right] \nn & + U_2 \sum\left[ f_{1,\sigma}^\dagger d_{xz,\sigma} d_{xz,\sigma'}^\dagger f_{1,\sigma'}+ f_{2,\sigma}^\dagger d_{yz,\sigma} d_{yz,\sigma'}^\dagger f_{2,\sigma'}\right]+ \bar{U}_2 \sum\left[ f_{1,\sigma}^\dagger d_{yz,\sigma} d_{yz,\sigma'}^\dagger f_{1,\sigma'}+ f_{2,\sigma}^\dagger d_{xz,\sigma} d_{xz,\sigma'}^\dagger f_{2,\sigma'}\right] \nn &+\dfrac{U_3}{2}\sum \left[ f_{1,\sigma}^\dagger d_{xz,\sigma} f_{1,\sigma'}^\dagger d_{xz,\sigma'}+f_{2,\sigma}^\dagger d_{yz,\sigma} f_{2,\sigma'}^\dagger d_{yz,\sigma'}\right] +\dfrac{\bar{U}_3}{2} \sum \left[ f_{1,\sigma}^\dagger d_{yz,\sigma} f_{1,\sigma'}^\dagger d_{yz,\sigma'}+ f_{2,\sigma}^\dagger d_{xz,\sigma} f_{2,\sigma'}^\dagger d_{xz,\sigma'}\right] \nn &+\dfrac{U_5}{2} \sum  \left[f_{1,\sigma}^\dagger f_{1,\sigma}f_{1,\sigma'}^\dagger f_{1,\sigma'}+f_{2,\sigma}^\dagger f_{2,\sigma}f_{2,\sigma'}^\dagger f_{2,\sigma'}\right]+ \tilde{U}_5 \sum \, f_{1,\sigma}^\dagger f_{1,\sigma}f_{2,\sigma'}^\dagger f_{2,\sigma'}+
 +\tilde{\tilde{U}}_5 \sum  \, f_{1,\sigma}^\dagger f_{2,\sigma}f_{2,\sigma'}^\dagger f_{1,\sigma'} \nn & +\dfrac{\bar{U}_5}{2} \sum \left[f_{1,\sigma}^\dagger f_{2,\sigma}f_{1,\sigma'}^\dagger f_{2,\sigma'}+f_{2,\sigma}^\dagger f_{1,\sigma}f_{2,\sigma'}^\dagger f_{1,\sigma'}\right].
 \label{Interaction in orbital basis app}
\end{align}
In Eq.~(\ref{Interaction in orbital basis app}), the momentum arguments of the fermion operators $\bk_i , i = 1,2,3,4$ are omitted and the summation is over spin indices $\sigma,\sigma'$ and momenta $\bk_i$, subject to the momentum conservation. The interaction parameters, $U_i,\tilde{U}_i,\tilde{\tilde{U}}_i,\bar{U}_i$ represent specific scattering process. For example, $U_4$ and $\tilde{U}_4$ stand for intra-orbital and inter-orbital density-density interaction near the $\Gamma$ point, respectively, and so on.

Next, we convert the interaction (\ref{Interaction in orbital basis app}) into the band basis using the form factor relation (\ref{orbital to band}): $d_{xz,\bk}=\sin\theta_\bk\, h_{\bk}$ and  $d_{yz,\bk}=\cos\theta_\bk\, h_{\bk}$ where $h_\bk$ is the annihilation operator for the fermions at the hole pocket. $\theta_\bk$ is the polar co-ordinate of the momentum $\bk$ near the $\Gamma$ point and measured from $ \Gamma-X$ axis. For the electron pockets, the form factors are unity as we consider them mono-orbital.  This induces angle-dependent interactions between bands enforced by the orbital contents of the hole Fermi pocket. A straightforward calculation gives $H_{\text{int}}=\sum_{i,j}H_{i,j}$, where
  $H_{i,j}$ is the interaction between the pockets $i$ and $j\epsilon\{h,Y,X\}$ with the following expression
\begin{align}
 \label{interaction1 in band basis app}
H_{h,h}&=\dfrac{1}{2} \sum_{\bk,\bp,\bq} h_{\bk,\sigma}^\dagger h_{\bk+\bq,\sigma}h_{\bp,\sigma'}^\dagger h_{\bp-\bq,\sigma'} V_{h,h}^{\text{den}}(\bk,\bk+\bq;\bp,\bp-\bq),\\
H_{h,Y}&= \sum_{\bk,\bp,\bq} f_{1,\bk,\sigma}^\dagger f_{1,\bk+\bq,\sigma} h_{\bp,\sigma'}^\dagger h_{\bp-\bq,\sigma'} V_{h,Y}^{\text{den}}(\bp,\bp-\bq)+\sum_{\bk,\bp,\bq}f_{1,\bk,\sigma}^\dagger h_{\bk+\bq,\sigma} h_{\bp,\sigma'}^\dagger f_{1,\bp-\bq,\sigma'} V_{h,Y}^{\text{ex}}(\bk+\bq,\bp) \nn & + \sum_{\bk,\bp,\bq}\left[f_{1,\bk,\sigma}^\dagger h_{\bk+\bq,\sigma} f_{1,\bp,\sigma'}^\dagger h_{\bp-\bq,\sigma'} V_{h,Y}^{\text{ph}}(\bk+\bq,\bp-\bq)+h.c\right] ,\\
H_{h,X}&= \sum_{\bk,\bp,\bq} f_{2,\bk,\sigma}^\dagger f_{2,\bk+\bq,\sigma} h_{\bp,\sigma'}^\dagger h_{\bp-\bq,\sigma'} V_{h,X}^{\text{den}}(\bp,\bp-\bq)+\sum_{\bk,\bp,\bq}f_{2,\bk,\sigma}^\dagger h_{\bk+\bq,\sigma} h_{\bp,\sigma'}^\dagger f_{2,\bp-\bq,\sigma'} V_{h,X}^{\text{ex}}(\bk+\bq,\bp)\nn & +\sum_{\bk,\bp,\bq} \left[f_{2,\bk,\sigma}^\dagger h_{\bk+\bq,\sigma} f_{2,\bp,\sigma'}^\dagger h_{\bp-\bq,\sigma'} V_{h,X}^{\text{ph}}(\bk+\bq,\bp-\bq)+h.c\right],\\
H_{Y,Y}&= \dfrac{1}{2}\sum_{\bk,\bp,\bq} f_{1,\bk,\sigma}^\dagger f_{1,\bk+\bq,\sigma} f_{1,\bp,\sigma}^\dagger f_{1,\bp-\bq,\sigma} V_{e,e}^\text{den},\\
H_{X,X}&= \dfrac{1}{2}\sum_{\bk,\bp,\bq} f_{2,\bk,\sigma}^\dagger f_{2,\bk+\bq,\sigma} f_{2,\bp,\sigma}^\dagger f_{2,\bp-\bq,\sigma} V_{e,e}^\text{den},\\
H_{X,Y}&= \sum_{\bk,\bp,\bq} f_{1,\bk,\sigma}^\dagger f_{1,\bk+\bq,\sigma} f_{2,\bp,\sigma}^\dagger f_{2,\bp-\bq,\sigma} V_{X,Y}^\text{den}+ \sum_{\bk,\bp,\bq} f_{1,\bk,\sigma}^\dagger f_{2,\bk+\bq,\sigma} f_{2,\bp,\sigma}^\dagger f_{1,\bp-\bq,\sigma} V_{X,Y}^\text{ex} \nn & + \sum_{\bk,\bp,\bq} \left[f_{1,\bk,\sigma}^\dagger f_{2,\bk+\bq,\sigma} f_{1,\bp,\sigma}^\dagger f_{2,\bp-\bq,\sigma} V_{X,Y}^\text{ph}+h.c\right].
\label{interaction2 in band basis app}
\end{align}
In Eqs. (\ref{interaction1 in band basis app}-\ref{interaction2 in band basis app}), we group interactions into density-density, exchange and pair-hopping by labeling them "den", "ex" and "ph" respectively. They are given by the following expression (we suppress the momentum dependence for simplicity but follow the same convention as in (\ref{interaction1 in band basis app}-\ref{interaction2 in band basis app})) \begin{align}
 \label{Vhh_a}
V_{h,h}^{\text{den}}=& U_4  \Big\{\sin\theta_\bk \sin\theta_{\bk+\bq}\sin\theta_\bp\sin\theta_{\bp-\bq}+\cos\theta_\bk \cos\theta_{\bk+\bq}\cos\theta_\bp\cos\theta_{\bp-\bq}\Big\}\nn & +\tilde{U}_4\Big\{\sin\theta_\bk \sin\theta_{\bk+\bq}\cos\theta_\bp\cos\theta_{\bp-\bq}+\cos\theta_\bk \cos\theta_{\bk+\bq}\sin\theta_\bp\sin\theta_{\bp-\bq}\Big\}\nn & +\tilde{\tilde{U}}_4\Big\{\sin\theta_\bk \cos\theta_{\bk+\bq}\cos\theta_\bp\sin\theta_{\bp-\bq}+\cos\theta_\bk \sin\theta_{\bk+\bq}\sin\theta_\bp\cos\theta_{\bp-\bq}\Big\}\nn & + \bar{U}_4\Big\{\sin\theta_\bk \cos\theta_{\bk+\bq}\sin\theta_\bp\cos\theta_{\bp-\bq}+\cos\theta_\bk \sin\theta_{\bk+\bq}\cos\theta_\bp\sin\theta_{\bp-\bq}\Big\} ,\\
V_{h,Y}^{\text{den}}&= U_1 \, \sin\theta_\bp \sin\theta_{\bp-\bq}+\bar{U}_1\, \cos\theta_\bp \cos\theta_{\bp-\bq},\\ V^{h,X}_{\text{den}}&= U_1 \, \cos\theta_\bp \cos\theta_{\bp-\bq}+\bar{U}_1\, \sin\theta_\bp \sin\theta_{\bp-\bq},  \\
V_{h,Y}^{\text{ph}}&= \dfrac{U_3}{2} \, \sin\theta_{\bk+\bq} \sin\theta_{\bp-\bq}+\dfrac{\bar{U}_3}{2}\, \cos\theta_{\bk+\bq} \cos\theta_{\bp-\bq},\\  V^{h,X}_{\text{ph}}&= \dfrac{U_3}{2} \, \cos\theta_{\bk+\bq} \cos\theta_{\bp-\bq}+\dfrac{\bar{U}_3}{2}\, \sin\theta_{\bk+\bq} \sin\theta_{\bp-\bq}, \\
V_{h,Y}^{\text{ex}}&=U_2 \, \sin\theta_{\bk+\bq} \sin\theta_{\bp}+\bar{U}_2\, \cos\theta_{\bk+\bq} \cos\theta_{\bp},\\ V_{h,X}^{\text{ex}}&=U_2 \, \cos\theta_{\bk+\bq} \cos\theta_{\bp}+\bar{U}_2\, \sin\theta_{\bk+\bq} \sin\theta_{\bp}
\end{align}
 \begin{align}
V_{e,e}^\text{den} &=U_5,\, V_{X,Y}^\text{den}= \tilde{U}_5,\, V_{X,Y}^\text{ex}=\tilde{\tilde{U}}_5,\, V_{X,Y}^\text{ph}=\bar{U}_5/2
\label{Vhx_a}
 \end{align}
\section{Derivation of the nematic instability equation}
\label{Appendix B}
\begin{figure}
 \centering
    \includegraphics[width=.72\textwidth]{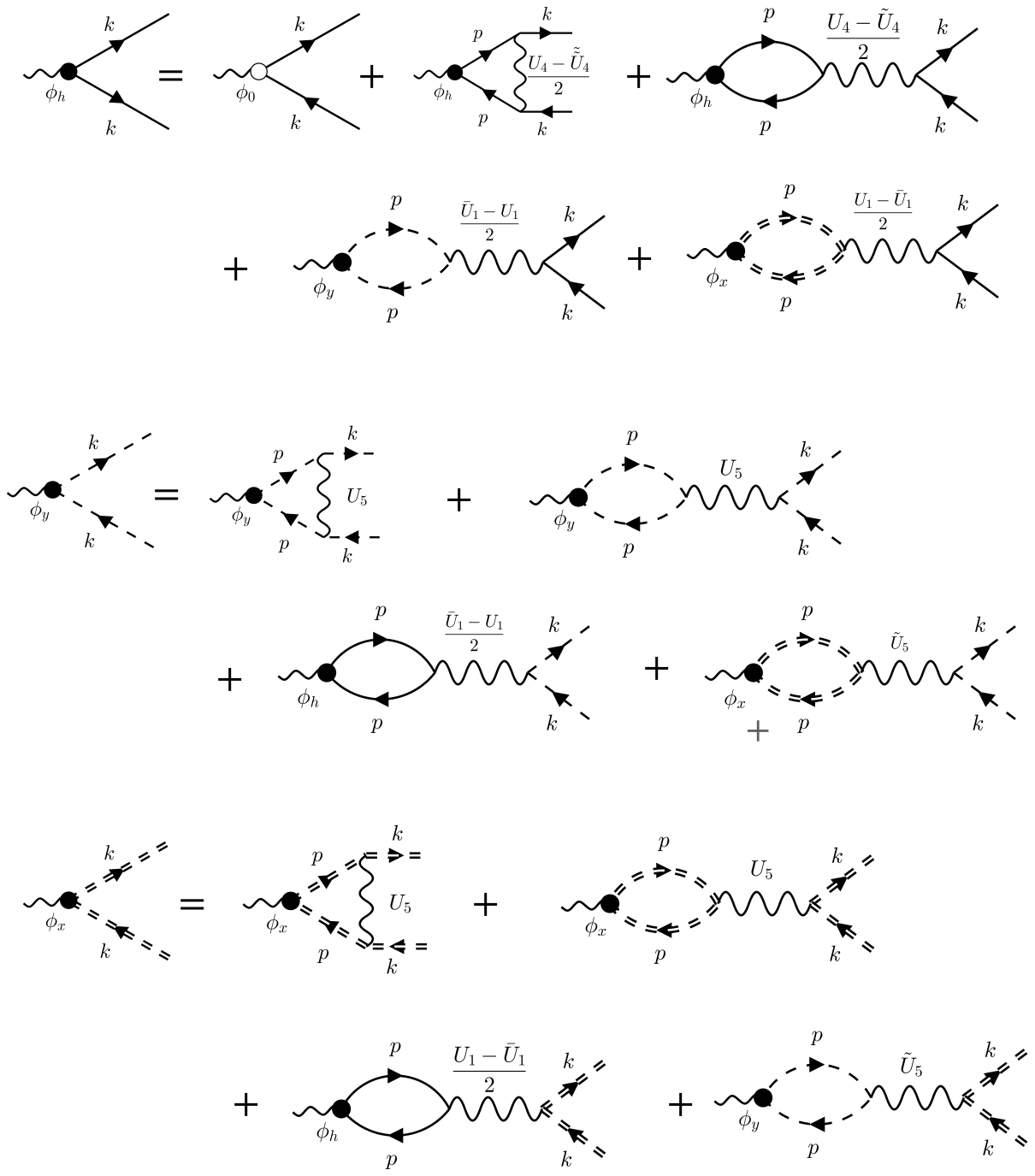}
\caption{Set of ladder equations for the nematic order vertex for hole pocket($\phi_h)$, $X-$pocket($\phi_x$) and $Y-$ pocket($\phi_y$). We label only the d-wave component of the interactions. The  solid line, dashed line and double dashed line represents Green's function for the hole pocket, $X-$ pocket and $Y-$ pocket fermion. The curvy line represents interaction between fermions. Black dot represent the nematic vertex while the void dot represent the bare vertex for the hole pocket only. We keep both the external fermion momentum $\bk$ same.}.
\label{nematic order diagram}
 \end{figure}

We derive the mean field self-consistent equations for the nematic order parameters $\phi_h, \phi_e (=\phi_x-\phi_y)$ defined in Eqs. (\ref{phih equation1}-\ref{phie equation1}) in Sec.\ref{Nematic susceptibility section}. They are obtained by adding up the Hartree and Fock diagrams (depicted in Fig. \ref{nematic order diagram}) for the hole and electron bands and turn out to be
 \begin{align}
 \phi_h \cos 2\theta_\bk  =\phi_0\, \cos 2\theta_\bk- \phi_h &\bigintss_p G_p^h G_{p}^h\,
\cos 2\theta_\bp  \,V_{h,h}^{\text{den}}\left(\bk,\bp;\bp,\bk\right) +2 \phi_h \bigintss_p G_p^h G_{p}^h
\,\cos2\theta_\bp \,V_{h,h}^{\text{den}}\left(\bk,\bk;\bp,\bp\right) \nn & +2 \phi_y \bigintss_p G_p^Y G_{p}^Y
\,V_{h,Y}^{\text{den}}\left(\bk,\bk;\bp,\bp\right)+2 \phi_x \bigintss_p G_p^X G_{p}^X
\,V_{h,X}^{\text{den}}\left(\bk,\bk;\bp,\bp\right),
\label{hole order eq}
 \end{align}
 \begin{align}
 \phi_y&=-\phi_y \bigintss_p G^Y_p G^Y_p\, U_5+2\, \phi_y\bigintss_p G^Y_p G^Y_p \, U_5+2\,\phi_h \bigintss_p G^h_p G^h_p \, \cos2\theta\bp\, V_{h,Y}^{\text{den}}(\bk,\bk;\bp,\bp)+2 \phi_x\,\bigintss_p G^X_p G^X_p \, \tilde{U}_5,
 \label{y order eq}
 \end{align}
  \begin{align}
 \phi_x&=-\phi_x \bigintss_p G^X_p G^X_p\, U_5+2\, \phi_x\bigintss_p G^Y_p G^Y_p \,U_5 +2\,\phi_h \bigintss_p G^h_p G^h_p \, \cos2\theta\bp\, V_{h,X}^{\text{den}}(\bk,\bk;\bp,\bp)+2 \phi_y\,\bigintss_p G^Y_p G^Y_p \,  \tilde{U}_5.
 \label{x order eq}
 \end{align}
Here $G^i(\bp,\omega)=1/i \omega-\xi_i(\bp)$ is the Green's function for the pocket $i$ with $4-$ momentum $p=(\bp,\omega)$. $\bigintss_p$ stands for $T \sum_{\Omega_n} \bigintss \dfrac{d^2\bp}{(2\pi)^2}$, where $\Omega_n$ is the fermionic Matsubara frequency($\Omega_n=(2n+1)\pi T$), $\bp$ is the lattice momentum and $T$ is the temperature. Because orbital order is a zero momentum order, we only keep the low momentum transfer density-density interaction and ignore large momentum transfer exchange interaction in Eqs. (\ref{hole order eq}-\ref{x order eq}). Since nematic orders $\phi_h$ and $\phi_e$ are d-wave order, only the d-wave component of the density interactions( proportional to $\cos2\theta$ terms) will contribute in Eqs.~(\ref{hole order eq}-\ref{x order eq}).
 Using Eqs.~(\ref{Vhh_a}-\ref{Vhx_a}), one finds the $s-$ and $d-$ wave components of the density interactions
 \begin{align}
  V_{h,h}^{\text{den}}\left(\bk,\bp;\bp,\bk\right)&=\dfrac{U_4+\tilde{\tilde{U}}_4}{2}  +\dfrac{U_4-\tilde{\tilde{U}}_4}{2} \cos2\theta_\bk\cos2\theta_\bp+\dfrac{\tilde{U}_4+\bar{U}_4}{2}\sin2\theta_\bk \, \sin2\theta_\bp,
  \label{vhh_b}
 \end{align}
 \begin{align}
  V_{h,h}^{\text{den}}\left(\bk,\bk;\bp,\bp\right)&=\dfrac{U_4+\tilde{U}_4}{2} +\dfrac{U_4-\tilde{U}_4}{2} \cos2\theta_\bk\cos2\theta_\bp+\dfrac{\tilde{\tilde{U}}_4+\bar{U}_4}{2}\sin2\theta_\bk \, \sin2\theta_\bp,
 \end{align}
 \begin{align}
  V_{h,Y}^{\text{den}}\left(\bk,\bk;\bp,\bp\right)&=\dfrac{U_1+\bar{U}_1}{2}-\dfrac{U_1-\bar{U}_1}{2} \cos2\theta_\bp,
 \end{align}
 \begin{align}
  V_{h,X}^{\text{den}}\left(\bk,\bk;\bp,\bp\right)&=\dfrac{U_1+\bar{U}_1}{2}+\dfrac{U_1-\bar{U}_1}{2} \cos2\theta_\bp.
  \label{vhx_b}
 \end{align}.
 Using Eqs.~(\ref{vhh_b}-\ref{vhx_b}) into Eqs.~(\ref{hole order eq}-\ref{x order eq}), we find a set of two coupled self-consistent equations for $\phi_h$ and $\phi_e$,
 \begin{align}
\phi_h&= \phi_0 -\phi_h\,  U^d_h \, \Pi_h^d -\phi_e \, U^d_{he}\, \Pi_e \label{phih equation1 a app}\\
\phi_e&=-\phi_e\,  U^d_{e} \Pi_e-2 \phi_h \,  U^d_{he} \Pi_h^d.
\label{phie equation1 app}
 \end{align}

$U_h^d=(U_4-2\tilde{U}_4+\tilde{\tilde{U}}_4)/2$ and
$U_e^d =U_5-2 \tilde{U}_5$ are the d-wave component of the intra-pocket density-density interactions for hole and electron pockets, respectively while $U_{he}= (U_1-\bar{U}_1)$ is the d-wave component of the inter-pocket density-density interaction between hole and electron pockets. $\Pi_h^d =-\bigintss G_p^h \, G_p^h \cos^22\theta_\bp,$ and $ \Pi_e=-(1/2) \bigintss_p \left(G_p^X \, G_p^X + G_p^Y \, G_p^Y\right)$ are the polarization bubbles for the hole and the electron pockets. Within our convention, the polarization bubbles are positive $\Pi_h^d,\Pi_e>0$.
\section{Specific heat coefficient $\gamma(t)$}
\label{Appendix C}
\subsection{Derivation of $\gamma^\text{SC}(t)$ for NFMS}
We derive the temperature dependence of the superconducting component of the specific heat coefficient labeled as $\gamma^\text{sc}(T)$ for the NFMS at temperature $T$ and $T_c$ is the superconducting transition temperature. This component of the specific heat comes from the section of the Fermi surface where the superconducting gap does not vanish: $\theta_\bk<\theta_0(t)$ and other symmetric parts. It is given by the following expression (see Eq.~(\ref{scaled sp heat 1}))
\begin{align}
\gamma^\text{SC}(t)&=\dfrac{8}{t^3} \bigintss_{0}^\infty dx \bigintss_0^{\theta_0(t)} \dfrac{d\theta}{2\pi} \sech^2\Bigr(\dfrac{\sqrt{x^2+|\bar{\Delta}(\theta)|^2}}{2\,t}\Bigr) \Bigr[x^2+|\bar{\Delta}(\theta)|^2-\dfrac{t}{2} \dfrac{\partial |\bar{\Delta}(\theta)|^2}{\partial t}\Bigr]=\bigintss_0^{\theta_0(t)} \dfrac{d\theta}{2\pi} f_\text{CV}(t_\theta)
\label{sc sp heat 1}
\end{align}
where to simplify the notation we set $\theta_\bk = \theta$.
$t=T/T_c, \bar{\Delta}(\theta)= \Delta(\theta)/T_c$ and $\Delta(\theta)$ is the solution of the NFMS gap equation
 \begin{align}
1=g \, \cos^2 2\theta \bigintssss_0^\Lambda d\xi  \dfrac{\tanh\Bigr(\dfrac{\sqrt{\xi^2+|\Delta(\theta)|^2}}{2 T}\Bigr)}{\sqrt{\xi^2+|\Delta(\theta)|^2}},
\label{int4 appendix}
\end{align}
where $g$ is the pairing strength and $\Lambda$ is the upper cut-off for the pairing.  To progress further, we make an observation that if we scale the gap function $\Delta(\theta)$ and temperature $T$ by the local transition temperature $\bar{T}_c(\theta)=T_c\,\exp(-\tan^22\theta/g)$ (below this temperature $\Delta(\theta)$ becomes non-zero) such that $t_\theta=T/\bar{T}_c(\theta)$ and $\tilde{\Delta}(\theta)= \Delta(\theta)/\bar{T}_c(\theta)$, the local gap equation (\ref{int4 appendix}) transforms into an angle independent ordinary s-wave gap equation for $\tilde{\Delta}(t_\theta)$
\begin{align}
    \bigintssss_0^{\bar{\Lambda}} d\xi  \dfrac{\tanh \dfrac{\xi}{2}}{\xi}=\bigintssss_0^{\bar{\Lambda}} d\xi  \dfrac{\tanh\Bigr(\dfrac{\sqrt{\xi^2+|\tilde{\Delta}(t_\theta)|^2}}{2 t_\theta}\Bigr)}{\sqrt{\xi^2+|\tilde{\Delta}(t_\theta)|^2}}.
    \label{int5 appendix}
\end{align} where $\bar{\Lambda}=\Lambda/T_c(\theta)$ is the renormalized upper cut-off. This greatly simplifies the situation as the solution of the NFMS gap equation (\ref{int4 appendix}) can be borrowed from ordinary s-wave solution (\ref{int5 appendix}) below the local temperature $t_\theta$. Performing the same set of transformations for the specific heat coefficient, we find
\begin{align}
     \gamma^\text{SC}(t)&=8 \bigintss_0^{\theta_0(t)}\dfrac{d\theta}{2\pi}\, \left[\dfrac{1}{t_\theta^3}\bigintss_{0}^\infty dx \sech^2\Bigr(\dfrac{\sqrt{x^2+|\tilde{\Delta}(t_\theta)|^2}}{2 t_\theta}\Bigr) \Bigr[x^2+|\tilde{\Delta}(t_\theta)|^2-\dfrac{t_\theta}{2} \dfrac{\partial |\tilde{\Delta}(t_\theta)|^2}{\partial t_\theta}\Bigr]\right]
     \label{gamma sc appendix}
 \end{align}
 The term inside the bracket $[\cdots]$ in Eq.~(\ref{gamma sc appendix}) is nothing but the specific heat coefficient of an ordinary s-wave superconductor $\gamma_\text{s}(t)$ which we compute and plot in Fig. \ref{specific heat fig app}(b) as a function of temperature.  This completes the proof for Eq.~(\ref{sc cv})
 \begin{align}
     \gamma^\text{SC}(t)=8 \bigintss_0^{\theta_0(t)}\dfrac{d\theta}{2\pi}\, \gamma_\text{s}(t_\theta)=\dfrac{\sqrt{g}}{\pi} \bigintss_t^1 \dfrac{\gamma_s(x)}{x\, (1+g\, \log \dfrac{x}{t})\,\sqrt{\log\dfrac{x}{t}}}\approx \sqrt{\dfrac{g}{\pi}}\, \gamma_s(t)\, \text{Erfi}[\sqrt{|\log t|}],
     \label{gamma sc}
 \end{align}
 where $\text{Erfi}(z)$ is the imaginary error function $\text{erf(I\, z)}$ with $\text{erf}(z)=2/\sqrt{\pi}\int_0^z dt \, e^{-t^2}$.

\subsection{Low temperature dependence of specific heat coefficients for a generic nodal superconductor}
The response of a superconductor at low temperature is dictated by gap structure near its minima. We consider a generic nodal superconductor where the gap amplitude varies in power-law variation ($\delta\theta $) with deviation from the nodal point such as $|\Delta(\delta \theta)|= A\, |\delta \theta|^\alpha$ where $A$ is a parameter. Specific examples of such situations are: ordinary s-wave with $\alpha=0$, ordinary d-wave with $\alpha=1$, and NFMS with $\alpha=\infty$. We derive how the low temperature behavior of specific heat coefficient $\gamma(t)$ depends on the generic exponent $\alpha$. We ignore the last derivative term of Eq.~(\ref{scaled sp heat 1}) as it is negligible at low temperature, and express the specific heat coefficient as
 \begin{align}
     \gamma(t)& =\dfrac{1}{t^3}\bigintss_0^{2\pi} \dfrac{d\theta}{2\pi} \bigintss_{\bar{\Delta}(\theta)}^\infty dE \dfrac{\sech^2 \dfrac{E}{2\, t}\, E^3}{\sqrt{E^2-|\bar{\Delta}(\theta)|^2}}\approx \dfrac{2 \sqrt{2}}{t^3} \bigintss_0^{2\pi} \dfrac{d\theta}{2\pi} |\bar{\Delta(\theta)}|^{5/2} \bigintss_{\Delta(\theta)}^\infty dE \dfrac{e^{-E/t}}{\sqrt{E-|\bar{\Delta}(\theta)|}} \\ & \approx \bigintss_0^{2\pi} \dfrac{d\theta}{2\pi}\, e^{-S(\bar{\Delta}(\theta)/t)}.
     \label{low t expansion}
 \end{align}
In Eq.~(\ref{low t expansion}), we define the action $S=\left(\dfrac{|\bar{\Delta}(\theta)|}{t}-\dfrac{5}{2}\log \dfrac{|\bar{\Delta}(\theta)|}{t}\right)$. The minimum of the action $S$ comes where $\bar{\Delta}(\theta)=5\,t/2$, which at low temperature corresponds to the region where the nodes. Employing the power-law behavior in this regime, we compute $\gamma(t)$ within saddle point approximation. A straightforward calculation gives $ \gamma(t)\propto t^{1/\alpha}$. This shows that more the suppressed gap is near the nodes, steeper the specific heat coefficient $\gamma(t)$ is at low temperature.  Translating this into the three classes of superconductor, we get the qualitatively correct results at low temperature (verified from numerics): for s-wave($\alpha=0$): $\gamma_s(t)=0$, d-wave($\alpha=1$): $\gamma_d(t)\propto t$ and NFMS ($\alpha=\infty$): $\gamma_\text{NFMS}(t)\approx $ constant.

\section{Angular profile of field induced specific heat coefficient $\delta\gamma(H,\phi)$}
\label{Appendix D}
We discuss the angular variation of magnetic field induced specific heat coefficient $\delta\gamma(H,\phi)$ as a function of in-plane field direction $\phi$ (defined with respect to the $\Gamma-X$ axis) for both ordinary d-wave and NFMS. $\delta\gamma(H,\phi)$ is defined as $\theta$ angular integration over the integrand $G$, Eq.~\eqref{Sp mag 1} and expressed as
 \begin{align}
     G(\theta,\phi,H)=1/\sqrt{1+\dfrac{1}{\bar{H}} \dfrac{ f^2(\theta)}{\sin^2(\theta-\phi)}}
     \label{G integrand}
 \end{align} where $\bar{H}=H/H_0$ and $f(\theta)$ is the gap function ($\cos 2 \theta$ for d-wave and $\exp(-\tan^22\theta/g)$ for NFMS). We particularly focus on the two extreme limits where some calculation on qualitative ground is possible: low field $H << H_0$ and large field $H>> H_0$ and analyze our numerical results from Fig. \ref{sp heat with Magnetic field app}. We first focus on the d-wave and qualitatively derive the results of Refs. (\citep{volovik1993superconductivity,vekhter1999anisotropic,vorontsov2006nodal})  For convenience, we keep $0\leq\phi\leq\theta_c$ for the rest of the analysis where $\theta_c=\pi/4$. At small magnetic field ($H/H_0 \ll 1$), the integrand $G$ is negligibly small for almost everywhere on the $\theta$ space, except near the cold spots where
 \begin{align}
     0 \leq |f_d(\theta)| \leq c\, \sqrt{\bar{H}}\, |\sin(\theta-\phi)|
     \label{width eqution}
 \end{align} satisfies with $c=O(1)$ is some chosen factor. Assuming the integrand $G$ can be approximated as a Lorentzian function peaked at these $4$ cold spots: $\theta_{l}=\theta_c+l \pi/2, l=0-3$ in the limit of low field strength $\bar{H}\ll 1$, we write it near these points as:
 \begin{align}
     G(\theta,\mathbf{H})= \dfrac{\Gamma_d^2(\theta_{l},\mathbf{H})}{(\theta-\theta_{l})^2+\Gamma_d^2(\theta_{l},\mathbf{H})},
 \end{align} where $\Gamma_d(\theta_l)$ is the width of the Lorentzian function peaked at the cold spot $\theta_l$ and an explicit function of the magnetic field $\mathbf{H}$.  we derive the expression for the width of the Lorentzian from Eq.~(\ref{width eqution}) by expanding the gap function $f_d(\theta)$ near the cold spots and find
 \begin{align}
 \label{width at c}
    \Gamma_d(\theta_c) &=c \sqrt{\bar{H}} \dfrac{|\sin(\theta_c-\phi)|}{|f_d'(\theta_c)|}\\
    \Gamma_d(\theta_c+\pi/2) &=c \sqrt{\bar{H}} \dfrac{|\cos(\theta_c-\phi)|}{|f_d'(\theta_c)|}
    \label{width at cpi}
 \end{align} where $\Gamma_d(\theta_c+\pi)=\Gamma_d(\theta_c)$, $f_d'$ is the derivative of gap function at the cold spot, and we suppress the $\mathbf{H}$ dependence of $\Gamma$ for the simplicity of the notation. At $\phi=0$, the width $\Gamma$ is same for all $4$ cold spots. As the field direction $\phi$ deviates from $0$, the width around the cold spots at $\theta_c$ and $\theta_c+\pi$ shrinks while the width around the other two cold spots increases. From Eqs. (\ref{width at c}-\ref{width at cpi}), one finds that the $\Gamma_d(\theta_c)$ decreases much faster compared to $\Gamma_d(\theta_c+\pi/2)$ as $\phi$ approaches $\theta_c$.  At $\phi=\theta_c$, only the cold spots at $\theta_c+\pi/2$ and $\theta_c+3\pi/2$ contributes. This makes $\delta\gamma(H,\phi)$ vary with $\phi$ as,
 \begin{align}
 \delta\gamma(H,\phi)=\Gamma_d(\theta_c)+\Gamma_d(\theta_c+\pi/2)=c\sqrt{\dfrac{\bar{H}}{f_d'}} \left(|\sin(\theta_c-\phi)|+|\cos(\theta_c-\phi)|\right)
 \end{align} which is functionally same as (\citep{volovik1993superconductivity,vekhter1999anisotropic,vorontsov2006nodal}), and qualitatively capture the numerical result (see Fig. \ref{sp heat with Magnetic field app}c). For the NFMS, one has to be more careful. Following the same line of logic, one finds in this case $f_\text{NFMS}'(\theta_c)=0$ (in fact all derivatives are zero at the cold spot) and does not translate the same result like the d-wave. In fact, we argue that the non-monotonic nature (presence of a local maxima at the nodal spots, Fig. \ref{sp heat with Magnetic field} of $\delta\gamma$ at low field strength for NFMS is the direct consequence of the exponentially suppressed gap structure ($f_\text{NFMS}^n(\theta_c)=0$ where $n$ represents the order of derivative) at the cold spot. Like the d-wave scenario, for NFMS, at low field strength, most of the contribution to the integrand $G$ comes near the nodal regions and with changing $\phi$ towards the cold spot at $\theta_c$, the width of the integrand $G$ around the cold spot at $\theta_c$ decreases while the width around the cold spot at $\theta_c+\pi/2$ increases. For small $\phi$, the first effect dominates over the second one, and makes $\delta\gamma$ decline. But for NFMS, there is an additional effect: as $\phi$ getting closer to the cold spot ($\theta_c$), the shrinking of the integrand $G'$s width around $\theta_c$ slows down and the second effect wins at some intermediate value of $\phi^*$. This can be seen from the behavior of the integrand $G(\theta,\phi)$ near the angle $\theta=\phi$ where it declines through the divergence of the function
 \begin{align}
     \lim_{\theta\to  \phi}\dfrac{f(\theta)}{\sin(\theta-\phi)}= \lim_{\theta\to  \phi}\dfrac{f(\phi)}{\theta-\phi}.
 \end{align}
      As $\phi$ approaches $ \theta_c$, the exponentially suppressed residue $f(\phi)=\exp(-\tan^22\phi/g)$ dominates over the divergence of $1/(\theta-\phi)$, and the angular window within which the integrand $G$ vanishes near $\theta=\phi$ becomes progressively smaller. This makes the first effect  mentioned above weaker while the second effect (increment) does not get affected. As a result, $\delta\gamma$ increases between the $\phi=\phi^*$ and $\theta_c$.

 Next, we discuss the opposite limit: large magnetic field strength $H/H_0 \gg 1$. In this limit, the integrand $G(\theta,\phi)$ of Eq.~(\ref{G integrand}) is roughly equal to $1$ for most of the range of variable $\theta$, except near $\theta=\phi$ and $\phi+\pi$ where it deviates significantly from $1$ and vanishes at those points. Because cold spots does not play any role in this large field strength limit (compared to the low field strength case where the physics is dictated by the cold spots), the analysis for the d-wave and NFMS does not need separate formulation. We expand $G$ near $\theta=\phi$ (and $\phi+\pi$ which gives identical contribution) upto linear order in deviation  $G(\delta\theta,\phi)= \dfrac{\sqrt{\bar{H}}}{|f(\phi)|}\, \delta\theta$ and approximate the width of this regime $\alpha$ using the condition $G(\alpha,\phi)=1$. This is justified as the slope $\sqrt{H}/|f(\phi)|$ is large. This leads to the expression
 \begin{align}
     \delta\gamma(H,\phi)=1-4\dfrac{\sqrt{\bar{H}}}{|f(\phi)|}\bigintss_0^\alpha \dfrac{d\delta \theta}{2\pi}  \delta\theta \, =1-\dfrac{\sqrt{\bar{H}}}{|f(\phi)|\, \pi}\alpha^2=1-\dfrac{|f(\phi)|}{\sqrt{\bar{H}}\pi}
     \label{N at large H}
 \end{align}

Eq.~(\ref{N at large H}) captures the numerical result (Fig. \ref{sp heat with Magnetic field app}(b,d)) qualitatively. Since $f(\phi)= \cos 2\phi (\text{d-wave})$ or $\exp(-\tan^22\phi/g)$ (NFMS) decreases from anti-nodal points and vanishes at the nodal points, residual DOS, $\delta\gamma(H,\phi)$ is minimum at the anti-nodal points and maximum at the nodal points.
\section{Comparison with s and d-wave superconductors}
\label{Appendix E}
We outline the differences a nematic fluctuation mediated superconductor (NFMS) has with other BCS superconductors. We particularly focus on an isotropic s-wave superconductor :$\Delta(\bk,T)=\Delta_s(T)$ and a point nodal d-wave superconductor: $\Delta(\bk,T)=\Delta_d(T)\, \cos2\theta_\bk$ at temperature $T$. The temperature evolution of the gap amplitudes $\Delta_{s,d}(T)$ below the transition temperature $T_c$ can be found by solving the gap equation: \begin{align}
    1&= g \, \bigintss_{0}^{\Lambda} d\xi \bigintss_0^{2\pi} \dfrac{d\theta}{2\pi} \, f_i^2(\theta)\dfrac{\tanh \dfrac{\sqrt{\xi^2+\Delta^2_i(T) f^2_i(\theta)}}{2 T}}{\sqrt{\xi^2+\Delta^2_i(T) f^2_i(\theta)}},
    \label{Gap Equation App}
\end{align} where $g$ is the pairing strength, $f_s(\theta)=1$, $f_d(\theta)=\cos2\theta$ and $\Lambda$ is the upper cut-off for the pairing. We numerically solve the gap equation (\ref{Gap Equation App}) for the s- and d-wave, use the solution to compute observables and compare them with NFMS.

 \subsection{Gap Structure}
\begin{figure}[H]
 \centering
   \subfigure[]{\includegraphics[width= 0.45 \textwidth]{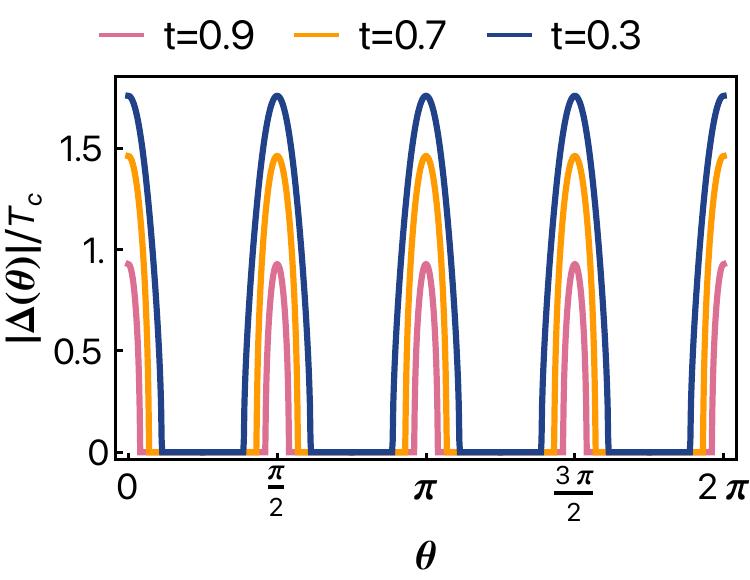}}
    \subfigure[]{\includegraphics[width= 0.45 \textwidth]{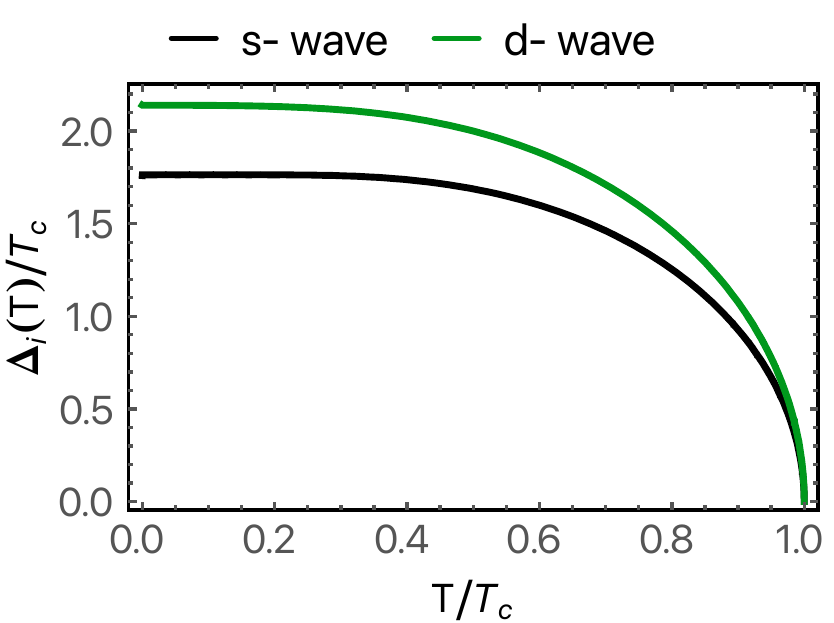}}\hspace{1 cm}
\caption{(a) Angular variation of gap amplitude $\Delta(\theta)$ of NFMS for a set of temperatures $T$. (b) Temperature dependence of the gap amplitude of a s-wave ($\Delta_s$) and d-wave ($\Delta_d$) superconductor, respectively. In all three cases, the gap is scaled by the corresponding transition temperature $T_c$. }.
\label{evolution of gap app}
 \end{figure}

 For the conventional s- and d-wave, only the gap amplitudes $\Delta_{s,d}(T)$ changes while for the NFMS, both the gap amplitude and angular variation evolves with temperature below the transition temperature $T_c$. We show this in Fig. \ref{evolution of gap app} (a,b) where we plot the full temperature dependence of $\Delta_{s,d}(T)$ below $T_c$ (a) and plot gap structure for NFMS for a set of temperatures (b). We also outline the differences in the functional dependence of gap structure near $T\approx T_c$ and $T=0$ for these three superconductors in Table. \ref{Table gap app}.
     \begin{table}[H]
\begin{center}
\begin{tabular}{ |c|c|c| }
 \hline
 $\Delta_i(T)/T_c$ & $T \approx T_c$ & $T=0$ \\
 \hline
  s-wave &  $\sqrt{\dfrac{8 \pi^2}{7\, \text{Zeta}(3)}}\, \left(1-\dfrac{T}{T_c}\right)^{1/2} $ &  $\approx 1.76$\\
 \hline
 d-wave & $\sqrt{\dfrac{4}{3}} \sqrt{\dfrac{8 \pi^2}{7\, \text{Zeta}(3)}}\, \left(1-\dfrac{T}{T_c}\right)^{1/2} $ &  $\approx 2.14$ \\
 \hline
 NFMS & \begin{tabular}{@{}c@{}} zero everywhere on the Fermi surface  \\ except near the hot spots with a width $\theta_0(T)$ \\ and maximum amplitude $\sqrt{\dfrac{8 \pi^2}{7\, \text{Zeta}(3)}}\, \left(1-\dfrac{T}{T_c}\right)^{1/2} $  \end{tabular}  &  $\approx 1.76 \exp^{-\tan^22\theta/g}$ \\
 \hline
\end{tabular}
\caption{Dependence of gap functions for s-wave, d-wave and NFMS on temperature in different limits: (i) near the transition temperature $T\approx T_c$, (ii) at zero temperature. }
\label{Table gap app}
\end{center}
\end{table}

\subsection{Specific heat}
\begin{figure}[H]
  \centering
 \subfigure[]{\includegraphics[width = 0.45 \textwidth]{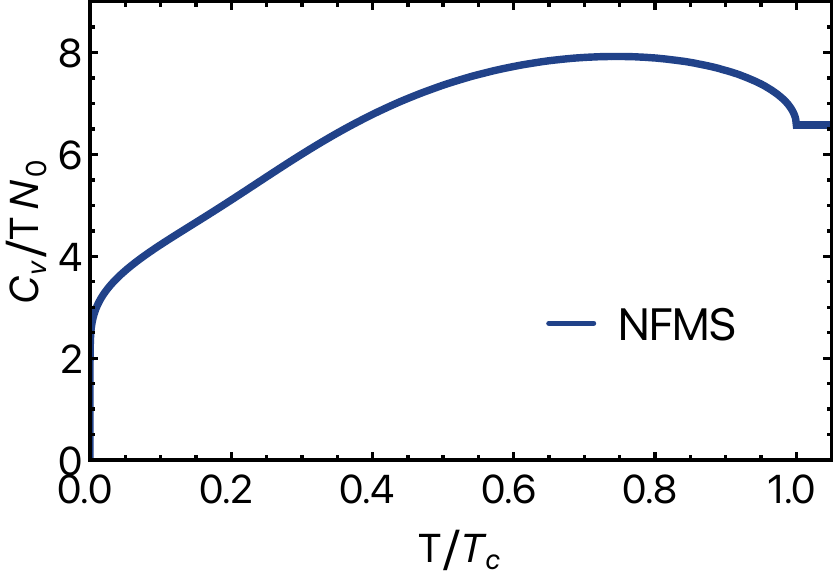}} \hspace{1 cm}
 \subfigure[]{\includegraphics[width = 0.45 \textwidth]{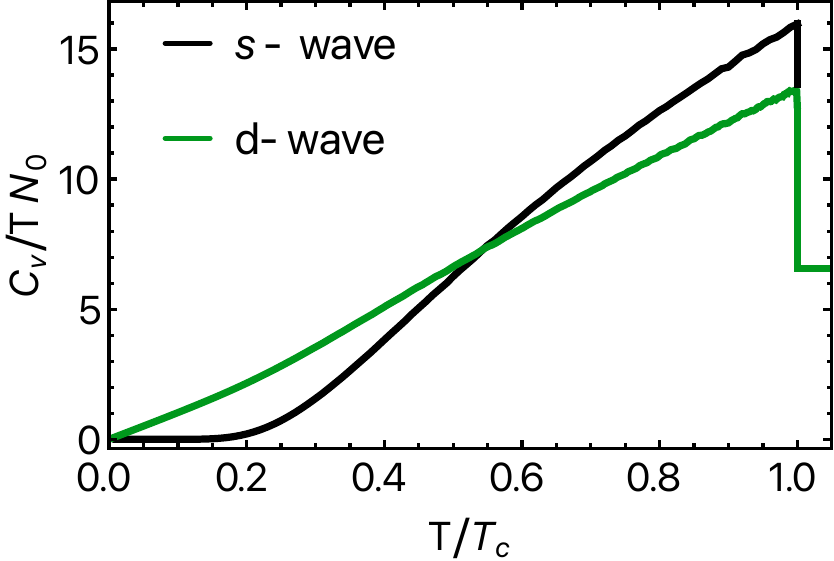}}
  \caption{ Temperature dependence of specific heat coefficient, $C_v/T\, N_0$ for NFMS (a), and  s- and d-wave superconductors (b) below the transition temperature $T_c$.}
\label{specific heat fig app}
 \end{figure}

We compute the specific heat coefficient $\gamma(T)$ (Eq.~\eqref{scaled sp heat 1}) for the s- and d-wave as a function of temperature and plot them along with the result for NFMS in Fig. \ref{specific heat fig app}(a,b). The most prominent differences among these three superconductors lie for the specific heat jump at the transition point and functional behavior at low temperature. As we pointed out before, for NFMS there is no specific heat jump at $T_c$, while for s-wave and d-wave, there is a finite jump with magnitude $\approx 1.43$ and $0.95$ approximately. At low temperature, for s-wave, $\gamma(t)$ is exponentially suppressed while for d-wave it increases linearly with $T$. We outline these differences quantitatively in Table. \ref{Table cv app}.
\begin{table}[H]
\begin{center}
\begin{tabular}{ |c|c|c| }
 \hline
 $\gamma(T)$ & jump at $T =T_c$ & behavior at $T\ll T_c$ \\
 \hline
  s-wave &  $ \dfrac{8 \pi^2}{7\, \text{Zeta}(3)} \approx 1.43 $&  $(\Delta_s(0)/T)^{5/2}\, \exp(-\Delta_s(0)/T) $\\
 \hline
 d-wave & $ \dfrac{2}{3} \dfrac{8 \pi^2}{7\, \text{Zeta}(3)} \approx 0.95 $ & $\propto T $ \\
 \hline
 NFMS & 0 & $\propto 1/\sqrt{g\log T/T_c}$ \\
 \hline
\end{tabular}
\caption{Behavior of specific heat coefficient $\gamma(T)$ in different temperature regimes: at the transition temperature $T=T_c$ and at low temperature for s-wave, d-wave and NFMS.}
\label{Table cv app}
\end{center}
\end{table}

\subsection{Specific heat in a magnetic field}
\begin{figure}
 \centering
    \subfigure[]{\includegraphics[width= 0.42 \textwidth]{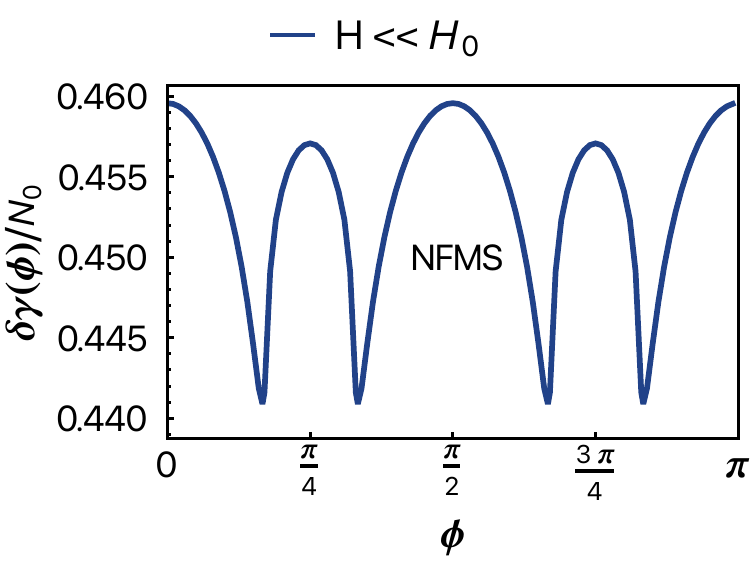}}\hspace{1 cm}
   \subfigure[]{\includegraphics[width= 0.42 \textwidth]{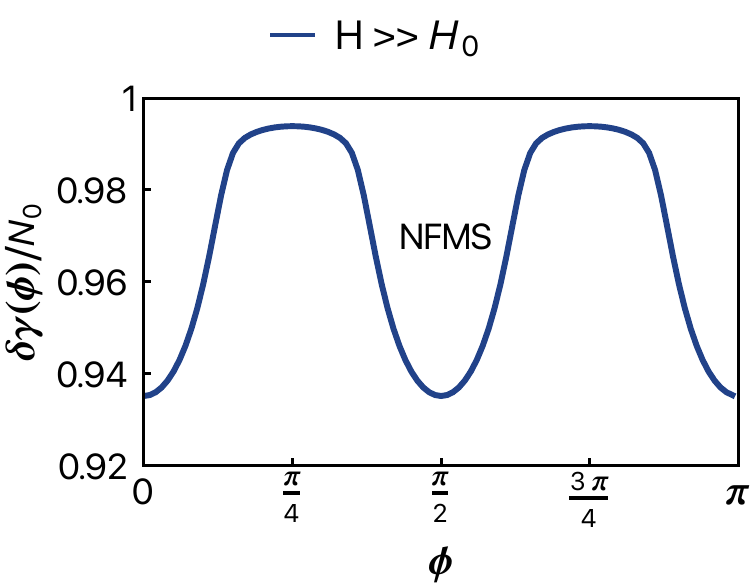}}
    \subfigure[]{\includegraphics[width= 0.42 \textwidth]{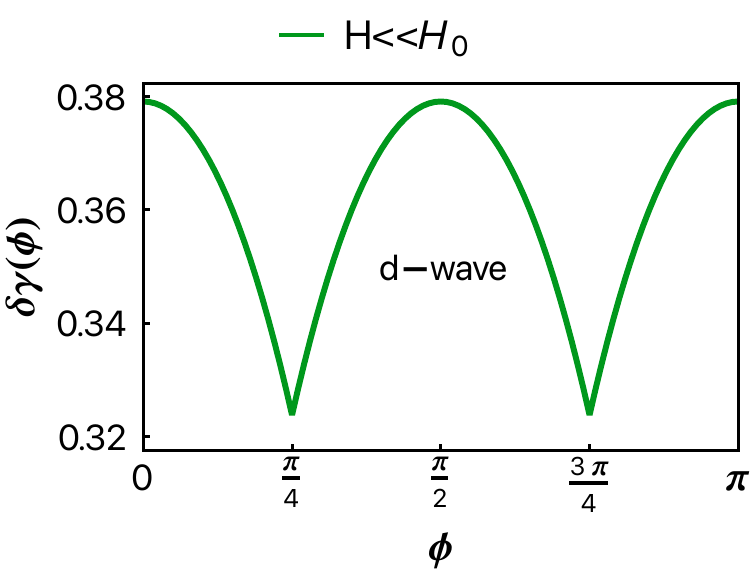}}\hspace{1 cm}
   \subfigure[]{\includegraphics[width= 0.42 \textwidth]{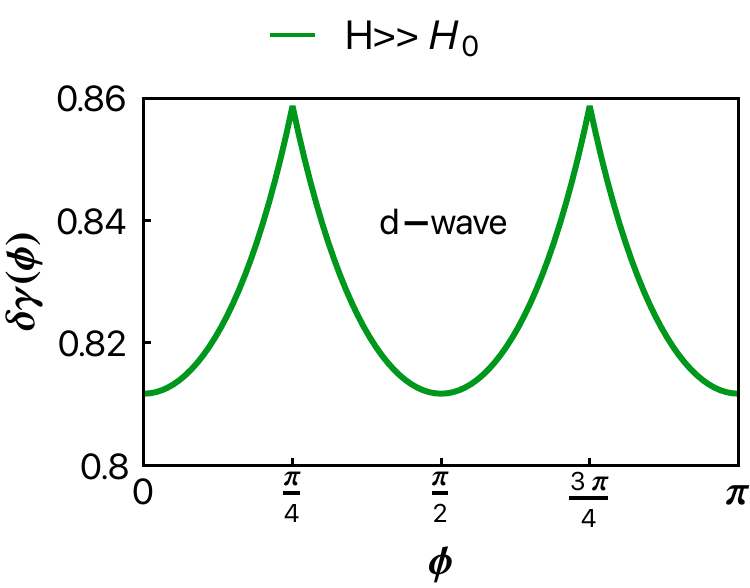}}
\caption{Field induced modulation of the specific heat coefficient $\delta\gamma(H,\phi)$ with varying direction $\phi$ of an in-plane magnetic field $\mathbf{H}$ counted from
  $\Gamma-X$ axis for a NFMS
  at a small field strength $H\ll H_\text{0}$ (a) and large field strength $H\gg H_\text{0}$ (b).
 and a d-wave superconductor at (c) a small field strength $H\ll H_\text{0}$, (d) large field strength $H\gg H_\text{0}$.}
\label{sp heat with Magnetic field app}
 \end{figure}

\begin{figure}[H]
 \centering
    \subfigure[]{\includegraphics[width= 0.45 \textwidth]{QNFMS2.pdf}}\hspace{1 cm}
   \subfigure[]{\includegraphics[width= 0.45 \textwidth]{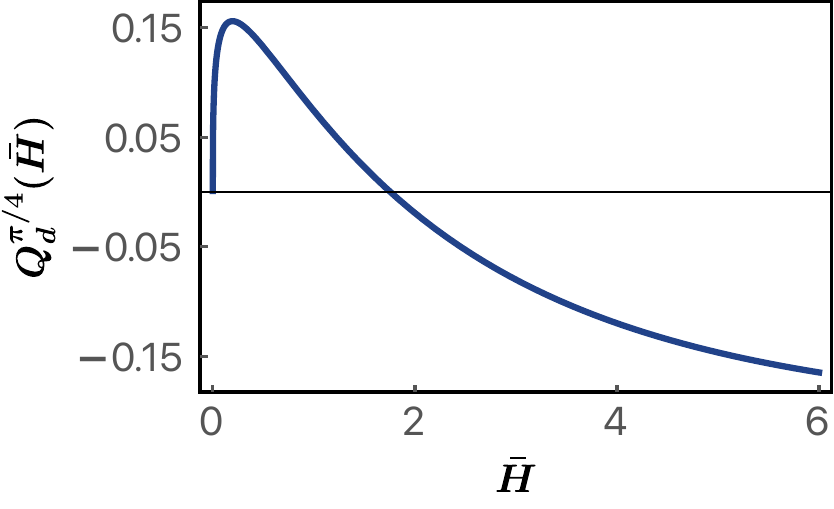}}
\caption{The field strength ($H$) dependence of (a) $Q^{\pi/4}_\text{NFMS}(H)$, coefficient of $(\phi-\pi/4)^2$ term in Taylor expansion of $\delta\gamma(H,\phi)$ for NFMS, and (b) $Q^{\pi/4}_\text{d}(H)$, coefficient of $|\phi-\pi/4|$ term in Taylor expansion of $\delta\gamma(H,\phi)$ for a d-wave gap near the field direction $\phi=\pi/4$.}
\label{QNFM plot}
 \end{figure}

We compare the field $\mathbf{H}=(H,\phi)$ dependence of the specific heat coefficient $\delta\gamma(H,\phi)$ at low temperature for the three classes of superconductors: s-wave, d-wave and NFMS using the Eq.~\eqref{Sp mag 1}. For an s-wave superconductor, the gap function is angle-independent and
 $\delta \gamma_s (H, \phi) = \dfrac{2\, N_0}{\pi} \arctan\sqrt{\bar{H}}$  is also angle-independent \citep{vekhter1999anisotropic}.
 For a $d-$wave superconductor, $f_d(\theta)=\cos 2\theta$ and $\delta \gamma_d (H, \phi)$ has a set of maxima and minima. This behavior is similar to the one in NFMS case, but there are two essential differences.  First, at $\phi =(\pi/4)(1+2n)$, $\delta \gamma_d (H, \phi)$  has a minimum at small $H <H_0$ and a maximum at $H >H_0$, while in our NFMS case, the maximum at
  $\phi =(\pi/4)(1+2n)$ persists for all $H$.  Second, near $\phi =(\pi/4)(1+2n)$,  $\delta \gamma_d (H, \phi)$ has a cusp-like behavior, i.e., is linear in a deviation from either a maximum or a minimum,  while in our case it is regular in deviations from $\phi =(\pi/4)(1+2n)$. We plot the $\phi$ dependence of $\delta\gamma$ for d-wave and NFMS in Fig. \ref{sp heat with Magnetic field app}.
The difference can be understood analytically.
 For a d-wave superconductor, the expansion in  $\epsilon =\phi -\pi/4$ yields
 \begin{equation}
 \delta \gamma_d (H, \phi) =  \frac{\sqrt{{\bar H}}}{\pi} \int_{-\pi/2}^{\pi/2}
 \frac{\sin{\theta}}{\sqrt{{\bar H} \sin^2{\theta}  + \sin^2{2\theta} + 2 \epsilon \sin{4\theta} + 4 \epsilon^2 \cos{4\theta}}}
 \end{equation}
 One can easily verify that  a formal expansion to order $\epsilon^2$ does not hold because the prefactor has a divergent contribution from  small $\theta$. Focusing on small $\theta \sim \epsilon$, introducing $\theta = |\epsilon| z$ and subtracting $\delta \gamma_d (H, \pi/4)$, we obtain
 \begin{equation}
 \delta \gamma_d (H, \phi) =\delta \gamma_d (H, \pi/4) + \left|\phi-\frac{\pi}{4}\right| Q_{d},
 \end{equation}
 where
  \begin{equation}
 Q_{d} ({\bar H}) = \frac{\sqrt{\bar{H}}}{\pi} \int_{0}^{\infty}  \left(\frac{ z}{\sqrt{{\bar H}z^2 + 4(z+1)^2}}  + \frac{ z}{\sqrt{{\bar H}z^2 + 4(z-1)^2}}- 2 \frac{1}{\sqrt{{\bar H} + 4}} \right)
 \end{equation}
 One can easily make sure that the integral converges, i.e., $Q_d$ is finite.  This implies that $\delta \gamma_d (H, \phi)$ has a cusp-like behavior at $\phi = \pi/4$ and related $\phi$.  We plot $Q_\text{NFMS}(\bar{H})$ (defined in Eq.~\eqref{qnfms2}) and $Q_d(\bar{H})$ in Fig.  \ref{QNFM plot}(a,b) and show their qualitatively different field strength dependence. We see that $Q_d ({\bar H})$ is positive for small ${\bar H}$, i.e., for $H < H_0$, but changes sign and becomes negative for large ${\bar H}$, i.e., for $H > H_0$. This explains why $\delta \gamma_d (H, \pi/4)$  transforms from a minimum to a maximum as $H$ increases. On other hand, $Q_\text{NFMS}(\bar{H})$ is negative for all $H$ and makes $\delta\gamma_\text{NFMS}(H,\pi/4)$ to be maximum always.
\subsection{Spin Susceptibility:}
\begin{figure}[H]
 \centering
   \subfigure[]{\includegraphics[width= 0.45 \textwidth]{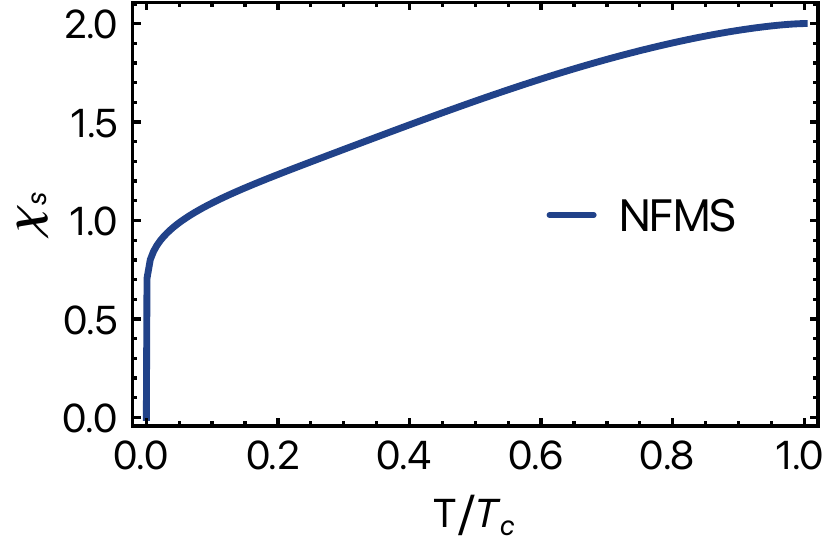}}\hspace{1 cm}
   \subfigure[]{\includegraphics[width= 0.45 \textwidth]{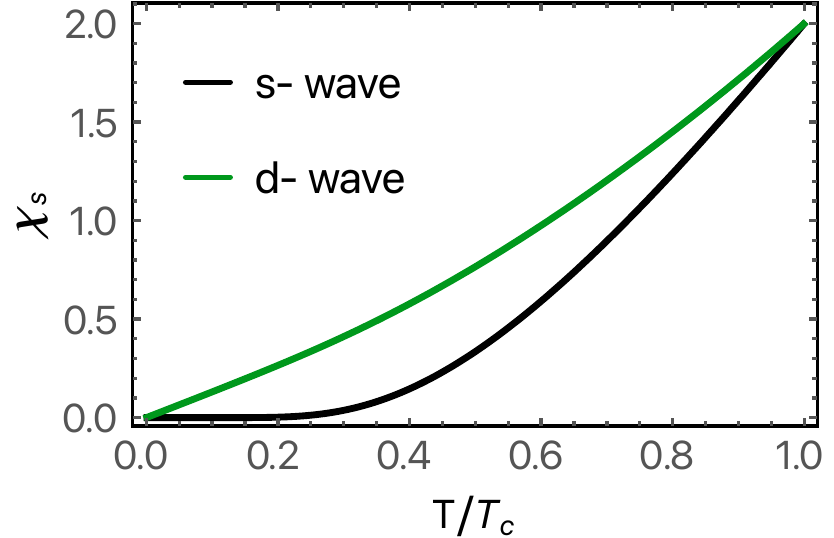}}
   \caption{Uniform static spin susceptibility $\chi_s$ as a function of temperature $T$ for NFMS (a), and s-wave and d-wave (b) below the transition temperature $T_c$.}
\label{Spin Susceptibility fig app}
   \end{figure}

We compute uniform spin susceptibility $\chi_s(T)$ (Eq.~\eqref{susceptibility eq}) for the s- and d-wave as a function of temperature $T$ and plot them along with the result for NFMS in Fig. \ref{Spin Susceptibility fig app}(a,b). Near $T_c$, $\chi_s$ for s- and d-wave varies roughly linearly with temperature deviation $T_c-T$, while for NFMS, it varies with a power of $3/2$. At low temperature $T\ll T_c$, qualitatively different nodal structures give distinct temperature dependence for these three classes of superconductors. We outline these results quantitatively in Table.  \ref{Table spin susceptibility app}.
\begin{table}[H]
\begin{center}
\begin{tabular}{ |c|c|c| }
 \hline
 $\chi_s(T)/2\,N_0$ & $T \approx T_c$ & $T\ll T_c$ \\
 \hline
  s-wave &   $1-2\left(1- \dfrac{T}{T_c}\right)$ &   $ \left(\dfrac{2 \pi \Delta_0}{T}\right)^{1/2} \exp{-\dfrac{\Delta_s(0)}{T}}$ \\
 \hline
 d-wave & $1-\dfrac{4}{3}\left(1- \dfrac{T}{T_c}\right)$ & $\sqrt{2} \dfrac{T}{\Delta_d(0)}$ \\
 \hline
  NFMS & $ 1-(1-T/T_c)^{3/2}$ & $1-1/\sqrt{\log T/T_c}$ \\
  \hline
\end{tabular}
\caption{Dependence of spin susceptibility $\chi_s(T)$ on temperature $T$ for a s-wave, d-wave and NFMS in different regimes: (i) near the transition temperature $T\approx T_c$ and (ii) at low temperature $T\ll T_c$.}
\label{Table spin susceptibility app}
\end{center}
\end{table}
\subsection{Dynamical density of states and tunneling conductance}
\begin{figure}
 \centering
    \subfigure[]{\includegraphics[width= 0.45 \textwidth]{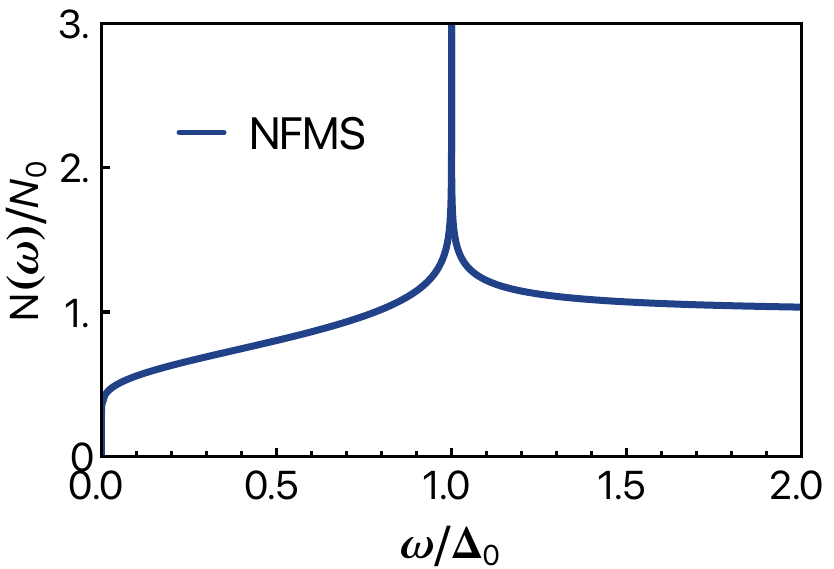}}\hspace{1 cm}
    \subfigure[]{\includegraphics[width= 0.45 \textwidth]{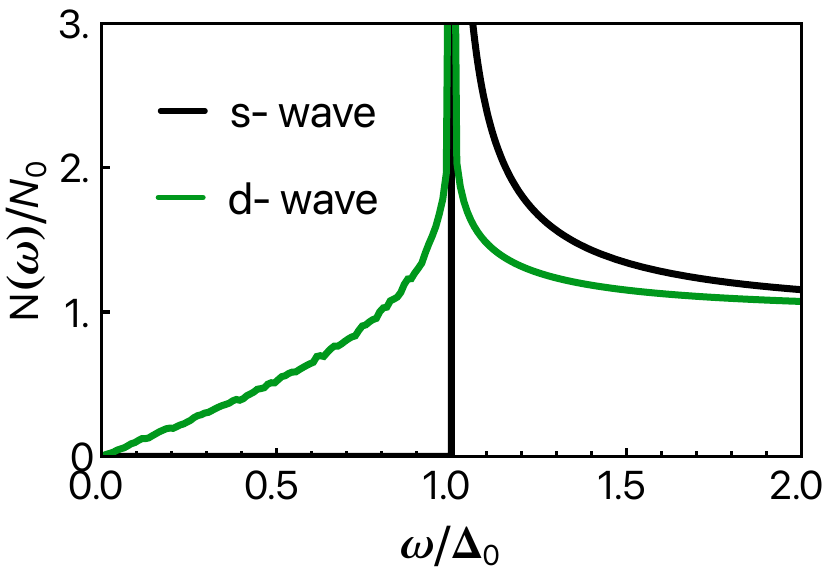}}
   \caption{Density of states, $N(\omega, 0)$) at zero temperature in units of $N_0$  as a function of energy $\omega$ for NFMS (a), and s-wave and d-wave superconductors (b). $\Delta_0$ is the gap amplitude.}
\label{DOS figure app}
   \end{figure}

\begin{figure}
 \centering
 \subfigure[]{\includegraphics[width= 0.45 \textwidth]{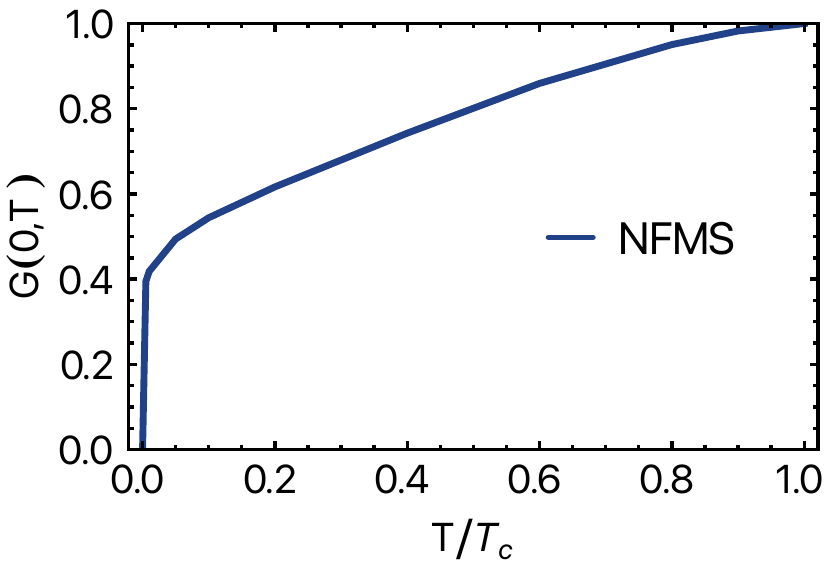}}
  \subfigure[]{\includegraphics[width= 0.45 \textwidth]{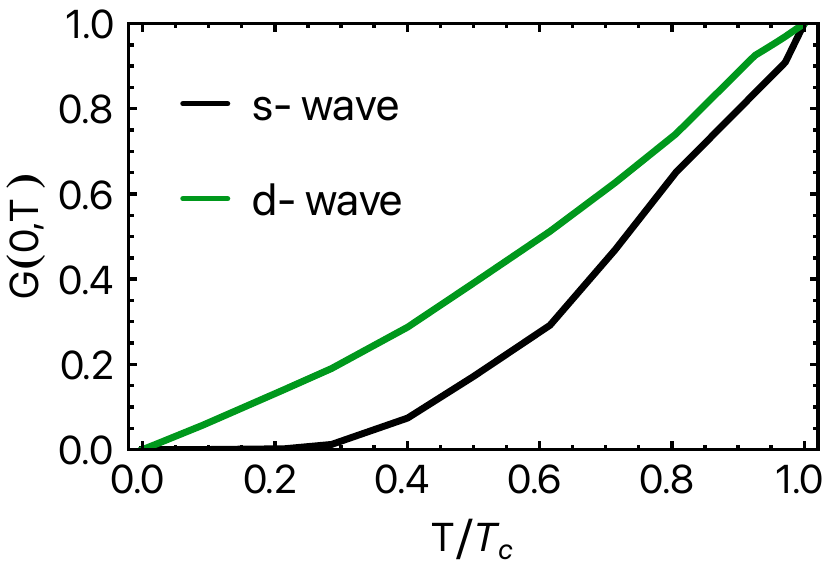}}
   \caption{Zero bias tunneling conductance $G(0,T)$ as a function of $T/T_c$ for (a) NFMS and  (b) s-wave and d-wave.}
\label{Zerobias figure app}
   \end{figure}

We compute density of states at zero temperature, $N(\omega, 0)\equiv N(\omega)$ (Eq.~\eqref{DOS1})  for the s- and d-wave as a function of energy $\omega$ and plot them along with the result for NFMS in Fig. \ref{DOS figure app}(a,b). For an isotropic s-wave gap, density of states has a simple expression $N(\omega)/N_0=\Theta(|\omega|-\Delta_0) \omega/\sqrt{\omega^2-\Delta_0^2}$. For d-wave, $N(\omega)\propto \omega$ at low energy because of the point nodal nature of the gap near the cold spots. On the other hand, close to $\omega =\Delta_0$, it behaves as $N(\omega)\propto \log(\omega/\Delta_0-1)$. For NFMS, the low energy behavior of density of state $N(\omega)\propto 1/\sqrt{\Delta_0/\omega}$, is exotic because of the exponentially small nature of the gap near the cold spots. Close to $\omega =\Delta_0$, density of states for NFMS is similar to the d-wave case as in both the cases gap structure near the hot spots are similar. However, in all cases (s-,d- and NFMS), $N(\omega)$ asymptotically approaches $N_0$ as $\omega$ goes to infinity. We outline these results quantitatively in Table.  \ref{Table density of states app}.
\begin{table}[H]
\begin{center}
\begin{tabular}{ |c|c|c| }
 \hline
 $N(\omega)$ & $\omega \ll \Delta_0$ & $\omega \sim \Delta_0+$ \\
 \hline
  s-wave &   $0 $ &   $ 1/\sqrt{\omega-\Delta_0}$ \\
 \hline
 d-wave & $\propto \omega$ & $\propto \log(\omega/\Delta_0-1)$ \\
 \hline
  NFMS & $ \propto 1/\sqrt{\log(\Delta_0/\omega)}$ & $\propto \log(\omega/\Delta_0-1)$ \\
  \hline
\end{tabular}
\caption{Approximate expression for $N(\omega)$ for a s-wave, d-wave and NFMS at small energy $\omega \ll \Delta_0$ and close to $\omega =\Delta_0$.}
\label{Table density of states app}
\end{center}
\end{table}
Next, we compute the zero bias tunneling conductance $G(0,T)$ (Eq.~\eqref{Zero Bias}) for s- and d-wave as a function of temperature $T$ and plot them along with the result for NFMS in Fig. \ref{Zerobias figure app}(a,b). For an isotropic s-wave gap, $G(0,T)$ is exponentially suppressed at low temperature $T\ll T_c$ because of the fully developed gap, while for d-wave it increases linearly with $T$ owing to the low-energy excitations coming from the point nodes. On the other hand, for NFMS, at low temperature $G(0,T)$ behaves the same way specific heat coefficient $\gamma(T)$ behaves. It vanishes at $T=0$, but very weakly as $1/\sqrt{T_c/T}$, and when extrapolated from low but finite temperature, it looks like a constant offset.
\subsection{Superfluid stiffness: }
\begin{figure}
 \centering
    \subfigure[]{\includegraphics[width= 0.45 \textwidth]{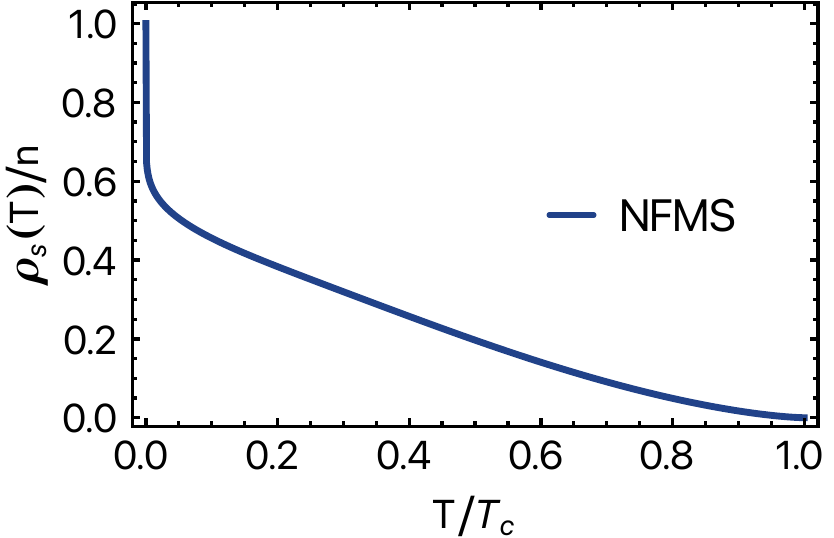}}
   \subfigure[]{\includegraphics[width= 0.45 \textwidth]{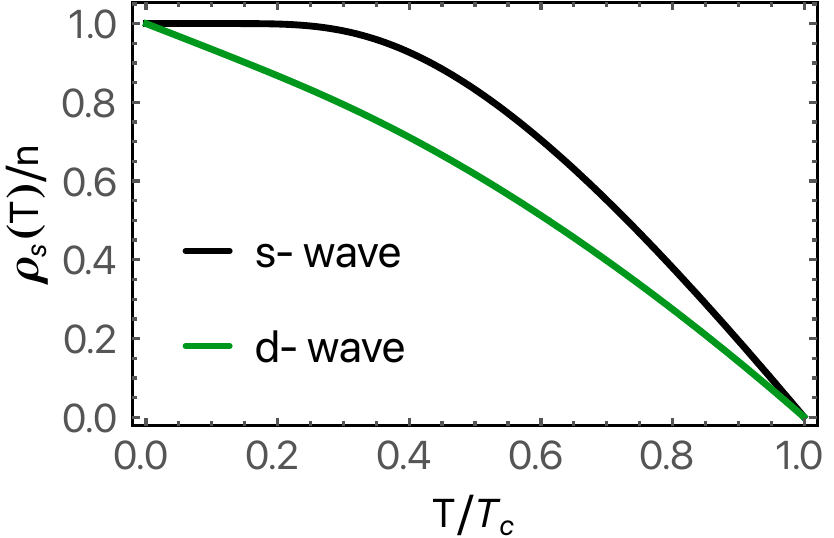}}
\caption{(a) Temperature variation of the superfluid density, $\rho_s$ normalized by the fermion density $n$ for NFMS (a), and s-wave and d-wave (b) gap below the corresponding transition temperature $T_c$.}
\label{SF pure fig app}
 \end{figure}

We compute superfluid density $\rho_s(T)$ (Eq.~\eqref{SF eq}) for the s- and d-wave as a function of temperature $T$ and plot them along with the result for NFMS in Fig. \ref{SF pure fig app}(a,b). We find that near $T_c$, $\rho_s$ for s- and d-wave varies roughly linearly with temperature deviation $T_c-T$, while for NFMS, it varies with a power of $3/2$. At low temperature $T\ll T_c$, qualitatively different nodal structures give distinct temperature dependence for these three classes of superconductors. We outline these results quantitatively in Table.  \ref{Table superfluid density}.
 \begin{table}[H]
\begin{center}
\begin{tabular}{ |c|c|c| }
 \hline
 $\rho_s(T)/n$ & $T \approx T_c$ & $T\ll T_c$ \\
 \hline
  s-wave &   $2\left(1- \dfrac{T}{T_c}\right)$ &   $ 1-\left(\dfrac{2 \pi \Delta_0}{T}\right)^{1/2} \exp{-\dfrac{\Delta_s(0)}{T}}$ \\
 \hline
 d-wave & $\dfrac{4}{3}\left(1- \dfrac{T}{T_c}\right)$ & $1-\sqrt{2} \dfrac{T}{\Delta_d(0)}$ \\
 \hline
 NFMS & $(1-T/T_c)^{3/2}$ & $1/\sqrt{g \log T/T_c}$ \\
 \hline
\end{tabular}
\caption{Theoretical expression for superfluid density $\rho_s(T)$ at different temperature regimes for NFMS, s-wave and d-wave gap: (i) near the transition temperature $T\approx T_c$ (ii) at low temperature $T\ll T_c$. }
\label{Table superfluid density}
\end{center}
\end{table}

 \subsubsection{Raman Intensity}
\begin{figure} [H]
 \centering
    \subfigure[]{\includegraphics[width= 0.47 \textwidth]{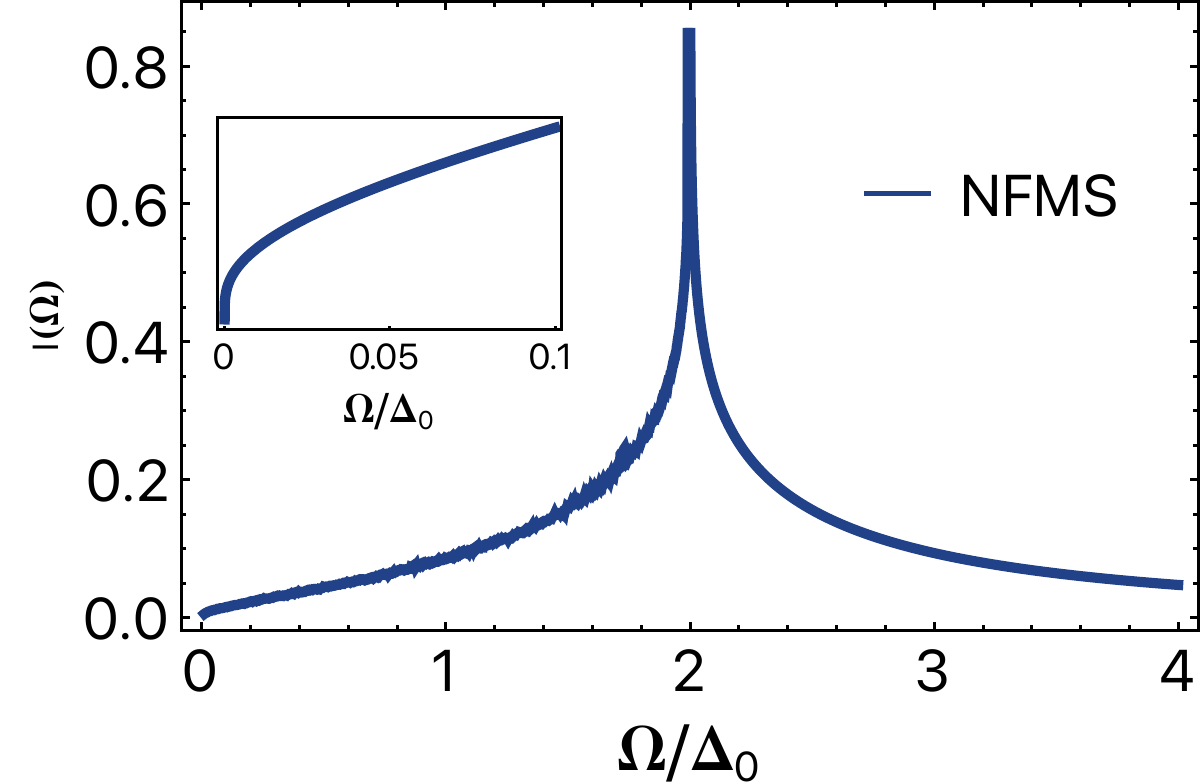}}\hspace{1 cm}
   \subfigure[]{\includegraphics[width= 0.47 \textwidth]{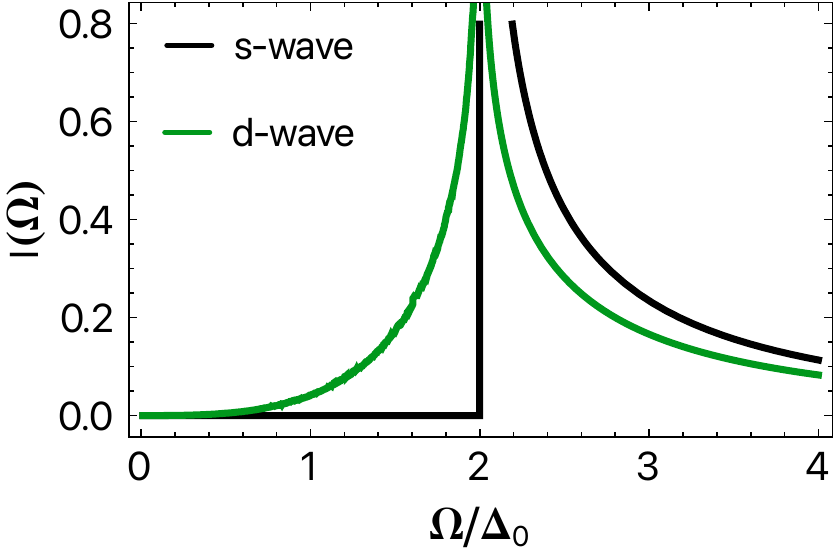}}
\caption{Frequency dependence of Raman Intensity $I(\Omega)$ in the $B_{1g}$ channel for NFMS (a) , and s- and d-wave superconductors (b). The low frequency behavior for NFMS is shown in the inset.}
\label{Raman figure App}
 \end{figure}

We compute Raman Intensity $I(\Omega)$ in the B$_{1g}$ channel as a function of external frequency $\Omega$ (Eq.~\eqref{Raman 2}) at zero tempertaure for the s- and d-wave and plot them along with the result for NFMS in Fig. \ref{Raman figure App}(a,b). At low frequency $\Omega \ll \Delta_0$ ($\Delta_0$ is the maximum gap amplitude), distinctly different nature of low-energy excitations give qualitatively different frequency dependence for $I(\Omega)$ for s-wave, d-wave and NFMS. On the other hand, near the cut-off $\Omega \gtrsim 2\Delta_0$, d-wave and NFMS behave similar way but very different from the s-wave. We outline this in Table. \ref{Table Raman Intensity}.
  \begin{table}[H]
\begin{center}
\begin{tabular}{ |c|c|c| }
 \hline
 $I(\Omega)$ & $\Omega \ll \Delta_0 $ & $\Omega \approx 2\Delta_0$ \\
 \hline
  s-wave &   $0$ &   $ \propto 1/\sqrt{\dfrac{\Omega}{2}-\Delta_0}$ \\
 \hline
 d-wave & $\propto \Omega^3$ & $\propto \log (\Omega-2\Delta_0)$ \\
 \hline
 NFMS & $ \propto \left(\log \dfrac{2\Delta_0}{\Omega}\right)^{5/2}$ & $\propto \log (\Omega-2\Delta_0) $ \\
 \hline
\end{tabular}
\caption{Approximate expression of Raman Intensity $I(\Omega)$ in $B_{1g}$ channel for NFMS, s-wave and d-wave  gap in different frequency regimes: (i)$\Omega \ll \Delta_0$ and (ii) $\Omega=2 \Delta_0$, where $\Delta_0$ is the corresponding gap amplitude.}
\label{Table Raman Intensity}
\end{center}
\end{table}
\section*{REFERENCES}
\bibliographystyle{apsrev4-2.bst}
\bibliography{biblio}

\end{document}